	   \documentclass[classic, hyperref,watermark,histinit,frontispiece,plain]{teipel-thesis-en} %%moder pone el formato lindo, pero los simbolos griegos no
\graphicspath{{figures/}}
\usepackage[scientific-notation=true]{siunitx}
\usepackage[english]{babel}
\usepackage{graphicx, color}
\graphicspath{{figures/}}
\usepackage{amsmath}
\usepackage{amscd}
\usepackage{youngtab}
\usepackage{titlesec}
\usepackage{listings}
\usepackage{enumerate}
\usepackage{bm}
\usepackage{amssymb}
\usepackage{amsthm}
\usepackage{syntonly}
\usepackage{fancyhdr}
\usepackage{cancel}
\usepackage{siunitx}
\usepackage{slashed}
\usepackage{bbm}
\usepackage{dsfont}
\usepackage{multirow}
\usepackage{gensymb}
\usepackage{cite} % to collet the references
\usepackage{fancyhdr}
\usepackage{dsfont}
\usepackage{float}
\usepackage{amsfonts}
\usepackage{booktabs}
\usepackage{siunitx}

\usepackage{fancyhdr}
	\UniversityName{University of Antioquia}
	\SchoolName{Physics Institute}
	\DepartmentName{Faculty of Exact and Natural Sciences \\
						{\large{Nuclear Physics Group (GFN-UdeA)}} }
	\UniversityLocation{Medellín}
	\ThesisType{\\Undergraduate Thesis}
	\title{ Structure of Heavy Nuclei Based on\\ Nucleon Quartets }
	\subtitle{Algebraic Approach in the Proxy-SU(3) Scheme}
	\authorName{Alejandro Restrepo Giraldo} 
	\supervisor{Dr. José Patricio Valencia Valencia}
	\supervisorTitle{}

	\thesisPlaceDate{Medellín, 2023}
	\ackPlaceDate{Medellín, 2023}
	\examinationDate{ }
	\approvalStatement{}
	\chaptercolor{brown!75!magenta}
	\appendixcolor{brown!80!magenta}
    	\hyperlinkcolor{red!55!brown}
    	\titlecolor{white}
    	\titlebackgroundcolor{brown!85!magenta} 
     \coverpageimage{0.045}{figures/Oval3}
     \vspace{2.0 cm}
	\UniversityLogoPath{figures/logo-udea}
	\UniversityBWLogoPath{figures/logo-udea.eps}
	\LogoScaleFactor{1.1}
	\draftIndication{}%poner algo debajo de superviso en la primera hoja
	\setFrontispiece[0.3]{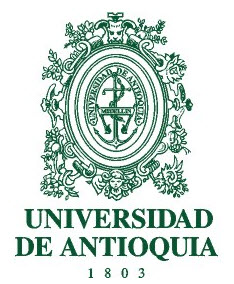}{University of Antioquia} %figura detras de la primera página ---> GUITARRA
\begin{document}
\maketitle
\beginfrontmatter

% Abstract (content in `abstract.tex')
	\begin{abstract}

In this thesis several topics concerning nuclear structure and electromagnetic interactions of heavy nuclei are reviewed. These comprehend the deformed single-particle shell model, nuclear collective motion, symmetry breaking and approximate symmetry restoring in atomic nuclei, particularly the proxy-$SU(3)$ scheme. The relations between these theoretical frameworks are stated and it is shown how they contribute to a better and broader understanding of atomic nuclei. All such theory developed previously by several authors is applied to the semimicroscopic algebraic quartet model of atomic nuclei in the proxy-$SU(3)$ scheme to obtain previously unknown irreducible representations, deformation parameters, nuclear excited states and electromagnetic transition probabilities for certain heavy isotopes. The experimental data available for such nuclear region is currently very poor,  so in preparation for future experimental efforts, the theoretical background is developed. In the appendices are shown explicit calculations, theoretical frameworks, and the use of specialized software required for the development of this thesis.

\begin{keywords}
Heavy nuclei, proxy-$SU(3)$, quartet symmetry, nuclear structure.
\end{keywords}

\end{abstract}

% Dedication
	\thesisDedication{It’s the job that’s never started as takes longest to finish\\
 - J. R. R. Tolkien}
% Acknowledgements (content in `acknowledgements.tex')
	%\include{acknowledgements}
% Table of Contents
	\tableofcontents
% List of Figures
	%\listoffigures
% List of Illustrations
	%\listofillustrations
% List of Tables
	%\listoftables
% Preface (content in `preface.tex') 
	%\include{preface}
	
\beginmainmatter

%%%%%%%%%%%%%%%%%%%%%%%%%%%%%%%%%%%%%%%%%%%%%%%%%%%%%
%% INCLUDE YOUR CHAPTERS/SECTIONS HERE
%%
% Introduction
	\chapter{Introduction} 

\InitialCharacter{S}ince the discovery of the atomic nucleus, several generations of scientists have been dedicated to its understanding and broad applications. From the point of view of physics, the theoretical and experimental richness of this particular system lies in the fact that three out of the four fundamental interactions influence the nuclear dynamics, which are evidenced as the many radioactive decay mechanisms, nuclear reactions, electromagnetic interactions, nuclear shapes, and many more \cite{prussin, Krane, takigawa}. However, such richness brings high complexity to its study due to its many body nature added to the fact that there is no precise knowledge of the residual nucleon-nucleon interaction, the combination of collective, single particle dynamics and radioactive decay instability, makes phenomenological modeling the best option for the understanding of the atomic nucleus.

The wide variety of nuclear species is reflected in the variety of nuclear models, which are usually designed to explain the characteristics of specific nuclear regions \cite{greiner1996nuclear}. There is also great diversity in the theoretical background, which might be founded on mathematical principles, classical mechanics, quantum theory, or combinations of them. Even in such simplified schemes, complexity remains, in some cases, involving computational costs that cannot be covered currently and only certain considerations and impositions can help resolve \cite{frank2019symmetries}.

The main interest of this thesis are heavy nuclei where the number of particles is larger and the collective phenomena dominate the nuclear dynamics. This region is relevant because it provides information that helps understand deeply the structure of atomic nuclei, their shapes, and their limits of existence. That is why several experimental efforts are being made to synthesize new chemical elements and isotopes \cite{Oganessian}, therefore, in preparation of such results, the theoretical background is developed for a particular kind of isotopes: those composed of nucleon quartets \cite{CSEH2015213}. A quartet is defined as two protons and two neutrons which can couple to a particular permutational representation of $[4]$ as will be explained later. 

The organization of the thesis starts from general aspects to more particular topics of the nuclear models. First, in chapter two the Nilsson model is explained, covering the three different regimes of single-particle nuclear anisotropic potential, reviewing some of its main results, presenting some derivations, explaining labeling of states and energy levels dependence with deformation. Then the introduction of a macroscopic model for nuclear excitations is done, which was essential for the explanation of nuclear spectra in the beginning of gamma ray spectroscopy. It is limited to collective rotations, where some implications of the symmetries in the spatial nuclear shape are stated. Finally, an application of these two models shows how they can be complemented for deeper understanding of the atomic nuclei.

In chapter three, the extensively applied topic of symmetries in other areas of physics is introduced to the atomic nuclei, in particular the $SU(3)$ algebra. General algebraic topics are introduced related to unitary algebras, algebra chains and symmetries in the many-body quantum systems. It is shown the relation between the quantum rotor and $SU(3)$ generators, deformation parameters and $SU(3)$ irreducible representation labels which are finally applied to one of the first algebraic models of atomic nuclei: the Elliot model.

Since the spin-orbit interaction is not negligible in the atomic nuclei and it becomes stronger the heavier the nucleus gets, it implies that $SU(3)$ symmetry is badly broken. However, multiple efforts have been made in order to restore it in subspaces of the original states proving to be fruitful. The most recent of them is shown in chapter four called the proxy-$SU(3)$ scheme. It relies in the fact that particular states called deShalit-Goldhaber pairs posses a high spatial overlap and thus can be substituted in order to restore the symmetry through the truncation of the original model space. Some of its corroborations comprehend the prolate over oblate dominance and the similarity between the truncated states with Nilsson levels. 

In chapter five, the theory exposed in the previous sections is applied to heavy isotopes composed of nucleon quartets to calculate irreducible representations, predict nuclear spectra, deformation parameters and electromagnetic transition probabilities. The results are shown in tables and plots with the methodology, theoretical background and use of specialized software expanded in the appendices. Additionally some perspectives of the results presented are stated which will be studied by the author in the near future.

Part of this thesis is a condensation of topics which can be found scattered among many books and research papers, so one of its purposes is to present them in a more self complete way with their relations stated explicitly and how they contribute to a more complete understanding of the atomic nucleus. The reader is encouraged to review the references by himself or herself and to try to learn more about the specialized free software for educational or research purposes.

% Parts/Chapters
%	\part{Part A}

\chapter{Nilsson and Collective Models}
\InitialCharacter{T}he shell model is one of the most remarkable theoretical frameworks in nuclear physics, mainly because of its introduction of some quantum mechanical aspects to the atomic nuclei (spin-orbit, orbit-orbit interaction), its predictions of the magic numbers (2, 8, 20, 28, 50, 82, and 126) and the explanation of several nuclear electromagnetic spectra energies\cite{Mayer, heydeshell,Krane, prussin}. However, to maintain the angular momentum quantum numbers $(l, s, j, m_j)$ for single-particle states, the model imposes a strong assumption in its formulation: the spherical symmetry of the nuclear (and electric) mean field interaction potential $V(r)$. The proposed Hamiltonian has the form
\begin{equation}
    H = -\frac{\hbar^2}{2M}\nabla^2 + V(r) -\mathcal{C}(r)\boldsymbol{L}\cdot \boldsymbol{S} - \mathcal{D}\left(\boldsymbol{L}^2 -\left<\boldsymbol{L}^2\right>_N\right),
    \label{shellmodelh}
\end{equation}
where $\mathcal{C}(r)$ is a function dependent on $V(r)$ \cite{suhonen} and $\mathcal{D}$ is a constant parameter \cite{nilsson1995shapes}. Some choices for $V(r)$ that can be found in the literature consist on the harmonic oscillator, Woods-Saxon potential, infinite and finite square wells \cite{prussin, takigawa} (see appendix A for more details). This condition of isotropy imposed on $V(r)$ constrains the equi-potential surfaces and thus, the physical shape of the nucleus to be a sphere.

\begin{figure}[h]
    \centering 
\raisebox{0.2\height}{\includegraphics[width=0.3\textwidth]{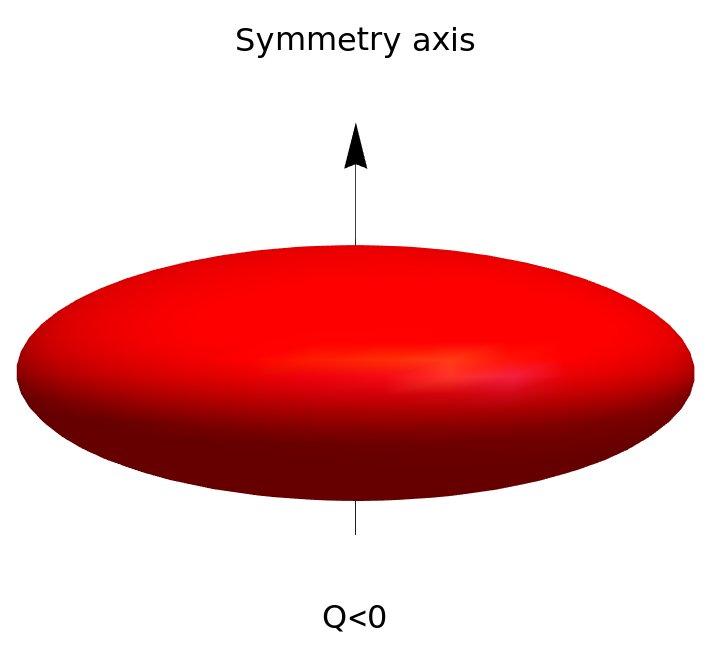}}
\hspace{.3in}
\raisebox{0.28\height}{\includegraphics[width=0.2\textwidth]{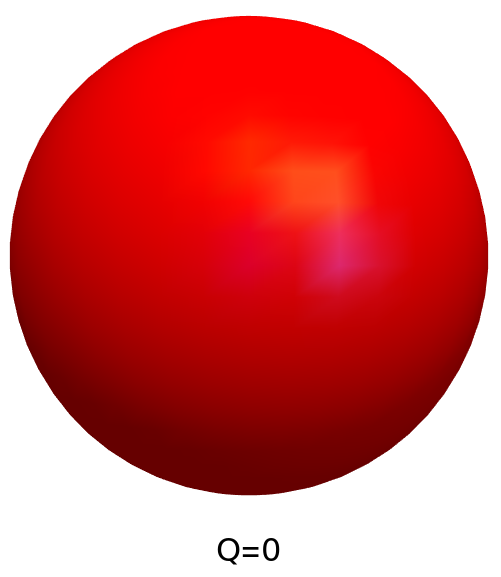}}
\hspace{.3in}
\includegraphics[width=0.15\textwidth]{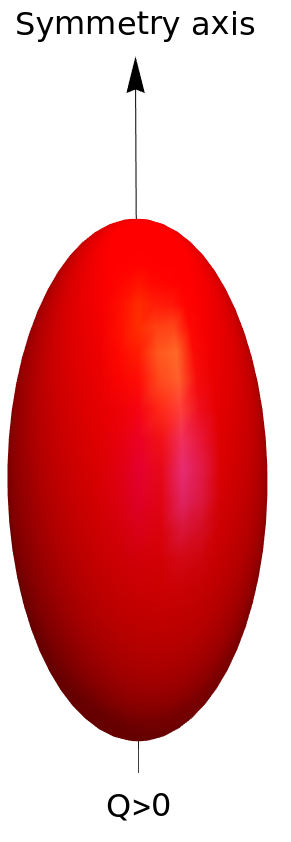}
    \caption{Axially symmetric quadrupole deformations of the nucleus. Left: The oblate shape corresponds to $\mathcal{Q}$<0.  Center: The spherical shape corresponds to $\mathcal{Q}$=0.  Right: The prolate shape corresponds to $\mathcal{Q}$ >0. Figure generated using Mathematica \cite{Mathematica}.}
    \label{Shape}
\end{figure}

\begin{figure}[h!]
    \centering 
\includegraphics[width=0.95\textwidth]{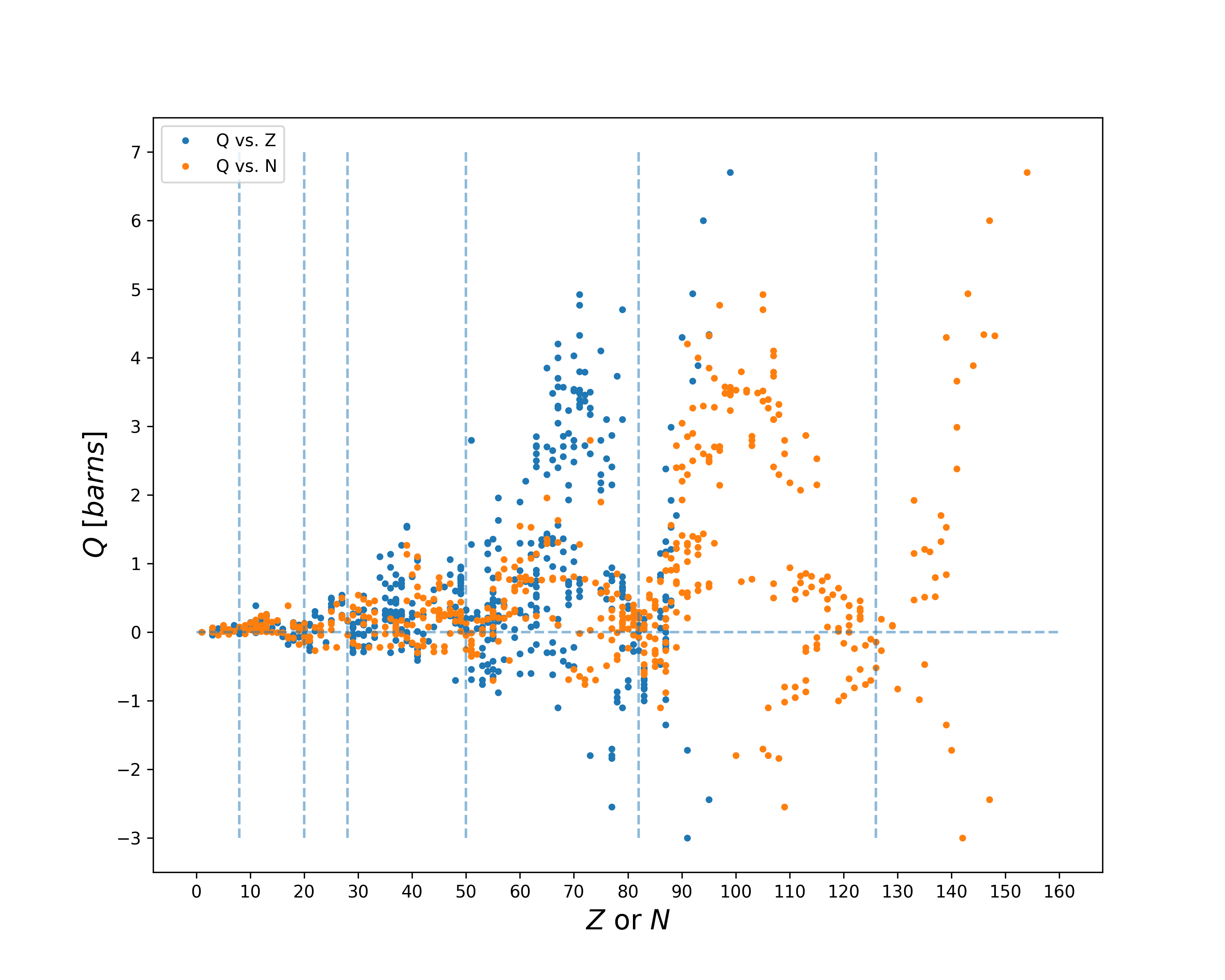}
    \caption{Experimental electric quadrupole moments for the ground states of several nuclei with respect to Z or N as indicated. The horizontal line indicates $\mathcal{Q}$ = 0 barns and the vertical lines the magic numbers. Data retrieved from \cite{STONE20161} and figure generated using Matplotlib \cite{Hunter:2007}.}
    \label{Quadr}
\end{figure}

In reality, as was observed by means of quadrupole electric moment $\mathcal{Q}$ measurements, most nuclei in their ground states are deformed \cite{STONE20161}. This particular multipole moment allows us to discern between spherical, prolate and oblate charge distributions along a symmetry axis fixed to the nuclei (see figure \ref{Shape}) which, according to the data shown in figure \ref{Quadr}, posses mostly prolate deformation. The figure also reveals that with greater proton ($Z$), neutron ($N$) numbers and the farther from closed shells these are, greater $\mathcal{Q}$ values and thus greater deformations are encountered, a fact that will be important for the model exposed in this thesis. 

Other difficulties in the shell model arose with the rapidly growing computational cost of its direct application \cite{VanIsacker2011, talmishell, successandfailure} and the discovery of collective (many nucleons) excitations that pointed to the existence of rich yet unaccounted nuclear dynamics\cite{Heyde_2016, Nakatsukasa_2016} observed through some significant deviations from single-particle shell model predictions in both electric and magnetic multipole moments \cite{nobellecture,Krane}. Even though, advances have been and are still being made in both of these mentioned aspects \cite{greiner1996nuclear,obertelli2021modern, Shimizu_2013, koonig, PhysRevC.104.054306 ,DJRowe_1985}, it was clear that the shell model alone required some modifications and even completely new theory in order to describe the nucleus more consistently and computationally feasible.  This section is dedicated to introduce an extension and a complementary theoretical framework developed for this purpose, which is necessary to understand the dynamics and phenomenology of the nuclear region studied along with the theoretical handling of it. 

%---------------------------------------------

\section{Nilsson Model}
The shell model determines independent single-particle states (some improvements consider residual interactions between two and three particles  \cite{suhonen, takigawa}) under a spherically symmetric mean field. As was already stated, evidence shows that nuclei away from closed shells are deformed, an aspect that influences the motion of nucleons and creates the need for considering nuclear shape and orientation as dynamical variables as well \cite{bohr1998nuclear, greiner1996nuclear, rowe2010fundamentals}. The most general wavefunction is then 

\begin{equation}
    \Psi_{nucleus} = \chi_{nucleons}\cdot\psi_{vibr.}\cdot\mathcal{D}_{rot.},
    \label{general}
\end{equation}
where $\chi_{nucleons}$ comprehends the intrinsic motion of protons and neutrons, $\psi_{vibr.}$ the vibrations about the equilibrium shape and $\mathcal{D}_{rot.}$ the collective rotation as a whole (see next section). The extension proposed by Sven G. Nilsson \cite{nilsson1955, nilsson1995shapes, bohr1998nuclear} focuses on $\chi_{nucleons}$ continuing the single-particle motion scheme but with a non-isotropic potential which introduces dependence with spherical coordinate $\theta$ and the need to use different quantum numbers than the usual $\left|N l s j m_j \pi\right>$.

\subsection{Hamiltonian, Choice of Basis and Spectra}

\begin{figure}[h]
    \centering 
\includegraphics[width=0.5\textwidth]{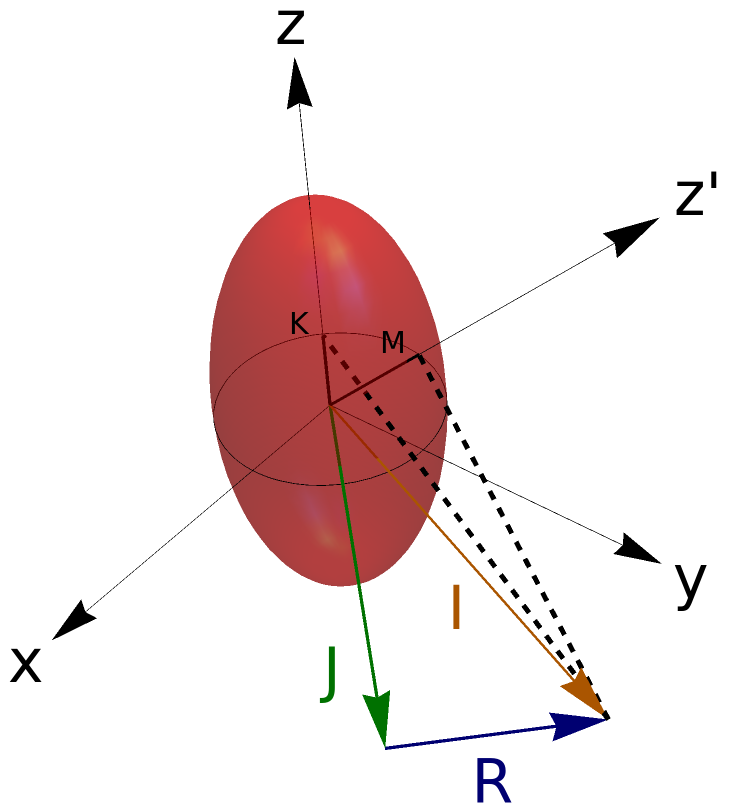}
    \caption{Angular momentum diagram. $\mathbf{J}$ indicates the intrinsic angular momentum, $\mathbf{R}$ arises from collective rotation and $\mathbf{I} = \mathbf{J}+\mathbf{R}$.  The quantities marked as $K$ and $M$ are the projections of  $\mathbf{I}$ over the $z$-axis fixed to the nucleus and the laboratory $z'$-axis respectively. Notice that for axially symmetric nuclei $\mathbf{R}$ is perpendicular to the $z$-axis, and thus $K$ corresponds to the projection of both $\mathbf{J}$ and $\mathbf{R}$. Figure generated using Mathematica. }
    \label{nilssoncoord}
\end{figure}

Based on \ref{shellmodelh}, the proposed Hamiltonian considers an axially symmetric harmonic oscillator potential of the form 
\begin{equation}
    H_N = -\frac{\hbar^2}{2M}\nabla^2 + \frac{M}{2}\left[\omega_{\perp}^2(x^2+y^2) +\omega_{\parallel}^2z^2  \right] -\mathcal{C}\boldsymbol{L}\cdot \boldsymbol{S} - \mathcal{D}\left(\boldsymbol{L}^2 -\left<\boldsymbol{L}^2\right>_N\right),
    \label{shellmodelhnilsson}
\end{equation}
where the coordinates $x$, $y$ and $z$ are fixed to the nucleus as shown in figure \ref{nilssoncoord},  $\omega_{\perp}$, $\omega_{\parallel}$ are the perpendicular and axial oscillator frequencies respectively and the average over each shell is $\left<\boldsymbol{L}^2\right>_N = N(N+3)/2$ \cite{nilsson1995shapes}. The deformation parameter $\epsilon$ is conveniently introduced as 
\begin{equation}
    \begin{aligned}
     &\omega_{\perp} = \omega_{0}(\epsilon)\left( 1+ \frac{1}{3}\epsilon\right),\\
     &\omega_{\parallel} = \omega_{0}(\epsilon)\left( 1- \frac{2}{3}\epsilon\right),\\
     &\omega_{0}(\epsilon) = \overset{0}{\omega}_{0}\left( 1 + \frac{1}{9}\epsilon^2+\mathcal{O}(\epsilon^3)\right),
    \end{aligned}
\label{omegas}
\end{equation}
and thus $\epsilon = (\omega_{\perp}-\omega_{\parallel})/\omega_{0}$ implying that $\epsilon >$ 0 correspond to prolate and $\epsilon <$ 0 to oblate shapes. The third equation arises from volume conservation $\omega_{\perp}^2\omega_{\parallel} = (\omega_{0}(\epsilon=0))^3=(\overset{0}{\omega}_{0} )^3$ and expresses the (weak) dependence of the oscillator frequency $\omega_{0}(\epsilon)$ with the non-deformed $\overset{0}{\omega}_{0}$. In this model $\mathcal{C}$ and $\mathcal{D}$ are taken as 
\begin{equation}
\begin{aligned}
    \mathcal{C} = 2\kappa\hbar\omega_0,\\
    \mathcal{D} = \kappa\mu\hbar\omega_0,
\end{aligned}
    \label{CD}
\end{equation}
where $\kappa$ and $\mu$ are parameters whose purpose is adjusting to Woods-Saxon energy levels and their values can be retrieved from \cite{BENGTSSON198514} for each $N$ (shell).

With all that defined, the main objective is finding the eigenstates, eigenvalues and their deformation dependence. Now three paths can be considered: small, medium and large deformations.

\subsubsection{Small Deformations}

\begin{figure}[h]
    \centering 
\includegraphics[width=0.7\textwidth]{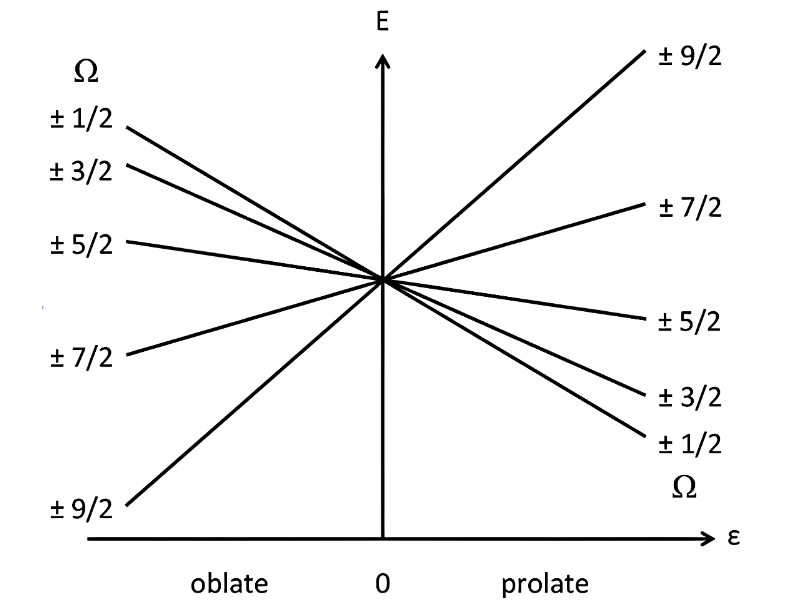}
    \caption{Effect of a small deformation on a $j = $9/2 level. Figure taken from \cite{Jenkins}.}
    \label{smalldef}
\end{figure}

This regime considers $|\epsilon| \sim $ 0.1. Using \ref{omegas}, the Hamiltonian \ref{shellmodelhnilsson} can be rewritten in spherical coordinates as
\begin{equation}
\begin{aligned}
    H_N = -\frac{\hbar^2}{2M}\nabla^2 +\frac{1}{2}M\omega_0^2 r^2  
    - \frac{2}{3}M\omega_0^2\epsilon r^2 P_2\left(\text{cos}(\theta)\right) 
    +\frac{1}{9}M\omega_0^2\epsilon^2r^2\left(P_2\left(\text{cos}(\theta)\right)+1\right)\\
    -\mathcal{C}\boldsymbol{L}\cdot \boldsymbol{S} - \mathcal{D}\left(\boldsymbol{L}^2 -\left<\boldsymbol{L}^2\right>_N\right),
\end{aligned}
\label{shellmodelhnilssonsmall}
\end{equation}
where $P_2\left(cos(\theta)\right)$ is the Legendre polynomial of degree 2. Notice that $\left[H_N, \boldsymbol{\pi} \right] = $ 0, so $\pi$ will be used as a label. Keeping only $\mathcal{O}(\epsilon)$ terms and treating them as a perturbation \cite{sakurai}
\begin{equation}
\begin{aligned}
    &H_N^{(small)} = H_0 + H',\\
    &H_0 =  -\frac{\hbar^2}{2M}\nabla^2 +\frac{1}{2}M\omega_0^2 r^2 -\mathcal{C}\boldsymbol{L}\cdot \boldsymbol{S} - \mathcal{D}\left(\boldsymbol{L}^2 -\left<\boldsymbol{L}^2\right>_N\right),\\
    &\epsilon H' = - \frac{2}{3}M\omega_0^2\epsilon r^2 P_2\left(\text{cos}(\theta)\right), 
\end{aligned}
\label{shellmodelhnilssonsmall1}
\end{equation}
the corrections to the energy eigenvalues of $H_0$ according to perturbation theory will be 
\begin{equation}
\begin{aligned}
    \left< Nlsj\Omega\pi\left|\epsilon H' \right|Nlsj\Omega\pi\right>=\frac{1}{6}\epsilon \hbar\omega_0\left(N+\frac{3}{2}\right)\frac{3\Omega^2-j(j+1)}{j(j+1)},
\end{aligned}
\label{correcsmall}
\end{equation}
where $\left|Nlsj\Omega\pi\right>$ is the coupled basis for $H_0$ ($m_j$ is denoted as $\Omega$). Notice the dependence on $\epsilon$ and $\Omega$; the former causes the inversion in the order of energy levels between prolate and oblate, while the later removes same-$j$ degeneracy but keeps a 2-fold one between $\pm\Omega$ states (see figure \ref{smalldef}).  The total energy will be
\begin{equation}
\begin{aligned}
    E(N,l, j) = \hbar\omega_0&\left[N+\frac{3}{2} - \kappa      
  \left\{
    \begin{array}{c}
        l\\
        -(l+1)
    \end{array}
\right\}          -\kappa\mu\left(l(l+1)-\frac{N(N+3)}{2}\right)  \right]\\ 
    &+\frac{1}{6}\epsilon \hbar\omega_0\left(N+\frac{3}{2}\right)\frac{3\Omega^2-j(j+1)}{j(j+1)}
    \hspace{3mm}
    \begin{cases}
        &\text{if } j=l+1/2.\\
        &\text{if } j=l-1/2.
    \end{cases}
\end{aligned}
\label{energysmall}
\end{equation}

\subsubsection{Large Deformations}
Before dealing with medium $\epsilon \sim$ 0.2-0.3 deformations, consider the case where the potential dominates and the $\mathbf{L}\cdot\mathbf{S}$, $\mathbf{L}^2$ terms are treated as perturbations. Thus
\begin{equation}
\begin{aligned}
    &H_N^{(large)} = H_0 + H',\\
    &H_0 =  -\frac{\hbar^2}{2M}\nabla^2 + \frac{M}{2}\left[\omega_{\perp}^2(x^2+y^2) +\omega_{\parallel}^2z^2  \right], \\    
    &H' = -\mathcal{C}\boldsymbol{L}\cdot \boldsymbol{S} - \mathcal{D}\left(\boldsymbol{L}^2 -\left<\boldsymbol{L}^2\right>_N\right). 
\end{aligned}
\label{shellmodelhnilssonlarge}
\end{equation}

A solution that diagonalizes $H_0$ can be found which will then be used to calculate $H'$ matrix elements according to perturbation theory. The eigenvalues of $L_z$ and $S_z$ will be denoted as $\Lambda$ and $\Sigma$ respectively. These are not conserved since $\left[H_N^{(large)}, L_z\right] \neq $ 0 and $\left[H_N^{(large)}, S_z\right] \neq $ 0 but, as is demonstrated below, they are required to establish the basis states. Now the task is finding appropriate labels for the eigenstates $\left|\psi\right>$ and that should be done by solving the time independent Schrödinger equation in position representation
\begin{equation}
\begin{aligned}
    &H_0\psi(\boldsymbol{r}) = E\psi(\boldsymbol{r}), 
\end{aligned}
\label{schrodinger}
\end{equation}
where $\psi(\boldsymbol{r}) = \left<\boldsymbol{r}|\psi\right>$. In order to separate the coordinates in a straightforward manner, define the stretched coordinates
\begin{equation}
\begin{aligned}
    &\xi= x\sqrt{\frac{M\omega_{\perp}}{\hbar}},\\
    &\eta=y\sqrt{\frac{M\omega_{\perp}}{\hbar}},\\    &\zeta=z\sqrt{\frac{M\omega_{\parallel}}{\hbar}},
\end{aligned}
\label{nexcoords}
\end{equation}
and their associate cylindrical coordinates
\begin{equation}
\begin{aligned}
    &\rho= \xi^2+\eta^2,\\
    &\phi=\text{arctan}\left(\eta/\xi\right),
\end{aligned}
\label{cylnexcoords}
\end{equation}
that allows \ref{schrodinger} to be rewritten as
\begin{equation}
\begin{aligned}
    &\left[\frac{1}{2}\hbar\omega_{\perp}\left(-\frac{1}{\rho}\frac{\partial}{\partial\rho}\rho\frac{\partial}{\partial\rho} - \frac{1}{\rho^2}\frac{\partial^2}{\partial\phi^2}+\rho^2\right) + \frac{1}{2}\hbar\omega_{\parallel} \left(-\frac{\partial^2}{\partial\zeta^2}+\zeta^2\right)\right]\psi = E\psi, 
\end{aligned}
\label{schrodingercylindrical}
\end{equation}
where $\boldsymbol{L}$ changes to the new coordinates under the notation $\boldsymbol{L}_t$. Proposing a variable separation 
\begin{equation}
\begin{aligned}
    \psi(\rho, \zeta, \phi) = U(\rho)Z(\zeta)\Phi(\phi),
\end{aligned}
\label{separation}
\end{equation}
it can be found that (details in \cite{nilsson1995shapes, QUENTIN1970365})
\begin{equation}
\begin{aligned}
    &\Phi(\phi) = e^{i\Lambda\phi},\\
    &Z(\zeta) = N_Z e^{-\zeta^2/2}\mathcal{H}_{n_{\parallel}}(\zeta),\\
    &U(\rho) =N_U \rho^{|\Lambda|}e^{-\rho^2/2}\mathcal{F}\left(-\frac{n_{\perp}-|\Lambda|}{2},|\Lambda|+1;\rho^2 \right),
\end{aligned}
\label{solutions}
\end{equation}
where $N_Z$, $N_U$ are normalization constants, $\mathcal{H}_{n_{\parallel}}(\zeta)$ is the Hermite polynomial and $\mathcal{F}$ the confluent hypergeometric function \cite{NIST}. These new quantum numbers $n_{\perp}$ and $n_{\parallel}$ are such that hold
\begin{equation}
\begin{aligned}
    &n_{\perp} = n_x + n_y;\hspace{5mm}n_{\perp} = 0,1,2,... \\
    &n_{\perp} = 2n_{\rho} + |\Lambda| \hspace{2mm}\text{with}\hspace{2mm}n_{\rho} = 0,1,2,... \\
    & \implies\hspace{2mm} |\Lambda| = n_{\perp}, n_{\perp}-2, n_{\perp}-4,...,0 \hspace{2mm}\text{or}\hspace{2mm} 1\\
    &N = n_{\perp} + n_{\parallel}; \hspace{6mm} N = 0,1,2,...\\
\end{aligned}
\label{properties}
\end{equation}

The parity can be found to be 
\begin{equation}
\begin{aligned}
\boldsymbol{\pi}\psi = (-1)^{n_{\parallel}+\Lambda}\psi = (-1)^{N}\psi =  (-1)^{l}\psi,
\end{aligned}
\label{properties}
\end{equation}
where $l$ is the spherical eigenvalue associated to $\textbf{L}^2$. 

\begin{figure}[h!]
    \centering 
\includegraphics[width=0.8\textwidth]{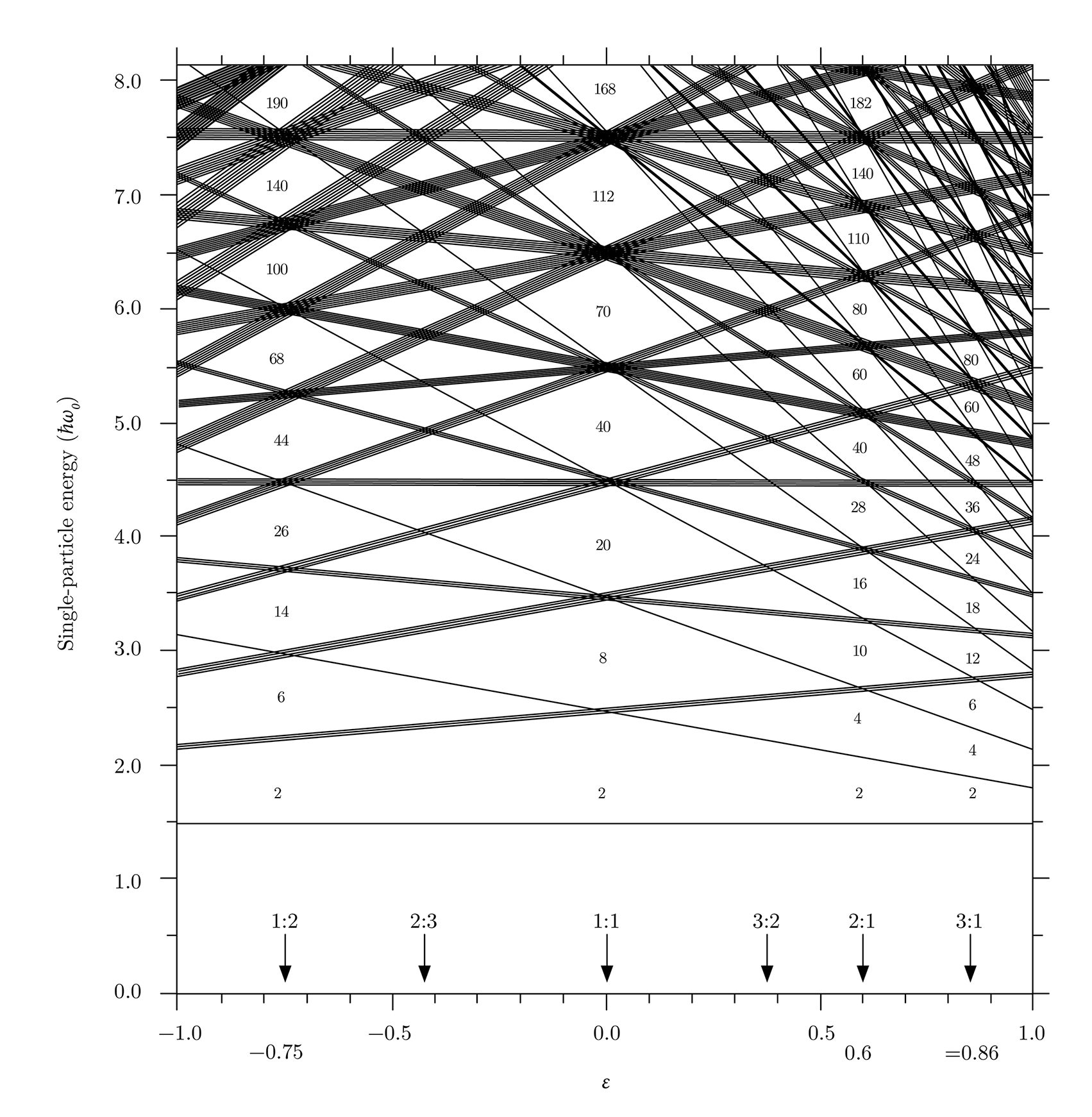}
    \caption{Anisotropic harmonic oscillator energy levels dependence on $\epsilon$ for large deformations. Notice the appearance of new shell structure and magic numbers for indicated values of $\omega_{\perp}:\omega_{\parallel}$. Figure taken from \cite{nilsson1995shapes}.}
    \label{restoring}
\end{figure}

The energy eigenvalue is 
\begin{equation}
\begin{aligned}
E(n_{\perp}, n_{\parallel}) = \hbar\omega_0\left(  N + \frac{3}{2} +(n_{\perp} - 2n_{\parallel})\frac{\epsilon}{3}  \right),
\end{aligned}
\label{energy}
\end{equation}
without including spin-orbit and orbit-orbit interactions yet. This result shows an important phenomenon seen in figure \ref{restoring} where for certain ratios $\omega_{\perp}:\omega_{\parallel}$, harmonic oscillator degeneracy is largely regained and new shell structure appears. This is be the basis for nuclear models and symmetries restoration introduced in \cite{ARIMA1969517, kota20203}. 

Perturbation theory dictates that only the following diagonal elements are needed
\begin{equation}
\begin{aligned}
&\left<\boldsymbol{L}_t\cdot\boldsymbol{S}\right> = \Lambda\Sigma,\\
&\left<\boldsymbol{L}_t^2\right> = \Lambda^2+2n_{\perp}n_{\parallel} + 2n_{\parallel}+n_{\perp},
\end{aligned}
\label{diagonal}
\end{equation}
they can be calculated using $\boldsymbol{L}_t\cdot\boldsymbol{S} = \frac{1}{2}(L_{t+}S_- - L_{t-}S_+)+L_{tz}S_z$ and $\boldsymbol{L}_t^2 = \frac{1}{2}\left(L_{t+}L_{t-}+L_{t-}L_{t+}\right)+L_{tz}^2$. The total energy will be
\begin{equation}
\begin{aligned}
E(n_{\perp}, n_{\parallel}, \Lambda,\Sigma) =& \hbar\omega_0\left(  N + \frac{3}{2}\right) + \frac{1}{3}\epsilon\hbar\omega_0(N-3n_{\parallel}) \\&- \mathcal{C}\Lambda\Sigma-\mathcal{D}\left(\Lambda^2+2n_{\perp}n_{\parallel}+2n_{\parallel}+n_{\perp}-\frac{N(N+3)}{2}\right).
\end{aligned}
\label{energylarge}
\end{equation}

Since same $j$ and $l$ degeneracy is lifted, these are no longer good quantum numbers for labeling the states and instead will be adopted the so called asymptotic basis $\left|Nn_{\parallel}\Lambda\Sigma\pi\right>$. Alternative options are $\left|Nn_{\parallel}\Lambda\Omega\pi\right>$,  $\left|n_{\perp}n_{\parallel}\Lambda\Omega\pi\right>$ and in spectroscopy is also found notation $\Omega^{\pi}[Nn_{\parallel}\Lambda\Sigma]$ or just $\Omega[Nn_{\parallel}\Lambda]$. Notice that $\Omega[Nn_{\parallel}\Lambda]$ will contain both $\pm\Omega$ total angular momentum projections.

\subsubsection{Medium Deformations}

This is the region of $\epsilon \sim$ 0.2-0.3 in which no term of the Hamiltonian \ref{shellmodelhnilsson} is to be treated as a perturbation and thus a different strategy will be used as shown next. Consider a Hamiltonian $H$ whose eigenstates are $\left| \psi_\alpha \right>$ and hold
\begin{equation}
\begin{aligned}
H\left| \psi_\alpha \right> = E_\alpha\left| \psi_\alpha \right>,
\end{aligned}
\label{schrod}
\end{equation}

Consider as well an orthogonal set of states $\left| \phi_\nu \right>$ that span the Hilbert space of the system then, it can be established
\begin{equation}
\begin{aligned}
\left| \psi_\alpha \right> = \sum_\nu \mathcal{S}_{\alpha\nu}\left| \phi_\nu \right>,
\end{aligned}
\label{span}
\end{equation}
where $\mathcal{S}$ is a unitary change of basis transformation. Replacing in \ref{schrod} and multiplying by $\left< \phi_\mu \right|$ 
\begin{equation}
\begin{aligned}
\sum_\nu \mathcal{S}_{\alpha\nu}\left< \phi_\mu \right| H\left| \phi_\nu \right> = \sum_\nu \mathcal{S}_{\alpha\nu} H_{\nu\mu}  = \sum_\nu E_\alpha \mathcal{S}_{\alpha\nu}\delta_{\mu\nu},
\end{aligned}
\label{replace}
\end{equation}
where hermiticity of $H$ has been used in $H_{\mu\nu} = H_{\nu\mu}$. In matrix form it can be written 
\begin{equation}
\begin{aligned}
\boldsymbol{\mathcal{S}} \boldsymbol{H} = \boldsymbol{E}\boldsymbol{\mathcal{S}},
\end{aligned}
\label{matrix}
\end{equation}
where $\boldsymbol{E}$ is the diagonal matrix 
\begin{equation}
\begin{aligned}
 \boldsymbol{E} = \begin{pmatrix}
E_1 & 0 & 0 &...\\
0 & E_2 & 0 &...\\
0 & 0 & E_3 &...\\
\vdots & \vdots & \vdots & \ddots 
\end{pmatrix},
\end{aligned}
\label{E}
\end{equation}
thus the energy eigenvalues of $H$ will be given by 
\begin{equation}
\begin{aligned}
\boldsymbol{E} = \boldsymbol{\mathcal{S}} \boldsymbol{H}\boldsymbol{\mathcal{S}}^{-1}.
\end{aligned}
\label{EigenvaluesofH}
\end{equation}
The problem is then calculating $\boldsymbol{H}$, guarantee that Det$(\boldsymbol{H}- \boldsymbol{E})$ = 0 for a non-trivial solution to exist and then diagonalize $\boldsymbol{H}$.

For that, the spherical uncoupled basis $\left|Nl\frac{1}{2}\Lambda\Sigma\pi\right>$ is chosen as the  $\left| \phi_\nu \right>$ states. Considering stretched coordinates \ref{nexcoords} and their corresponding spherical ones
\begin{equation}
\begin{aligned}
    & \varrho = \sqrt{\xi^2 +\eta^2+\zeta^2},\\
    &\vartheta = \text{arcos}\left(\frac{\zeta}{\sqrt{\xi^2 +\eta^2+\zeta^2}}\right),\\    
    &\varphi=\text{arctan}\left(\eta/\xi\right),
\end{aligned}
\label{spherstretched}
\end{equation}
the Hamiltonian \ref{shellmodelhnilsson} can be separated into 
\begin{equation}
\begin{aligned}
    &H_N^{(medium)} =  H_d +H_{s.o.}+ H_{\epsilon},\\
    &H_d = \frac{1}{2}\hbar\omega_0\left(-\frac{\partial^2}{\partial\xi^2}-  \frac{\partial^2}{\partial\eta^2} -\frac{\partial^2}{\partial\zeta^2} +\varrho^2  \right)  - \mathcal{D} \left( \boldsymbol{L}_t^2 -\left<\boldsymbol{L}_t^2\right>_N \right), \\ 
    &H_{s.o.} = -\mathcal{C}\boldsymbol{L}_t\cdot \boldsymbol{S},\\
    &H_{\epsilon} = \frac{1}{6}\hbar\omega_0\epsilon\left( -\frac{\partial^2}{\partial\xi^2} -\frac{\partial^2}{\partial\eta^2} +2\frac{\partial^2}{\partial\zeta^2} -2\varrho^2P_2\left(\text{cos}(\vartheta)\right) \right).
\end{aligned}
\label{Hmedium}
\end{equation}
Notice that no $\mathcal{O}(\epsilon^2)$ survive. $H_d$ matrix elements are straightforward and diagonal 
\begin{equation}
\begin{aligned}
   \left<Nl\frac{1}{2}\Lambda\Sigma\pi\right| H_d \left|Nl\frac{1}{2}\Lambda\Sigma\pi\right> = \frac{1}{2}\hbar\omega_0\left(N+\frac{3}{2}\right)  -\mathcal{D}\left( l(l+1) - \frac{N(N+3)}{2} \right).
\end{aligned}
\label{matrixHd}
\end{equation}
For $H_{s.o.}$ one has diagonal and non-diagonal elements (recall $\boldsymbol{L}_t\cdot\boldsymbol{S} = \frac{1}{2}(L_{t+}S_- - L_{t-}S_+)+L_{tz}S_z$)
\begin{equation}
\begin{aligned}
   &\left<Nl\frac{1}{2}\Lambda\Sigma\pi\right| H_{s.o.} \left|Nl\frac{1}{2}\Lambda\Sigma\pi\right> = -\mathcal{C}\Lambda\Sigma,\\
   &\left<Nl\frac{1}{2}\Lambda-1\Sigma+1\pi\right| H_{s.o.} \left|Nl\frac{1}{2}\Lambda\Sigma\pi\right> = -\frac{1}{2}\mathcal{C}\sqrt{(l+ \Lambda)(l-\Lambda+1)},\\
   &\left<Nl\frac{1}{2}\Lambda+1\Sigma-1\pi\right| H_{s.o.} \left|Nl\frac{1}{2}\Lambda\Sigma\pi\right> = -\frac{1}{2}\mathcal{C}\sqrt{(l-\Lambda)(l+\Lambda+1)}.\\
\end{aligned}
\label{matrixHso}
\end{equation}
For $H_\epsilon$ it can be demonstrated \cite{nilsson1995shapes} that 
\begin{equation}
\begin{aligned}
   \left<Nl'\frac{1}{2}\Lambda'\Sigma'\pi\right|-\frac{\partial^2}{\partial\xi^2}-  \frac{\partial^2}{\partial\eta^2} +2\frac{\partial^2}{\partial\zeta^2}
   \left|Nl\frac{1}{2}\Lambda\Sigma\pi\right>& = \left<Nl'\frac{1}{2}\Lambda'\Sigma'\pi\right| \xi^2+\eta^2-2\zeta^2 \left|Nl\frac{1}{2}\Lambda\Sigma\pi\right> \\
   & = -2 \left<Nl'\frac{1}{2}\Lambda'\Sigma'\pi\right| \varrho^2 P_2\left(\text{cos}(\vartheta)\right)   \left|Nl\frac{1}{2}\Lambda\Sigma\pi\right>.
\end{aligned}
\label{equalityofme}
\end{equation}
Then 
\begin{equation}
\begin{aligned}
   \left<Nl'\frac{1}{2}\Lambda'\Sigma'\pi\right|H_{\epsilon}\left|Nl\frac{1}{2}\Lambda\Sigma\pi\right>& = -\frac{2}{3}\hbar\omega_0\epsilon\left(\int R_{Nl'}R_{Nl}\varrho^2d\varrho \right)\Upsilon\\
   \Upsilon=& \sqrt{\frac{2l+1}{2l'+1}}\left<l\Lambda20|l'\Lambda'\right>\left<l020|l'0\right>,
\end{aligned}
\label{matrixHepsilon}
\end{equation}
where $\Upsilon$ arises from application of spherical harmonic addition theorem \cite{rose2013elementary}, $R_{Nl}$ is the harmonic oscillator radial wavefunction and applying Clebsch-Gordan coefficients selection rules on projections it is obtained
\begin{equation}
\begin{aligned}
   \left<Nl'\frac{1}{2}\Lambda\Sigma\pi\right|H_{\epsilon}\left|Nl\frac{1}{2}\Lambda\Sigma\pi\right>& = -\frac{2}{3}\hbar\omega_0\epsilon\left(\int R_{Nl'}R_{Nl}\varrho^2d\varrho \right)\Upsilon\\
   \Upsilon=& \sqrt{\frac{2l+1}{2l'+1}}\left<l\Lambda20|l'\Lambda\right>\left<l020|l'0\right>.
\end{aligned}
\label{matrixHepsilonrules}
\end{equation}

\begin{figure}[h!]
    \centering 
\includegraphics[width=0.8\textwidth]{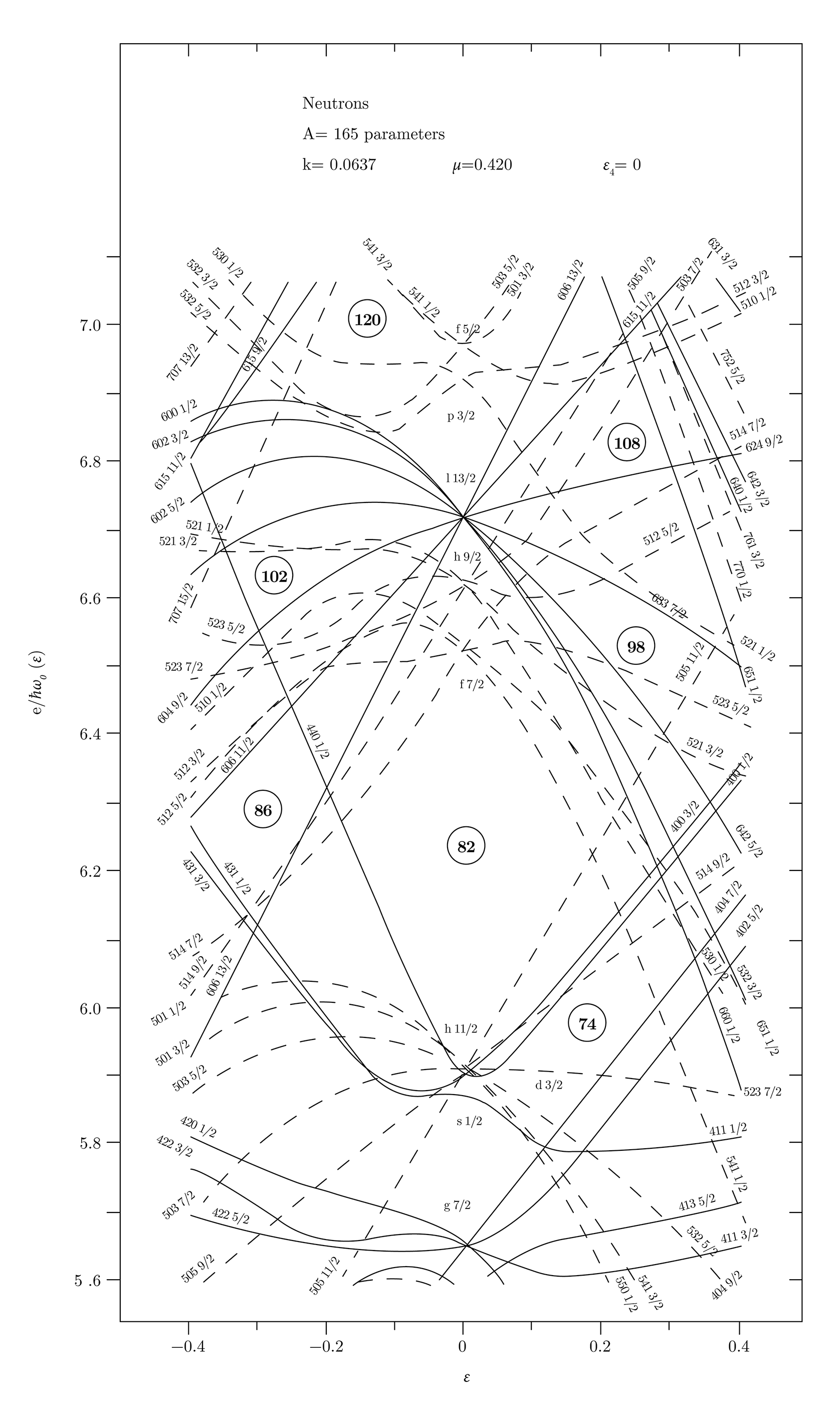}
    \caption{Energy levels dependence on $\epsilon$ for medium deformation in shells 50-82 and 82-126. Labels $Nn_{\parallel}\Lambda\Omega$ are adopted for each level. Figure taken from \cite{nilsson1995shapes}.}
    \label{nilssonmedium}
\end{figure}

The task now is to diagonalize the calculated matrix representation of $H_N^{(medium)}$ which is made by computer algorithms that essentially find matrix $\boldsymbol{\mathcal{S}}$. One of such programs was used to generate the front page and can be found in \cite{nilssonprogram}. A plot of some levels can be seen in figure \ref{nilssonmedium} and again new shell structure seem to emerge at non-zero deformation states.

Before finishing, it is worth mentioning that an operator method and generalization to higher axially symmetric multipole orders (octupole, hexadecapole,...) has been developed which can be found in \cite{nilsson1995shapes} along with the references there. A final aspect refers to the use of asymptotic labels ($N, n_{\parallel},\Lambda,\Omega$) in this region. Reference \cite{nilsson1995shapes} states that off diagonal elements of $H_N$ in the asymptotic basis are small and remain so in the intermediate deformation regime, a fact that supports preserving $N,n_{\parallel},\Lambda,\Omega$ quantum numbers for labelling states. Such practice has become almost a rule in spectroscopy as well. An interesting discussion of this issue related to this thesis can be found in \cite{proxykeeping} and some careful analysis about the change from spherical to Nilsson basis is in \cite{Sobhani2021}.

\subsection{Triaxial Nuclei}

Even though quadrupole deformation seems to be a good approximation, in many cases, as for not so strongly deformed and rotating nuclei (which causes the breaking of axial symmetry by the compression along the plane perpendicular to $\boldsymbol{R}$) it is necessary to extend the original Nilsson model to include triaxially deformed nuclei\cite{nilsson1995shapes, Larsson}. The Hamiltonian is now
\begin{equation}
    H_{N.T.} = -\frac{\hbar^2}{2M}\nabla^2 + \frac{M}{2}\left[\omega_{x}^2x^2+\omega_{y}^2y^2 +\omega_z^2z^2  \right] -\mathcal{C}\boldsymbol{L}\cdot \boldsymbol{S} - \mathcal{D}\left(\boldsymbol{L}^2 -\left<\boldsymbol{L}^2\right>_N\right).
    \label{triaxH}
\end{equation}

A new parameter $\gamma$ that accounts for the departure of axial symmetry is introduced in  
\begin{equation}
\begin{aligned}
    &\omega_x = \omega_0(\epsilon, \gamma)\left[1-\frac{2}{3}\epsilon \hspace{1mm}\text{cos}\left(\gamma+\frac{2\pi}{3}\right)\right], \\
    &\omega_y = \omega_0(\epsilon, \gamma)\left[1-\frac{2}{3}\epsilon \hspace{1mm}\text{cos}\left(\gamma-\frac{2\pi}{3}\right)\right], \\
    &\omega_z = \omega_0(\epsilon, \gamma)\left[1-\frac{2}{3}\epsilon \hspace{1mm}\text{cos}\left(\gamma\right)\right],
\end{aligned}
\label{omegastriaxial}
\end{equation}
where $\omega_0(\epsilon, \gamma)$ dependence with gamma can be obtained from volume conservation condition $\omega_x\omega_y\omega_z = \left(\overset{0}{\omega}_{0}\right)^3$ whose explicit value can be found in \cite{Larsson}. More details about $\gamma$ can be found in the next section, but for now focus on Hamiltonian \ref{triaxH}. This can be easily solved in Cartesian coordinates, however, in order to use angular momentum quantum numbers it can be rewritten in stretched coordinates \ref{nexcoords} and their associate spherical coordinates \ref{spherstretched} as 
\begin{equation}
\begin{aligned}
    H_{N.T.} =& \frac{1}{2}\hbar\omega_0\left(-\frac{\partial^2}{\partial\xi^2}-  \frac{\partial^2}{\partial\eta^2} -\frac{\partial^2}{\partial\zeta^2} +\varrho^2  \right) + \frac{1}{6}\epsilon\hbar\omega_0\hspace{0.5mm}\text{cos}(\gamma)\left[ \xi^2+\eta^2-2\zeta^2 -\frac{\partial^2}{\partial\xi^2}-\frac{\partial^2}{\partial\eta^2}+2\frac{\partial^2}{\partial\zeta^2}\right] \\
    & +\frac{1}{2\sqrt{3}}\epsilon\hbar\omega_0\hspace{0.5mm}\text{sin}(\gamma)\left[ \xi^2-\eta^2-\frac{\partial^2}{\partial\xi^2}+\frac{\partial^2}{\partial\eta^2}\right]-\mathcal{C}\boldsymbol{L}_t\cdot \boldsymbol{S} - \mathcal{D} \left( \boldsymbol{L}_t^2 -\left<\boldsymbol{L}_t^2\right>_N \right).  
   \end{aligned} 
    \label{triaxHstretched}
\end{equation}

Notice that second and third terms matrix elements can be computed in a similar manner as indicated in \ref{equalityofme}. Applying this just mentioned, one can consider an equivalent Hamiltonian $ H'_{N.T.}$ which can be transformed to a slightly more manageable form as 
\begin{equation}
\begin{aligned}
    H'_{N.T.} = \frac{1}{2}\hbar\omega_0\left(-\frac{\partial^2}{\partial\xi^2}-  \frac{\partial^2}{\partial\eta^2} -\frac{\partial^2}{\partial\zeta^2} +\varrho^2  \right) -\frac{2}{3}\hbar\omega_0\varrho^2\epsilon\sqrt{\frac{4\pi}{5}}&\left( \text{cos}(\gamma)Y_{20} -\frac{\text{sin}(\gamma)}{\sqrt{2}}\left(Y_{22}+Y_{2-2}\right)   \right) \\
    &-\mathcal{C}\boldsymbol{L}_t\cdot \boldsymbol{S} - \mathcal{D} \left( \boldsymbol{L}_t^2 -\left<\boldsymbol{L}_t^2\right>_N \right).  
   \end{aligned} 
    \label{triaxHstretchedhandled}
\end{equation}

It can be separated as 
\begin{equation}
\begin{aligned}
    &H'_{N.T.} = H_d + H_{s.o.} + H_{\epsilon},\\
    & H_d = \frac{1}{2}\hbar\omega_0\left(-\frac{\partial^2}{\partial\xi^2}-  \frac{\partial^2}{\partial\eta^2} -\frac{\partial^2}{\partial\zeta^2} +\varrho^2  \right) - \mathcal{D} \left( \boldsymbol{L}_t^2 -\left<\boldsymbol{L}_t^2\right>_N \right),  \\
    &H_{s.o.} = -\mathcal{C}\boldsymbol{L}_t\cdot \boldsymbol{S}, \\
    &H_{\epsilon} = -\frac{2}{3}\hbar\omega_0\varrho^2\epsilon\sqrt{\frac{4\pi}{5}}\left( \text{cos}(\gamma)Y_{20} -\frac{\text{sin}(\gamma)}{\sqrt{2}}\left(Y_{22}+Y_{2-2}\right)   \right). 
   \end{aligned} 
    \label{spearationtriaxial}
\end{equation}

In the spherical basis, matrix elements of $H_d$ and $H_{s.o.}$ are identical to those of \ref{matrixHd} and \ref{matrixHso}. Those of $H_\epsilon$ are
\begin{equation}
\begin{aligned}
   \left<Nl'\frac{1}{2}\Lambda'\Sigma\pi\right|H_{\epsilon}\left|Nl\frac{1}{2}\Lambda\Sigma\pi\right>& = -\frac{2}{3}\hbar\omega_0\epsilon\left(\int R_{Nl'}R_{Nl}\varrho^2d\varrho \right) \sqrt{\frac{2l+1}{2l'+1}}\left<l020|l'0\right> 
\\& \times \left(\text{cos}(\gamma)\left<l\Lambda20|l'\Lambda'\right>-\frac{\text{sin}(\gamma)}{\sqrt{2}}\left(\left<l\Lambda22|l'\Lambda'\right>+\left<l\Lambda2-2|l'\Lambda'\right>\right)\right).\\
\end{aligned}
\label{matrixHepsilon}
\end{equation}

The next step would be the process of diagonalization performed by a computer program. Graphs of the respective energy levels can be found in the references. Now we leave single-particle models and focus on a macroscopic aspect of atomic nuclei.

\section{Collective Rotation of Nuclei}

The discovery of rotational motion in nuclei appeared as an impossibility of the shell model to describe certain excitation energies, large deviations of some multipole moments predictions, electromagnetic transitions and the clear indication of the participation of several nucleons in these. Nowadays, many experimental evidence supports the existence of this dynamics and energy levels are organized in specific collective sequences of angular momentum called rotational (or vibrational) bands. Furthermore, it is accepted that shape coexistence, a phenomenon where a specific isotope will posses different nuclear shapes at low excitation energies occurs essentially in all nuclei\cite{Heyde_2016}, causing the existence of rotational dynamics and effects that allow the synthetization of carbon in the stars and thus the existence of life \cite{hoyle1954nuclear}.

\subsection{Nuclear Surface and Quadrupole Deformation}

In this model the nucleus is thought as a structureless, constant charge and mass density object with a sharp surface defined by the expansion 
\begin{equation}
    \begin{aligned}
    R(\theta, \phi,t) = R_0\left(1+\sum_{\lambda=0}^{\infty}\sum_{\mu=-\lambda}^{\lambda}\alpha_{\lambda\mu}^*(t)Y_{\lambda\mu}\left(\theta,\phi\right)\right),
   \end{aligned} 
    \label{Rexpansion}
\end{equation}
where $R_0 \approx r_0 A^{1/3}$ is the non deformed nuclear radius, $Y_{\lambda\mu}\left(\theta,\phi\right)$ are the spherical harmonics and $\alpha_{\lambda\mu}^*(t)$ are the shape parameters which behave as spherical tensors of angular momentum $\lambda$. Other properties of $R(\theta,\phi,t)$ and $\alpha_{\lambda\mu}$ are explained in \cite{greiner1996nuclear}. The monopole $\lambda = $ 0 term accounts for compression of nuclear matter which is not relevant for low energy spectra and the dipole $\lambda = $ 1 correspond to translations which do not contribute to the internal energy of the nucleus. The lowest order of physical interest is the quadrupole excitation $\lambda =$ 2 which was already mentioned and will be discussed more below. Other relevant moments are octupole $\lambda =$ 3 which deforms the nucleus somewhat like a "pear" for $\mu = $ 0 \cite{pear, Butler} and hexadecapolar $\lambda =$ 4 which is the highest moment relevant to nuclear theory. In fact, for a moment to be relevant in a certain nucleus it must hold $\lambda < A^{1/3}$.

Consider a static pure quadrupole deformation, then \ref{Rexpansion} reduces to 
\begin{equation}
    \begin{aligned}
    R(\theta, \phi) = R_0\left(1+\sum_{\mu=-2}^{2}\alpha_{2\mu}^*Y_{2\mu}\left(\theta,\phi\right)\right).
   \end{aligned} 
    \label{R2}
\end{equation}

Rewriting in Cartesian components of the unit vector in the direction $(\theta, \phi)$ 
\begin{equation}
    \begin{aligned}
    &\Xi = \text{sin}(\theta)\text{cos}(\phi),\\
    &H = \text{sin}
    (\theta)\text{sin}(\phi),\\
    &Z = \text{cos}(\theta),
   \end{aligned} 
    \label{unit}
\end{equation}
and then transforming to the principal axis coordinate system $(\Xi', H', Z')$ it is obtained
\begin{equation}
    \begin{aligned}
    R(\Xi', H', Z') = R_0\left(1 + \alpha_{\Xi'}\Xi'^2 + \alpha_{H'}H'^2 + \alpha_{Z'}Z'^2\right).  
   \end{aligned} 
    \label{R2Principal}
\end{equation}

The relations with the five spherical deformation parameters  in the rotated frame $\alpha'_{\lambda\mu}$ are 
\begin{equation}
    \begin{aligned}
    &\alpha'_{20} = \sqrt{\frac{8\pi}{15}}\frac{1}{\sqrt{6}}(2\alpha_{Z'}-\alpha_{\Xi'}-\alpha_{H'})  \equiv a_0,\\
    &\alpha'_{2\pm 1} = 0 ,\\
    &\alpha'_{2\pm2} = \sqrt{\frac{2\pi}{15}}(\alpha_{\Xi'}-\alpha_{H'}) \equiv a_2,
   \end{aligned} 
    \label{relsphecart}
\end{equation}

Notice that equation \ref{R2Principal} describes an ellipsoidal shape where the semi-axes have lengths $1/\sqrt{\alpha_{i'}R_0}$ where $i' = \Xi', H', Z'$. Thus, quadrupole deformations describe triaxial nuclei. Notice also the change of dimensions from the dimensionless $\alpha_{\lambda\mu}$ to the inverse distance squared $\alpha_{i'}$.

Now focus on the definitions of $a_0$ and $a_2$. The former indicates the stretching along the $Z'$ axis with respect to $\Xi'$ and $H'$ while the later determines the difference in stretching along $\Xi'$ and $H'$ axes. The case $a_2 = $ 0 describes axial symmetry extensively worked by Nilsson. An alternative set of coordinates introduced by A. Bohr \cite{bohr1954} considers something similar to polar coordinates in $(a_0,a_2)$ space defined by 
\begin{equation}
    \begin{aligned}
    &a_0 = \beta\hspace{0.5mm}\text{cos}(\gamma),\\
    &a_2 = \frac{1}{\sqrt{2}}\beta\hspace{0.5mm}\text{sin}(\gamma),
   \end{aligned} 
    \label{betagamma}
\end{equation}
where the $\frac{1}{\sqrt{2}}$ factor has the purpose of making $\beta^2$ rotationally invariant by means of 
\begin{equation}
    \begin{aligned}
    \beta^2 = a_0^2+2a_2^2 = \sum_{\mu}|\alpha'_{2\mu}|^2 = \sum_{\mu}|\alpha_{2\mu}|^2 = \sqrt{5}\left[\alpha_2\times\alpha_2\right]^0_0.
   \end{aligned} 
    \label{betairrot}
\end{equation}

Another rotational invariant of relevance is 
\begin{equation}
    \begin{aligned}
    \beta^3\text{cos}(3\gamma) = -\sqrt{\frac{35}{2}} \left[\left[\alpha_2\times\alpha_2\right]^2\times\alpha_2\right]^0_0
   \end{aligned} 
    \label{betairrot2}
\end{equation}

The coordinate $\gamma$ is the same as the one introduced in the last section in \ref{omegastriaxial} whereas $\epsilon$ and $\beta$ are related by 
\begin{equation}
    \begin{aligned}
    \epsilon \approx \frac{3}{2}\sqrt{\frac{5}{4\pi}}\beta \approx \text{0.95}\beta,
   \end{aligned} 
    \label{betae}
\end{equation}

Solving for $\alpha_{k}$ in terms of $(\beta,\gamma)$ where $k = $ 1, 2, 3 correspond to $i' = \Xi', H', Z'$ it is obtained 
\begin{equation}
    \begin{aligned}
    \alpha_{k} = \sqrt{\frac{5}{4\pi}}\beta\hspace{0.5mm}\text{cos}\left(\gamma -\frac{2\pi k}{3}\right).
   \end{aligned} 
    \label{alphasinbetagamma}
\end{equation}

\begin{figure}[h]
    \centering 
\includegraphics[width=0.55\textwidth]{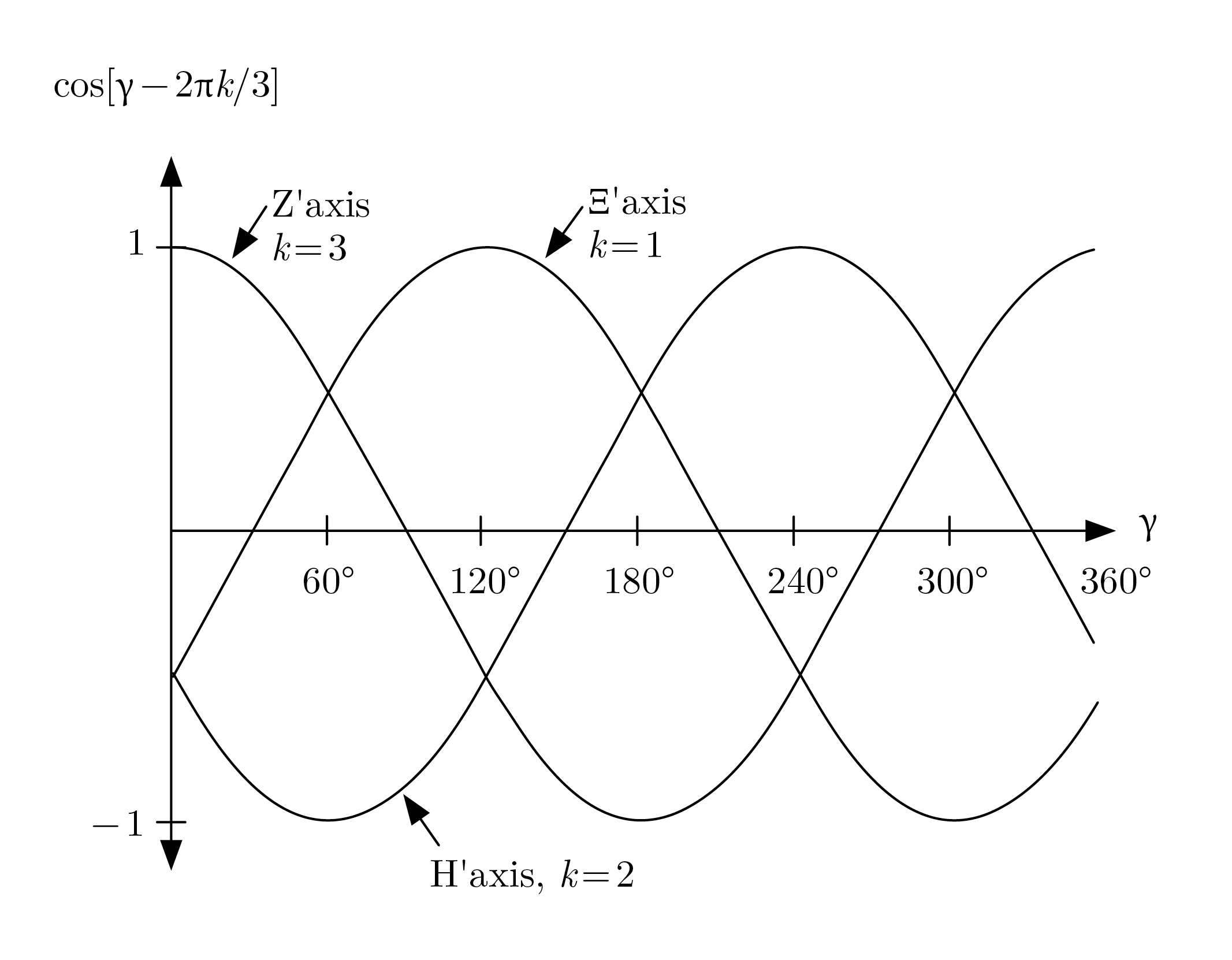}
    \caption{Plot of function \text{cos}$\left(\gamma -2\pi k/3\right)$. Figure taken from \cite{greiner1996nuclear}.}
    \label{gammafordiffk}
\end{figure}

\begin{figure}[h]
    \centering 
\includegraphics[width=0.6\textwidth]{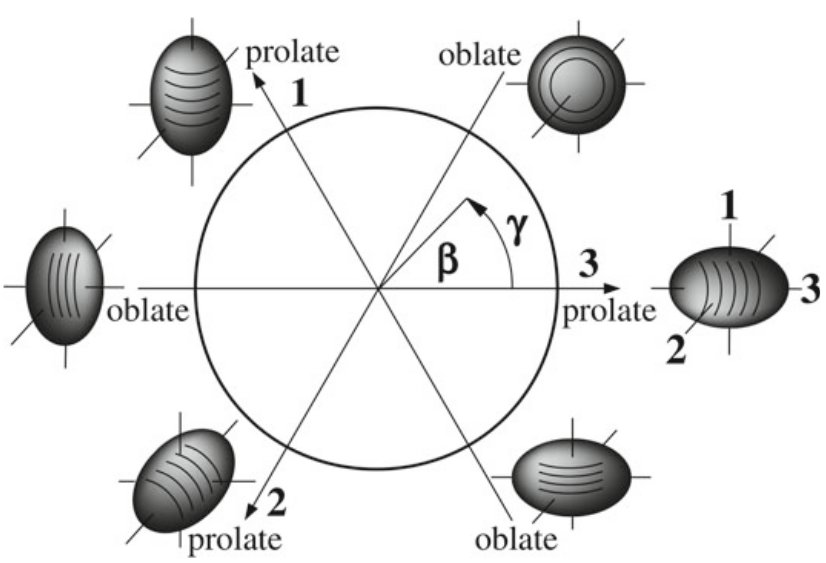}
    \caption{Plot of $(\beta,\gamma)$ plane showing the repeated nuclear shapes with different axes labelling each 60°. Figure taken from \cite{takigawa}.}
    \label{betagammaplaneshapes}
\end{figure}

Notice the similarity between \ref{alphasinbetagamma} and \ref{omegastriaxial}. Analyzing equation \ref{alphasinbetagamma} it can be seen that the various quadrupolar nuclear shapes are repeated each 60° but with a different labelling of the axes as can be seen in figure \ref{gammafordiffk}. Thus, these coordinates are somehow "degenerate" in the sense that same physical shapes can be represented by different $(\beta,\gamma)$ and only the range $\left[0\text{°}, 60\text{°}\right]$ in $\gamma$ is required to completely specify the quadrupole nuclear deformation. The particular values of $\gamma = $ 0° and $\gamma = $ 60° represent axially symmetric prolate and oblate shapes respectively, while all angles in between represent triaxial ellipsoids. It can be seen clearer in the graphical representation of figure \ref{betagammaplaneshapes}.

\subsection{Rotating Nuclei}
Due to the breaking of nuclear shape symmetry, the nuclei can rotate as a whole and thus richer dynamics arise. Now we focus on the wave function $\mathcal{D}_{rot.}$ of equation \ref{general} that describes the collective rotation of atomic nuclei. 

\subsubsection{Hamiltonian and Eigenfunctions}

Consider a rigid structureless extended body so that the internal angular momentum is $\boldsymbol{J} = $ 0 and it holds the equality between total and rotational angular momentum $\boldsymbol{I} = \boldsymbol{R}$. Its Hamiltonian for rotations about the principal axes $(\Xi',H',Z')$ is 
\begin{equation}
    \begin{aligned}
     H_{rot.} = \frac{R^2_{\Xi'}}{2\Theta_{\Xi'}} + \frac{R^2_{H'}}{2\Theta_{H'
     }} + \frac{R^2_{Z'}}{2\Theta_{Z'}},
   \end{aligned} 
    \label{hrotor}
\end{equation}
where the $\boldsymbol{R}$ components have the particularity of being body fixed angular momentum operators while the laboratory frame components will be denoted with prime as $\boldsymbol{R}'$ and the $\Theta$ denote moments of inertia about indicated axes. In quantum mechanics, rotations that leave the surface and thus the quantum state invariant are not possible. Then, structureless spheres cannot rotate around any direction nor axial ellipsoids around the symmetry axis. In the case of atomic nuclei, the deviations from these properties will depend on the involvement of internal degrees of freedom.

The possible Hamiltonians are the triaxial rotor of \ref{hrotor} and the axially symmetric about $Z'$ given by
\begin{equation}
    \begin{aligned}
     H^{ax.}_{rot.} = \frac{1}{2\Theta} \left(R^2_{\Xi'} + R^2_{H'} \right).
   \end{aligned} 
    \label{hrotoraxially}
\end{equation}

As was mentioned earlier, the projection over the body fixed symmetry axis will be denoted by $\hbar K$ associated to $R_{\Xi'}$ and over the laboratory frame as $\hbar M$ associated to $R_{z'}$. The scalar operator $\boldsymbol{R}'^2 = \boldsymbol{R}^2$ will have associated eigenvalue $\hbar^2 I(I+1)$.

The associated eigenfunctions $\phi$ can be obtained in any orientation by rotation of a known solution using Wigner matrices as 
\begin{equation}
    \begin{aligned}
     \phi_{IM}(\boldsymbol{\theta}) = \sum_{M'}\mathcal{D}^{(I)*}_{MM'}(\boldsymbol{\theta})\phi_{IM'}(\boldsymbol{0}),
   \end{aligned} 
    \label{hrotsol}
\end{equation}
where labels $I$ and $M$ where chosen because the energy will be independent on its orientation, $\boldsymbol{\theta}$ are the Euler angles and $\boldsymbol{0}$ represent all Euler angles equal to zero with $\phi_{IM'}(\boldsymbol{0})$ as the known solution.

\subsubsection{Symmetric Quadrupole Rotor}

A graphical representation of this case can be seen in figure \ref{nilssoncoord}. Since this particular wavefunction cannot rotate around the symmetry axis, $M'$ is conserved and sum in \ref{hrotsol} drops out. Actually, due to the mathematical form of $\mathcal{D}^{(I)*}_{MM'}$ the associated eigenvalue of $J_{Z'}$ must be $M' = K$. Then, the associated normalized eigenfunction of \ref{hrotoraxially} will be
\begin{equation}
    \begin{aligned}
     \phi_{IMK}(\boldsymbol{\theta}) = \sqrt{\frac{2I+1}{8\pi^2}} \mathcal{D}^{(I)*}_{MK}(\boldsymbol{\theta}).
   \end{aligned} 
    \label{hrotsol1}
\end{equation}

Equation \ref{hrotoraxially} can be rewritten as 
\begin{equation}
    \begin{aligned}
     H^{ax.}_{rot.} = \frac{1}{2\Theta} \left(\boldsymbol{R}^2 - R^2_{Z'} \right),
     \end{aligned} 
    \label{hrotred}
\end{equation}
which results in the energy 
\begin{equation}
    \begin{aligned}
     E = \frac{1}{2\Theta}\hbar^2\left( I(I+1) -K^2 \right).
     \end{aligned} 
    \label{hrotenergy}
\end{equation}

\begin{figure}[h]
    \centering 
\includegraphics[width=0.3\textwidth]{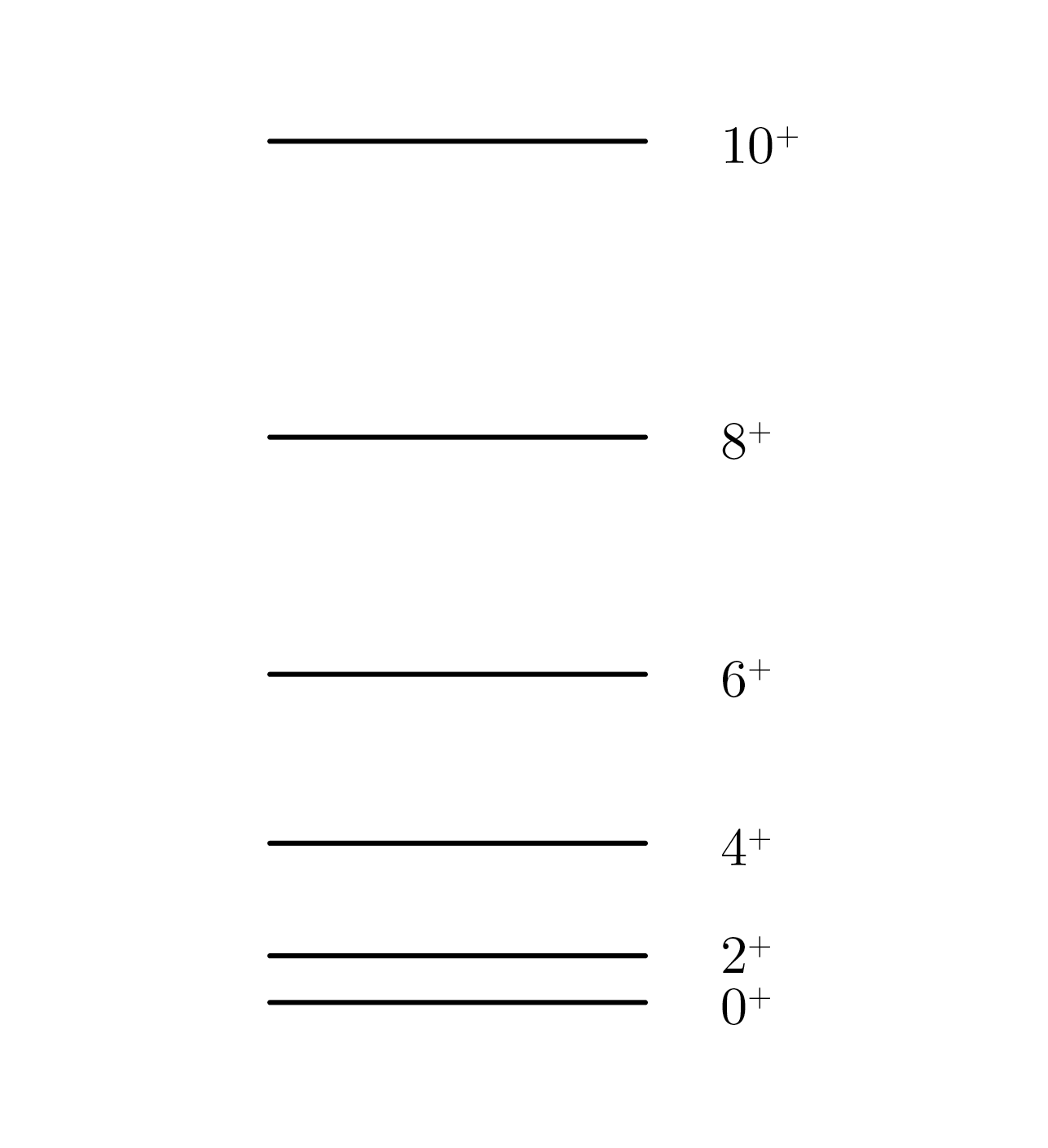}
    \caption{Spectrum of a rigid axially symmetric rotor. Levels of higher angular momenta are possible but not shown. Figure taken from \cite{greiner1996nuclear}.}
    \label{rotspectr}
\end{figure}

Because the nucleus cannot rotate around $Z'$, the following must hold
\begin{equation}
    \begin{aligned}
     \phi_{IMK}(\boldsymbol{\theta}) = e^{-\frac{i}{\hbar}\theta J_{Z'}}\phi_{IMK}(\boldsymbol{\theta}) = e^{-\frac{i}{\hbar}\theta K}\phi_{IMK}(\boldsymbol{\theta}),
     \end{aligned} 
    \label{kequalszero}
\end{equation}
implying that $K = $ 0. Another condition arises from the fact that an inversion of $Z'$ axis results in an equivalent description of the system, if this operation is represented by an operator $\mathcal{Z}$ such that $\mathcal{Z}^2 = \mathds{1}$ with eigenvalues $z=\pm 1$ it holds
\begin{equation}
    \begin{aligned}
     \mathcal{Z}\phi_{IM0}(\boldsymbol{\theta}) = \pm\phi_{IM0}(\boldsymbol{\theta}) =(-1)^I\phi_{IM0}(\boldsymbol{\theta}),
     \end{aligned} 
    \label{ieven}
\end{equation}
implying that $I$ can have even or odd values but not a mixture of them. Since reflection of all axes leave the state unchanged, the following property holds as well
\begin{equation}
    \begin{aligned}
     \boldsymbol{\pi}\phi_{IM0}(\boldsymbol{\theta}) = \pm\phi_{IM0}(\boldsymbol{\theta}) =\pi\phi_{IM0}(\boldsymbol{\theta}),
     \end{aligned} 
    \label{paritypositive}
\end{equation}
implying that the possible spectra are $I=0^{\pm},2^{\pm}, 4^{\pm},...$ or $I=1^{\pm}, 3^{\pm},5^{\pm},...$. The energy \ref{hrotenergy} is changed to 
\begin{equation}
    \begin{aligned}
     E = \frac{1}{2\Theta}\hbar^2 I(I+1) ,
     \end{aligned} 
    \label{hrotenergyc}
\end{equation}
and a typical spectrum with $z=\pi=+1$ and $I$ even can be seen in figure \ref{rotspectr}.

\subsubsection{Asymmetric Quadrupole Rotor}

The Hamiltonian \ref{hrotor} can be written as a diagonal and non-diagonal part as
\begin{equation}
    \begin{aligned}
     H_{rot.} = \frac{1}{4}\left(\frac{1}{\Theta_{\Xi'}} + \frac{1}{\Theta_{H'}} \right)\left(  \boldsymbol{J}^2 - J^2_{Z'}\right) + \frac{J^2_
    {Z'}}{2\Theta_{Z'}} + \frac{1}{2}\left(\frac{1}{\Theta_{\Xi'}} - \frac{1}{\Theta_{H'}} \right) \left(  J^2_{+} - J^2_{-}\right).
   \end{aligned} 
    \label{hrottriaxial}
\end{equation}

In a similar way as the last subsection, symmetries can be applied to condition $K$ and $I$. In this case inversion of $\Xi'$ and $H'$ axes denoted by $\mathcal{R}_{inv}$ must produce the same state and the relation  
\begin{equation}
    \begin{aligned}
     \phi_{IMK}(\boldsymbol{\theta}) = \mathcal{R}_{inv}\phi_{IMK}(\boldsymbol{\theta}) =e^{iK\pi}\phi_{IMK}(\boldsymbol{\theta}),
     \end{aligned} 
    \label{parityinversionxy}
\end{equation}
implies that $K$ must be even. Inversion of $Z'$ axis transforms $\mathcal{D}^{(I)*}_{MK}(\boldsymbol{\theta}) \rightarrow (-1)^I\mathcal{D}^{(I)*}_{M-K}(\boldsymbol{\theta})$ implying the need for symmetrization of $\phi_{IMK}(\boldsymbol{\theta})$ and after normalization it is obtained
\begin{equation}
    \begin{aligned}
     \phi_{IMK}(\boldsymbol{\theta}) = \sqrt{\frac{2I+1}{(1+\delta_{K0})18\pi^2}}\left( \mathcal{D}^{(I)*}_{MK}(\boldsymbol{\theta})+ (-1)^I\mathcal{D}^{(I)*}_{M-K}(\boldsymbol{\theta})
 \right).
     \end{aligned} 
    \label{inverionofz}
\end{equation}

Notice two aspects; for odd $I$ all $K = $ 0 states must vanish and $I = 1$ states are not allowed. The rule $I = K, K+1,K+2,K+3,...$ holds.

\subsubsection{Octupole Rotor}

\begin{figure}[h]
\centering 
\includegraphics[width=0.2\textwidth]{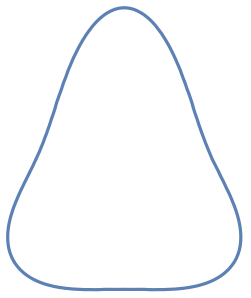}\hspace{5mm}
\includegraphics[width=0.25\textwidth]{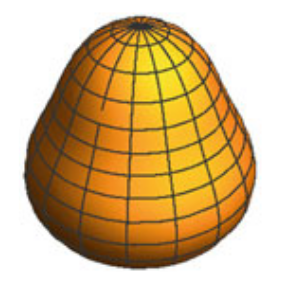}
    \caption{\textit{Left:} Equipotential shape of octupole deformed potential. Generated using Mathematica. \textit{Right:}
    Representation of octupole deformation of the atomic nuclei. Figure taken from \cite{kota20203}.}
    \label{pear}
\end{figure}

As was briefly mentioned before, the octupole $\lambda = $ 3 deformation combined with a quadrupole moment makes the nucleus resemble to a "pear" shape or to a "heart" shape \cite{heart} as shown in figure \ref{pear}. This shape respects the $K = $ 0 rule but no longer holds parity nor $Z'$ inversion axis symmetry \cite{obertelli2021modern, reflexionasymmetry}. However the operator 
\begin{equation}
    \begin{aligned}
\mathcal{S} = \boldsymbol{\pi}\mathcal{Z},
     \end{aligned} 
    \label{pizet}
\end{equation}
with eigenvalue 
\begin{equation}
    \begin{aligned}
s = \pi(-1)^I,
     \end{aligned} 
    \label{pizeteigenvalues}
\end{equation}
called simplex quantum number is conserved. That is expressed as 
\begin{equation}
    \begin{aligned}
     \phi_{IM0}(\boldsymbol{\theta}) = \mathcal{S}\phi_{IM0}(\boldsymbol{\theta}) = s\phi_{IM0}(\boldsymbol{\theta}),
     \end{aligned} 
    \label{pizeteapplication}
\end{equation}
which constrains the spectra to be of the form 
\begin{equation}
    \begin{aligned}
     &I^{\pi} = \begin{cases}
        &0^+,1^-,2^+,3^-,4^+,5^-,...\hspace{3mm}\text{if } s=+1\\
        &0^-,1^+,2^-,3^+,4^-,5^+,...\hspace{3mm}\text{if } s=-1.
        \end{cases}
     \end{aligned} 
    \label{spectraoctupole}
\end{equation}

This spectrum are observed in the nuclear region studied indicating a reflection asymmetry in their ground states.

A couple of statements are worth saying before closing this section. The vibrational dynamics will not be very relevant for this thesis so will not be discussed, however, it is very important for other nuclear regions and reference \cite{greiner1996nuclear} explains it with great detail. As is explained in \cite{Jenkins}, the experiments show that atomic nuclei do not behave as rigid rotors nor as irrotational fluids but as an intermediate.  This implies the need for additional considerations but in first approach these models will suffice for the purpose of this thesis.

\section{Application}
\begin{figure}[h!]
    \centering 
\includegraphics[width=0.6\textwidth]{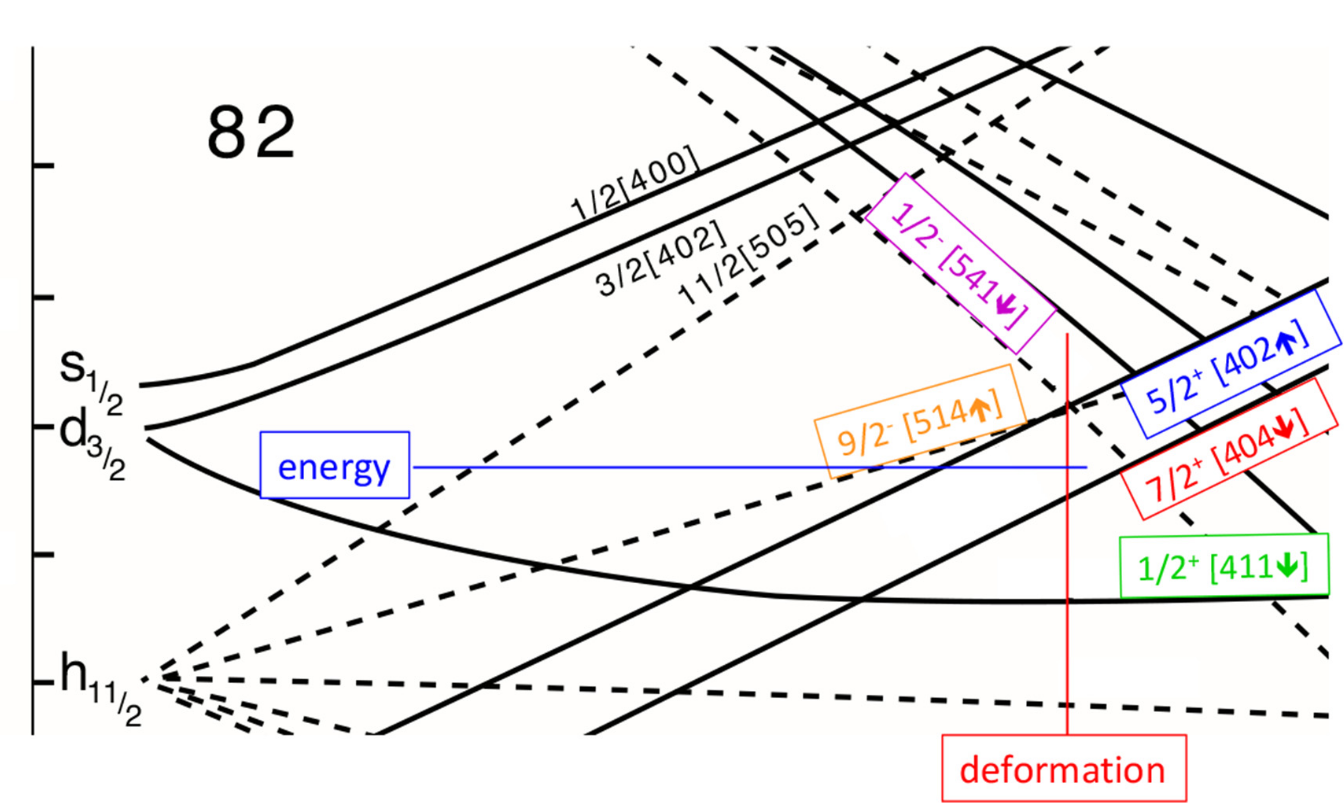}
    \caption{Nilsson energy levels for unpaired proton of $^{\text{175}}$Lu. Taken from \cite{Jenkins}}
    \label{Lulevels}
\end{figure}

Both models can be complemented to explain the electromagnetic spectra of several nuclei for which an example of core plus single-particle coupling with $\boldsymbol{J} \neq 0$ from \cite{Jenkins} is shown. Since Nilsson model focuses in single-particle orbitals, it is useful mainly for odd-$A$ nuclei where valence nucleons contributions are more easily recognizable. Consider the isotope $^{\text{175}}$Lu whose ground state quadrupole moment value is $\mathcal{Q} = $ 3.49 barns, so the nuclei can posses rotational motion. Calculating its energy and deformation parameter $\epsilon$, it can be found that its unpaired proton is in $\Omega^{\pi}[Nn_{\parallel}\Lambda\Sigma] = 7/2^{+}[404\downarrow]$ state. Figure \ref{Lulevels} shows this and adjacent energy levels the proton can be excited into.

\begin{figure}[h!]
    \centering 
\includegraphics[width=0.8\textwidth]{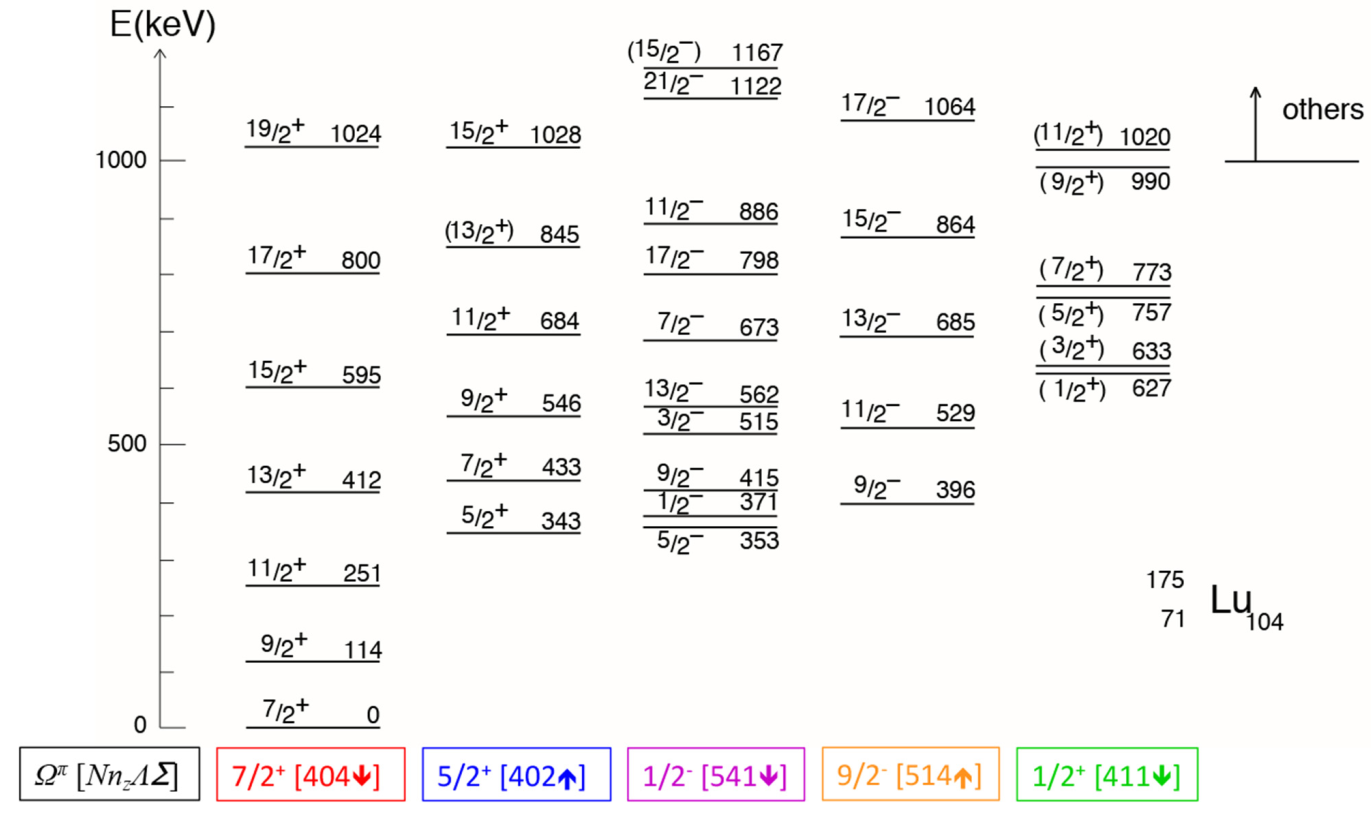}
    \caption{Rotational bands of $^{\text{175}}$Lu. Taken from \cite{Jenkins}}
    \label{Lubands}
\end{figure}

The Nilsson states allow the identification of band heads and organization of  bands due to the coupling of $\boldsymbol{R}$ and $\boldsymbol{J}$. This can be seen in figure \ref{Lubands} where the band heads are chosen as the single-particle states (or single hole for the case of $1/2^{+}[411\downarrow]$ where a proton is excited to $7/2^{+}[404\downarrow]$ state) according to Nilsson model and the rest of the band levels increase monotonically by 1 unit according to the particle plus core strong coupling \cite{greiner1996nuclear}. Details about the seemingly disorganized $1/2^{-}[541\downarrow]$ band can be found in \cite{Jenkins} and theoretical development of these phenomena in \cite{greiner1996nuclear}.

As was shown, the models combine to provide a more detailed description of nuclear spectra and dynamics. These will be the foundation for proxy-$SU(3)$ model and will allow greater understanding of spectra and transitions for the to be defined quartet heavy nuclei. Several more applications to nuclear decays, scattering and spectra can be found in \cite{nilsson1955,bohr1998nuclear, casten2000nuclear}. 

\chapter{Introduction to $SU(3)$ Symmetry in the Atomic Nucleus}
\InitialCharacter{A}s was already mentioned, analytical solutions to the Hamiltonians established in the last section are difficult to obtain and sometimes can get computationally expensive in spite of the simplifications to the many-body problem. To put this statement in a concrete example, the dimension of a basis consisting of $N$ valence neutrons and $Z$ valence protons distributed in $\Omega_N$ and $\Omega_Z$ states respectively, is given by the binomial coefficients poduct
\begin{equation}
\begin{aligned}
    \frac{\Omega_N!}{N!(\Omega_N-N)!}\frac{\Omega_Z!}{Z!(\Omega_Z-Z!)!},
\label{binomials}
\end{aligned}
\end{equation}
which for nuclear regions away from closed shells can rise up to orders of $10^{11}$ in the relatively low 28-50 shell\cite{isackerpaper}. This would imply the process of calculation and diagonalization of tremendously big matrix representations impossible to perform. Only \textit{a priori} considerations can help reduce this problem to most relevant matrix element contributions.

Due to this, a different theoretical framework has been developed taking advantage of the symmetries that the atomic nuclei possess to construct the observables and properties of the system. When speaking of symmetries, the competent branch of mathematics is group theory, particularly the Lie groups and algebras have been extensively applied. It will be seen that this framework is relevant for the atomic nuclei allowing progress on  the many-particle problem and the integration of the seemingly incompatible shell and liquid drop models in Elliot's $SU(3)$ along with its further revisions and extensions\cite{kota20203}. 

The purpose of this chapter is to provide a brief review of the concepts and applications of groups to the nuclear theory which will be crucial for the development and results of this thesis. The applications of $SU(3)$ to the atomic nuclei are too many to be treated here, so this chapter will focus on the quantum rotor and Elliot models exclusively. A certain knowledge on group theory and its applications to physics is assumed, so it is not intended to develop and show in detail several results, but to be a guide to the line of though followed. For detailed expositions a selected number of references are cited throughout the chapter.

\section{Symmetry in Quantum Mechanics}

In this section some concepts and notations will be briefly introduced concerning symmetries in quantum mechanics assuming knowledge in Lie groups and algebras. The main references for this section are \cite{frank2019symmetries, isackerpaper, iachello2014lie, hamermesh1989group}. A symmetry of a time independent system with respect to a Lie group $G$ with corresponding Lie algebra $\mathcal{G}$ having generators and parameters $g_k$ and $\alpha_k$ respectively, occurs if the Hamiltonian $H$ holds the commutation relation
\begin{equation}
\begin{aligned}
    [H, g_k] = 0 \implies  \left[H,e^{-i\alpha_k g_k}\right] = 0,
\label{algebra}
\end{aligned}
\end{equation}
for all $g_k$. In this way the notion of conserved quantities is expressed in terms of symmetries and thus the quantum numbers will be related directly to the Lie algebras labels. 

If $\mathcal{G}'$ is a subalgebra of $\mathcal{G}$, it is denoted as
\begin{equation}
\begin{aligned}
    \mathcal{G}\supset \mathcal{G'},
\label{chainoftwo}
\end{aligned}
\end{equation}
which are somehow related to the physical system (read next subsections) and whose irreducible representations are given by labels  $\Lambda$ and $\Lambda'$ respectively, then the quantum state will be denoted by 
\begin{equation}
\begin{aligned}
    \left|\Lambda \Lambda'\right>.
\label{labelsoftwo}
\end{aligned}
\end{equation}

A detailed description of the labels and Casimir operators of all relevant Lie algebras for physics can be found in \cite{iachello2014lie}. Equation \ref{chainoftwo} is called an algebra chain and can be generalized to further subalgebras in order to fully  characterize the associated states.

\subsection{Dynamical Symmetry Breaking}

Consider an algebra chain as in equation \ref{chainoftwo}. In many applications in physics, a symmetry $\mathcal{G}$ is too strong, so a symmetry breaking must be imposed without changing the state labels in order to describe the system appropriately. This can be achieved by imposing a Hamiltonian of the form 
\begin{equation}
\begin{aligned}
    H = \sum_m \left(\kappa_m \mathcal{C}_m[\mathcal{G}] + \kappa'_m \mathcal{C}_m[\mathcal{G'}]\right) ,
\label{dynamicalH}
\end{aligned}
\end{equation}
where $\mathcal{C}_m[\mathcal{G}]$ $(\mathcal{C}_m[\mathcal{G'}])$ denotes the Casimir operator of order $m$ of the algebra $\mathcal{G}$ $(\mathcal{G'})$ and $\kappa_m$ $(\kappa'_m)$ a proportionality coefficient. This way only
\begin{equation}
\begin{aligned}
    \left[H, g'_k\right] = 0,
\label{dynamicalHcommut}
\end{aligned}
\end{equation}
can be guaranteed.

For a more general chain 
\begin{equation}
\begin{aligned}
    \mathcal{G}_1\supset \mathcal{G}_2\supset...\supset\mathcal{G}_n,
\label{generalchain}
\end{aligned}
\end{equation}
it will be proposed analogously
\begin{equation}
\begin{aligned}
    H = \sum_{i,m} \kappa_{im} \mathcal{C}_{im}[\mathcal{G}_i].
\label{dynamicalHgeneral}
\end{aligned}
\end{equation}

The algebra $\mathcal{G}_1$ is called spectrum generating algebra while $\mathcal{G}_n$ is the symmetry algebra. The proposal \ref{dynamicalHgeneral} defines the  dynamical symmetry breaking. The energy eigenvalues will be given by 
\begin{equation}
\begin{aligned}
    H\left|\Lambda_1...\Lambda_n\right> = \left(\sum_{i,m} \kappa_{im} \left<\mathcal{C}_{im}[\mathcal{G}_i]\right>\right)\left|\Lambda_1...\Lambda_n\right> = E(\Lambda_1,...,\Lambda_n)\left|\Lambda_1...\Lambda_n \right>.
\label{dynamicalHgeneralenergy}
\end{aligned}
\end{equation}

A simple example can be thought with the algebra chain $SU(2) \supset SO(2)$, where the algebra $SU(2)$ with generators $T_k$ for $k=1,2,3$ and label $T$ describes isospin symmetry and the $SO(2)$ with only generator $T_3$ and label $M_T$  describes its projection. These can be used to describe the Coulomb interaction of isospin multiplets by means of a potential of the form
\begin{equation}
\begin{aligned}
&V = \kappa_0 + \kappa_1 T_3 + \kappa_3 T_3^2,\\
&\left<TM_T\right|V \left|TM_T\right>= \kappa_0 + \kappa_1 M_T + \kappa_3 M_T^2,
\label{VCOULOMB}
\end{aligned}
\end{equation}
where the presence of $T_3$ breaks $SU(2)$. This can be used to model atomic nuclei binding energy and mass in the so-called isobaric-multiplet mass equation\cite{frank2019symmetries}.

\subsection{Canonical Chains}

In order to uniquely characterize the basis, i.e., to provide a complete set of quantum numbers, a given algebra must be decomposed down to the lowest order non-trivial algebra as in the already introduced chains \ref{generalchain}. This brings two problems: the subalgebras decomposition determination and the corresponding irreducible representations (irreps) contained in a fixed irrep of a given spectrum generating algebra. For example $SU(3)$ can be decomposed as $SU(3)\supset SO(3) \supset SO(2)$ or  $SU(3)\supset [SU(2) \supset SO(2)]\otimes U(1)$. Concerning the second issue called the branching problem, lots of references and software have been developed \cite{DRAAYER1989279}. 

Consider a particular kind of chains called canonical chains of unitary and orthogonal algebras 
\begin{equation}
\begin{aligned}
&U(n)\supset U(n-1)\supset ... \supset U(1),\\
&SO(n)\supset SO(n-1)\supset ... \supset SO(2).
\label{canonical}
\end{aligned}
\end{equation}

The $U(n)$ irreps are denoted by $n$ numbers as $[\lambda_1, \lambda_2, ..., \lambda_n]$ holding $\lambda_1\geq\lambda_2\geq ...,\geq\lambda_n$, an equivalence to the corresponding $SU(n)$ irrep exists given by $[\lambda_1-\lambda_n, \lambda_2-\lambda_n, ..., \lambda_{n-1}-\lambda_n]$, then many results of canonical unitary chain can be extended to $SU(n)$.

These canonical chains are particularly useful mainly because of Ado's theorem which states that any Lie algebra is a subalgebra of $U(n)$, also for the fact that quantum many-particle systems under mean field interaction have $U(n)$ as spectrum generating algebra and because their branching problem is completely solved \cite{Gelfand1,Gelfand2}. However,  sometimes in physics non-canonical chains appear which bring more difficulties due to missing labels. For example, the non-canonical chain $SU(3)\supset SO(3) \supset SO(2)$ has four labels denoted as
\begin{equation}
\begin{aligned}
\left|(\lambda,\mu)LM\right>, 
\label{noncanonicallabels}
\end{aligned}
\end{equation}
while the canonical $SU(3)\supset U(2) \supset U(1)$ has five labels denoted as 
\begin{equation}
\begin{aligned}
\left|(\lambda,\mu)\hspace{0.5mm}[\lambda_1, \lambda_2]\hspace{0.5mm}[\lambda_0]\right>, 
\label{canonicallabels}
\end{aligned}
\end{equation}
thus there is a missing label in the non-canonical representation implying a multiplicity in the $SU(3)\supset SO(3)$ decomposition. This is corrected by the introduction of a label $K$ as 
\begin{equation}
\begin{aligned}
\left|(\lambda,\mu)KLM\right>.
\label{noncanonicallabelsK}
\end{aligned}
\end{equation}

This $K$ label is not the same as the projection of angular momentum introduced in the last chapter, and it will be used to organize states in rotational bands by the same values of $K$.

\subsection{Tensor Products}

In many cases in physics, several degrees of freedom must be coupled in order to consider all aspects of the system from a different perspective. This process is done by tensor product (also called Kronecker product) of the corresponding irreps and the determination of the irreps contained on it. For example two $U(n)$ representations can be coupled to a sum of the form
\begin{equation}
\begin{aligned}
[\lambda'_1, \lambda'_2, ..., \lambda'_n]\otimes[\lambda''_1, \lambda''_2, ..., \lambda''_n] = \sum\oplus [\lambda_1, \lambda_2, ..., \lambda_n]. 
\label{tensorprod}
\end{aligned}
\end{equation}

These computations are studied by Young calculus and can be performed by already developed computer software \cite{schurprogram}. It will be applied to coupling between proton and neutron degrees of freedom in the symmetry models of the atomic nuclei.

\subsection{Symmetries in Many-Body Quantum Systems}

For this section the formalism of second quantization is required. It introduces particle creation and destruction operators $c^{\dagger}_i$ and $c_i$ respectively in a state $i$. These operators will satisfy a certain algebra depending whether the system studied is composed of fermions or bosons. Defining commutation and anti-commutation relations for operators $u$ and $v$ in a single expression as 
\begin{equation}
\begin{aligned}
\left[u,v\right\}_q = uv-(-1)^{q}vu,
\label{commutationandanticommutation}
\end{aligned}
\end{equation}
with $q = $ 0 for bosons and $q = $ 1 for fermions, one can condense $c^{\dagger}_i$ and $c_i$ relations as
\begin{equation}
\begin{aligned}
&\left[c_i,c^{\dagger}_j\right\}_q = \delta_{ij},\\
&\left[c^{\dagger}_i,c^{\dagger}_j\right\}_q = \left[c_i,c_j\right\}_q = 0.
\label{candcdagger}
\end{aligned}
\end{equation}

The Fock state $\left|n_1...n_N\right>$ will be created from the vacuum $\left|0\right>$ as 
\begin{equation}
\begin{aligned}
\left|n_1...n_N\right> = \prod_i \frac{\left(c^{\dagger}_i\right)^{n_i}}{\sqrt{n_i!}}\left|0\right>,
\label{creation}
\end{aligned}
\end{equation}
where the total single-particle states quantity is $N$. The Hamiltonian must be rewritten in the second quantization formalism with the restriction of particles conservation. It is  
\begin{equation}
\begin{aligned}
H = \sum_i\epsilon_ic^{\dagger}_ic_i + \sum _{ijkl}\nu_{ijkl}c^{\dagger}_ic^{\dagger}_jc_kc_l + ...,
\label{Hsecond}
\end{aligned}
\end{equation}
where the first term accounts for one body interactions with parameter $\epsilon_i$, the second for two-body with parameter $\nu_{ijkl}$ and higher orders are omitted. Defining the $N^2$ operators $u_{ij} = c^{\dagger}_ic_j$ and using rules \ref{candcdagger}, the Hamiltonian can be rewritten as 
\begin{equation}
\begin{aligned}
H = \sum_{i}\epsilon_iu_{ii} -(-1)^q \sum_{ijl} \nu_{ijjl}u_{il} + (-1)^q\sum_{ijkl}\nu_{ijkl}u_{ik}u_{jl}+...
\label{Hsecondrewritten}
\end{aligned}
\end{equation}

If one considers the commutation relations of $u_
{ij}$, 
\begin{equation}
\begin{aligned}
\left[u_{ij},u_{kl}\right] = u_{il}\delta_{jk}-u_{kj}\delta_{il},
\label{uijcommutation}
\end{aligned}
\end{equation}
which are independent of fermionic or bosonic nature, it can be seen that they generate the $U(N)$ algebra. Then, it is noticed that only single-particle $\epsilon_i$ contribution of \ref{Hsecondrewritten} posses a $U(N)$ symmetry according to definition \ref{algebra} since this term is proportional to the Casimir operator of order one of such algebra. In conclusion, many-particle systems governed by mean field interactions will posses a $U(N)$ symmetry. As an example, consider the 3D harmonic oscillator
\begin{equation}
\begin{aligned}
H = \sum_{i=x,y,z}\left(\frac{p_i^2}{2m} + \frac{1}{2}m\omega^2x_i^2\right),
\label{harmonicsecondq}
\end{aligned}
\end{equation}
which can be rewritten as 
\begin{equation}
\begin{aligned}
H = \sum_{i=x,y,z}\left(u_{ii}+\frac{1}{2}\right) = n+\frac{3}{2} = \mathcal{C}_1[U(3)]+\frac{3}{2},
\label{harmonicsecondquii}
\end{aligned}
\end{equation}
and thus possesses a $U(3)$ symmetry. This small result will be crucial in the following sections.

Before starting with a more specific topic concerning nuclear structure, it is worth exposing the Racah form that consists in coupling the operators $b^{\dagger}_{lm}$ and $b_{lm}$ in the spherical basis which create and destroy respectively a particle with angular momentum $l$ and projection $m$ (state $\left|lm\right>$), so that its coupling transforms as a spherical tensor under $SO(3)$. This is defined as
\begin{equation}
\begin{aligned}
(b^{\dagger}_lb_{l'})^k_q = \sum_{m,m'}\left<lml'm'|kq\right>b^{\dagger}_{lm}\widetilde{b}_{l'm'},
\label{Racahform}
\end{aligned}
\end{equation}
where $\widetilde{b}_{lm} = (-1)^{l+m}b_{l-m}$ is called the adjoint of destruction operator $b_{lm}$. It will be extensively used in the following sections.

\section{$SU(3)$ Symmetry, Quantum Rotor and Elliott Model}

As was shown in the last section, many-particle quantum systems under mean field interaction posses a $U(N)$ symmetry, which with the development of shell model seemed like a potential application area. It was J. P. Elliott who showed that the nuclear shell model admits a $SU(3)$ symmetry and calculated the observables related to such algebra\cite{ELLIOTTMODEL1958}. More remarkably, he integrated them in a way that two body interactions are considered by means of quadrupole residual forces which allowed to account for collective rotational spectra within the independent particle picture. However, the model has some limitations given mainly by the strong nuclear spin-orbit interaction which breaks $SU(3)$ symmetry more notably in higher shell orbitals and requires further considerations in order to extend it. This section will focus in the main characteristics of $SU(3)$ required by the formulation of Elliott and then shows his model along with its limitations.

\subsection{Relevant Properties of $SU(3)$}

$SU(3)$ is the first non trivial group where the complications of a general group appear. Some of these are the  multiplicity in the couplings of representations $(\lambda_1,\mu_1)\otimes(\lambda_2,\mu_2)$, its different subgroup chains $SU(3)\supset SO(3)\supset SO(2)$ and $SU(3)\supset [SU(2) \supset SO(2)]\otimes U(1)$ which result in different coupling coefficients and the loss of one label in chain $SU(3)\supset SO(3)$ as explained in section 3.1.2. 

In general, the generators of $SU(N)$ are obtained from those of $U(N)$ of \ref{uijcommutation} only changing the diagonal ones by 
\begin{equation}
\begin{aligned}
\widetilde{u}_{ii} = u_{ii} -\frac{1}{n}\sum_{j=1}^N u_{jj}.
\label{diagonalsun}
\end{aligned}
\end{equation}

This causes the Casimir operator of first order to be zero
\begin{equation}
\begin{aligned}
C_1[SU(N)]=\sum_{i}\widetilde{u}_{ii}=0 .
\label{casimirzero}
\end{aligned}
\end{equation}

Returning to the particular case of $SU(3)$, it is of interest the eigenvalues of the Casimir operators of orders 2 and 3 as will be seen soon. The algorithm to obtain them is described in \cite{iachello2014lie} and in this thesis will be adopted up to constant factors as
\begin{equation}
\begin{aligned}
&\left<(\lambda,\mu)\right|C_2[SU(3)]\left|(\lambda,\mu)\right> =\left<C_2[SU(3)]\right>= \lambda^2+\mu^2+\lambda\mu+3(\lambda+\mu),\\
&\left<(\lambda,\mu)\right|C_3[SU(3)]\left|(\lambda,\mu)\right> = \left<C_3[SU(3)]\right>=(\lambda-\mu)(\lambda+2\mu+3)(2\lambda+\mu+3).
\label{casimir2and3}
\end{aligned}
\end{equation}

The chain of interest for this thesis is $SU(3)\supset SO(3)\supset SO(2)$, henceforth special attention will be given to it. The branching problem of the representation in the chain $SO(3)\supset SO(2)$ is simply $-L\leq M\leq L$ but the $SU(3)\supset SO(3)$ requires the addition of a missing label denoted as $K$. This imposes the rules
\begin{equation}
\begin{aligned}
&K = \text{min}(\lambda,\mu), \text{min}(\lambda,\mu)-2, \text{min}(\lambda,\mu)-4,...,0\hspace{1mm}\text{or}\hspace{1mm}1,\\
&L = \begin{cases}
        &K,K+1,K+2,K+3,...,K+\text{max}(\lambda,\mu),\hspace{3mm}\text{if}\hspace{1mm} K\neq 0\\
        &\text{max}(\lambda,\mu),\text{max}(\lambda,\mu)-2,\text{max}(\lambda,\mu)-4,...,0\hspace{1mm}\text{or}\hspace{1mm}1,\hspace{3mm}\text{if}\hspace{1mm} K=0,
        \end{cases}
\label{Krules}
\end{aligned}
\end{equation}
and the associated states to this chain will be denoted as $\left|(\lambda,\mu)KLM\right>$.

In the next section it will be shown that the generators of $SU(3)$ are the three components of the angular momentum $L^{(1,1),K=1,L=1}_q$ and the five of the quadrupole operator $Q^{(1,1),K=1,L=2}_q$ (see appendix E). These allow the definition of the tensor operators $T^{(\lambda,\mu)KL}_M$ with respect to $SU(3)$ as 
\begin{equation}
\begin{aligned}
\relax [G^P_q, T^{(\lambda,\mu)KL}_M] = \sum_{K',L',M'}\left<(\lambda,\mu)K'L'M'\right|G^P_q\left|(\lambda,\mu)KLM\right>T^{(\lambda,\mu)K'L'}_{M'},
\label{SU3TENSORS}
\end{aligned}
\end{equation}
where $G^P_q= L^{(1,1),K=1,L=1}_q$ or $G^P_q=Q^{(1,1),K=1,L=2}_q$ and the Wigner-Eckart theorem
\begin{equation}
\begin{aligned}
&\left<\gamma'(\lambda',\mu')K'L'M'\right|T_{M_0}^{\gamma_0(\lambda_0,\mu_0)K_0L_0}\left|\gamma(\lambda,\mu)KLM\right> =\\
& \sum_{\rho}\left<\rho\gamma'(\lambda',\mu') |||T^{\gamma_0(\lambda_0,\mu_0)}|||\gamma(\lambda,\mu)\right>\left<\rho(\lambda,\mu)KLM;(\lambda_0,\mu_0)K_0L_0M_0|(\lambda',\mu')K'L'M'\right>, 
\label{WESU3}
\end{aligned}
\end{equation}
where $\gamma$ indicates upper algebra labels and $\rho$ the multiplicity of the couplings. In appendix B the general expressions for arbitrary algebras of these results are shown. 

The Wigner (or Clebsch-Gordan) coefficients can be made independent of the projections $M$ by means of Racah factorization lemma (see appendix B). Exemplified in the coupling of two states as 
\begin{equation}
\begin{aligned}
&\left| \hspace{1mm}\{ \left|(\lambda_1,\mu_1)K_1L_1\right> \left|(\lambda_2,\mu_2)K_2L_2\right> \}\hspace{1mm} LM\right> =\\&\sum_{\rho,(\lambda,\mu),K,L}\left<\rho(\lambda_1,\mu_1)K_1L_1;(\lambda_2,\mu_2)K_2L_2||(\lambda,\mu)KL\right>\left|\rho(\lambda,\mu)KLM\right>,
\label{racahfactorizationSU3SO3}
\end{aligned}
\end{equation}

where the double-barred coefficients are called the reduced Wigner coefficients and hold similar properties \cite{kota20203}. This last result will be important in the calculations of electromagnetic reduced transition probabilities. They can be computed by the code in \cite{su3lib}.

\subsection{$SU(3)$ in the Harmonic Oscillator}

Consider the quantum isotropic harmonic oscillator for a single-particle
\begin{equation}
\begin{aligned}
&H_0 = \frac{p^2}{2m} + \frac{1}{2}m\omega^2 r^2 = \mathcal{N}+\frac{3}{2},
\label{qisotropicho}
\end{aligned}
\end{equation}
whose energy eigenvalue in spherical coordinates is given by the radial $n$, orbital angular momentum $l$ and principal quantum number $\eta$ as
\begin{equation}
\begin{aligned}
E_{nl}=2n+l&+\frac{3}{2}=\eta+ \frac{3}{2}\\
\eta &= 2n+l,\\
\implies\hspace{1mm}l=\eta,\eta-2&,\eta-4,...,0 \hspace{1mm}\text{or}\hspace{1mm}1
.
\label{qisotropiceigenvalue}
\end{aligned}
\end{equation}

Every particle in the shell $\eta$ carries one of the allowed orbital angular momentum $l$ values independent of its fermionic or bosonic nature. The degeneracy in such shell is $\frac{(\eta+1)(\eta+2)}{2}$ which indicate that the associated symmetry of the spatial part of the system is $U(\frac{(\eta+1)(\eta+2)}{2})$ and is generated by the Racah form of the operators $(b^{\dagger}_lb_{l'})^k_q$. The commutation relation of such algebra was computed in appendix C where the $b^\dagger_{lm}$ and $b_{lm}$ are bosonic operators since we are dealing only with the spatial part.

The spectrum generating algebra $U(\frac{(\eta+1)(\eta+2)}{2})$ can be decomposed into $SU(3)$ by the already mentioned branching problem. More explicitly as $U(\frac{(\eta+1)(\eta+2)}{2})\supset U(3)\supset SU(3)\supset SO(3) \supset SO(2)$. This algebra $SU(3)$ is relevant since it is generated by the well known single-particle three angular momentum components and a modified version of the five components of the quadrupole operator containing a linear momentum $p^2$ term as 
\begin{equation}
\begin{aligned}
&L^1_q = \left[\boldsymbol{r}\times\boldsymbol{p} \right]^1_q,\\
&Q^2_q = \sqrt{\frac{4\pi}{5}}\frac{m\omega}{\hbar}\left[r^2Y^2_{q}(\Omega)+p^2Y^2_{q}(\Omega_p)\right],
\label{qmodified}
\end{aligned}
\end{equation}
in order for the algebra to be constructed \cite{Castanos1988} and to ensure no mixing of different oscillator shell states. In several calculations the reduction $\frac{m\omega}{\hbar} = 1$ is used. These hold the following commutation relations, two of which are demonstrated in the appendix D  
\begin{equation}
\begin{aligned}
&\left[L^1_q,L^1_{q'}\right]= -\sqrt{2}\left<1\hspace{0.5mm}q\hspace{0.5mm}1\hspace{0.5mm}q'\hspace{0.5mm}|\hspace{0.5mm}1\hspace{0.5mm}q+q'\right>L^1_{q+q'},\\
&\left[L^1_q,Q^2_{q'}\right]= -\sqrt{6}\left<1\hspace{0.5mm}q\hspace{0.5mm}2\hspace{0.5mm}q'\hspace{0.5mm}|\hspace{0.5mm}2\hspace{0.5mm}q+q'\right>Q^2_{q+q'},\\
&\left[Q^2_q,Q^2_{q'}\right]=3\sqrt{10}\left<2\hspace{0.5mm}q\hspace{0.5mm}2\hspace{0.5mm}q'\hspace{0.5mm}|\hspace{0.5mm}1\hspace{0.5mm}q+q'\right>L^1_{q+q'}.
\label{commutatorsofSU3}
\end{aligned}
\end{equation}

The operators $L^1_q$ and $Q^2_q$ along with $\mathcal{N} = H-3/2$ form the $U(3)$ algebra. In order to consider the many particles in a given shell, the sumation over each particle must be considered in \ref{qmodified}. The relevant Casimir operators of $SU(3)$ are given by \cite{casimirsSU3}
\begin{equation} 
\begin{aligned}
&\mathcal{C}_2[SU(3)] = \frac{1}{4}\left(\boldsymbol{Q}^2\cdot\boldsymbol{Q}^2+3\boldsymbol{L}^1\cdot\boldsymbol{L}^1\right),\\
&\mathcal{C}_3[SU(3)] = \frac{1}{36}\sqrt{\frac{7}{2}}\left[\boldsymbol{Q}^2\times\boldsymbol{Q}^2\right]^2\cdot\boldsymbol{Q}^2-\frac{1}{4}\sqrt{\frac{5}{2}}\left[\boldsymbol{L}^1\times\boldsymbol{Q}^2\right]^1\cdot\boldsymbol{L}^1,
\label{casimirsql}
\end{aligned}
\end{equation}

where the products are given by the tensor couplings of equation \ref{Tensorcoupling}. Notice also that $\left[\boldsymbol{L}^1\times\boldsymbol{L}^1\right]^0_0 = \boldsymbol{L}^1\cdot\boldsymbol{L}^1 = \boldsymbol{L}^2$ is the Casimir operator of $SO(3)$. These results will allow the introduction of rotational spectra in the spherical shell model by means of dynamical symmetry breaking as shown in the next section.

\section{Quantum Rotor}

In references \cite{LESCHBER19871,CASTANOS198871, Castanos1988} is shown that it is possible to map the quantum rotor Hamiltonian $H_{rot.}$ of equation \ref{hrotor} in terms of $SU(3)$ operators. It is done by means of the $SU(3)\supset SO(3)$ integrity basis operators which are defined as the $SO(3)$ scalars constructed from $SU(3)$ operators. For three and four bodies these are \cite{kota20203}
\begin{equation} 
\begin{aligned}
&X_3 = \sqrt{3}\left[\boldsymbol{L}^1\times\boldsymbol{Q}^2\right]^1\cdot\boldsymbol{L}^1,\\
&X_4 = \sqrt{3}\left[\boldsymbol{L}^1\times\boldsymbol{Q}^2\right]^1\cdot\left[\boldsymbol{Q}^2\times\boldsymbol{L}^1\right]^1.
\label{integritybasisop}
\end{aligned}
\end{equation}

And then the Hamiltonian \ref{hrotor} can be expressed as
\begin{equation} 
\begin{aligned}
H_{int.} = aL^2+bX_3+cX_4,
\label{hrotrew}
\end{aligned}
\end{equation}
where the relation between constants $a,b,c$ and moments of inertia can be found in \cite{LESCHBER19871}. This result demonstrates the goodness of the description of nuclear rotations by means of $SU(3)$ algebra. In reference \cite{Castanos1988} is shown the comparisons of spectra and electromagnetic transitions between $H_{rot.}$, $H_{int.}$ and experimental data.

Another relevant result is the relation between surface deformation parameters $\beta$, $\gamma$ and labels $(\lambda,\mu)$ computed by means of equations \ref{betairrot},\ref{betairrot2} and \ref{casimirsql} resulting in \cite{ROWEBETAGAMMA}
\begin{equation} 
\begin{aligned}
&\beta^2 =\frac{16\pi}{5N_0^2}(\lambda^2+\mu^2+\lambda\mu),\\
&N_0=0.9A^{4/3},\\
&\gamma = \text{arctan}\left(\frac{\sqrt{3}\mu}{2\lambda+\mu}\right).
\label{betagammalambdamu}
\end{aligned}
\end{equation}

This result connects directly the collective model and the symmetry framework of the atomic nuclei since representations of the type $(0, 0)$ correspond to spherical, $(\lambda, 0)$ to prolate, $(0,\mu)$ to oblate and arbitrary $(\lambda, \mu)$ to triaxial shapes\cite{CASTENALGEBRA, CSEH201959}.

\subsection{Elliott Model}

J. P. Elliot was the first to realize the potential of $SU(3)$ symmetry in the nuclear structure\cite{ELLIOTTMODEL1958}, so he modelled the nuclear interaction under the assumption of Wigner spin-isospin $SU(4)$ symmetry \cite{Wignerspinisospin, frank2019symmetries} where protons and neutrons are on the same oscillator shell and interact by means residual two body quadrupole forces. The spectrum generating algebra decomposition is
\begin{equation} 
\begin{aligned}
U\left(4\cdot\frac{(\eta+1)(\eta+2)}{2}\right)\supset U\left(\frac{(\eta+1)(\eta+2)}{2}\right)\otimes SU(4),
\label{sgqofelliott}
\end{aligned}
\end{equation}

where the first product accounts for the spatial part and the second for the spin-isospin. The spatial part should be decomposed further as shown in the last section in order to account for the rotational spectrum. This implied the dynamical symmetry breaking of the spectrum generating algebra, thus, Elliott proposed the Hamiltonian
\begin{equation} 
\begin{aligned}
H_{E} = H_0 - \kappa \boldsymbol{Q}^2\cdot\boldsymbol{Q}^2,
\label{Helliott}
\end{aligned}
\end{equation}
where $H_0$ comes from \ref{qisotropicho} and $\kappa$ is a proportionality constant. Using \ref{casimirsql} it can be rewritten as 
\begin{equation} 
\begin{aligned}
H_{E} = H_0 - \kappa \left(4\mathcal{C}_2[SU(3)]-3\boldsymbol{L}^2\right),
\label{Helliottrewritten}
\end{aligned}
\end{equation}
with associated eigenvalue
\begin{equation} 
\begin{aligned}
E_E = \mathcal{N}+\frac{3}{2}- \kappa 4(\lambda^2+\mu^2+\lambda\mu+3(\lambda+\mu))-3\kappa L(L+1).
\label{Helliotteigenvalue}
\end{aligned}
\end{equation}

\begin{figure}[h]
    \centering 
\includegraphics[width=1\textwidth]{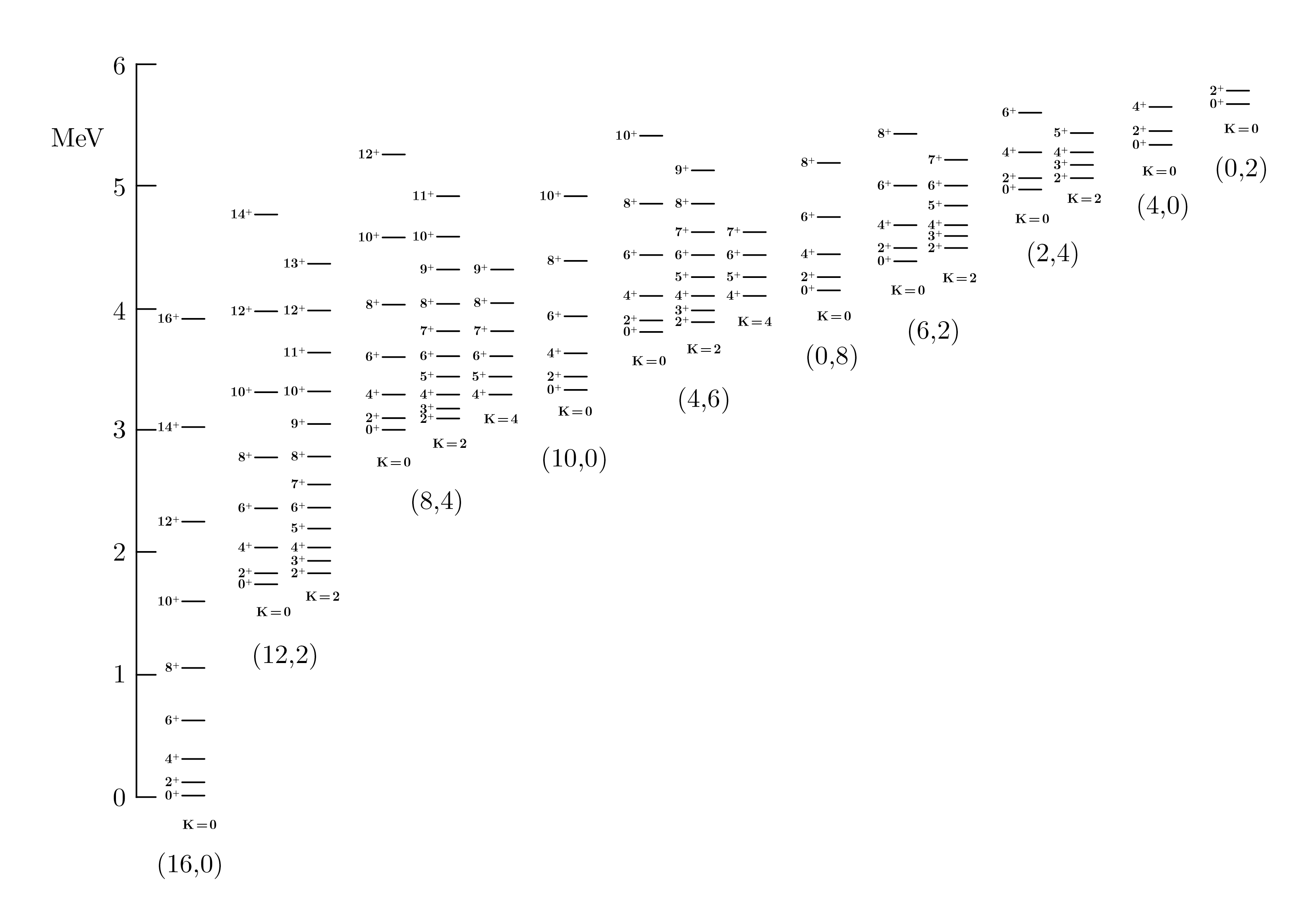}
    \caption{$SU(3)$ spectrum with associated irreps and values of $K$. Taken from \cite{talmi2017simple}. }
    \label{su3spectrum}
\end{figure}

For an attractive quadrupole interaction, the ground state is given by the representation corresponding to the maximum value of $2\lambda+\mu$ or equivalently, the highest value of second order Casimir operator. This representation is called the leading irrep and is the main bias for truncation of the full model spaces when calculating the energy eigenspectrum. The $\boldsymbol{L}^2$ term generates the rotational spectrum for fixed $\mathcal{N}$ and irrep $(\lambda,\mu)$ which along with the label $K$ that allows the identification of rotational bands incorporates collective rotation in the single-particle framework. An $SU(3)$ spectrum can be seen in figure \ref{su3spectrum} along with its associated representations.

Elliott model has been crucial for the introduction of algebraic approaches to the atomic nuclei because of how it connects seemingly incompatible aspects. Many corrections and extensions have been developed considering multi-shell mixing \cite{Isacker_2016} and trying to apply it to heavier elements where the strong spin-orbit interaction breaks the $SU(3)$ symmetry. This last issue has been worked along with Nilsson model in order to approximately restore $SU(3)$ in heavy nuclei in a few models\cite{kota20203}, one of which is introduced in the next chapter and that is the basis of this thesis: the proxy-$SU(3)$ symmetry.

 \chapter{The Proxy-$SU(3)$ Scheme}
\begin{figure}[h]
    \centering 
\includegraphics[width=0.3\textwidth]{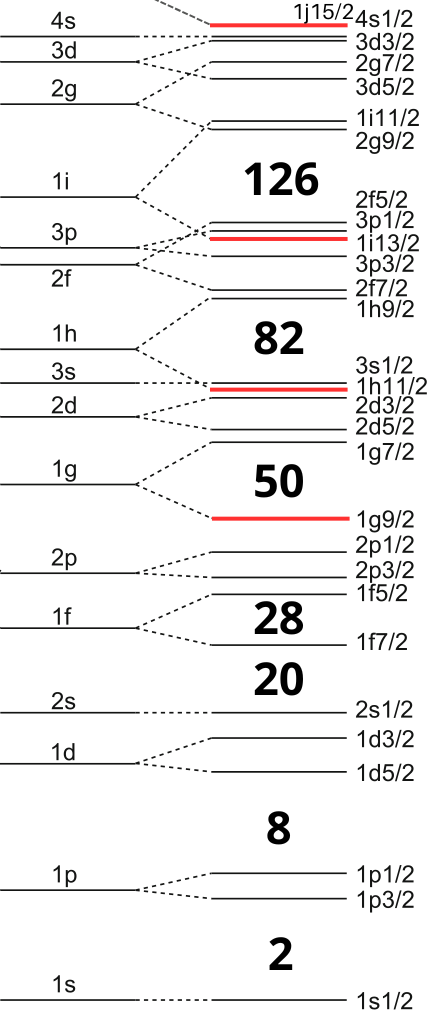}
    \caption{Mixing of levels by spin-orbit interaction starting from a Wood-Saxon potential. The intruder levels are shown in red. The same phenomenon appears in higher shells but are not shown. Figure adapted from \cite{takigawa}}
    \label{shellmixing}
\end{figure}

\InitialCharacter{I}t is well known that the spin-orbit interaction in the atomic nuclei is not negligible and influences directly its structure. In the shell model framework, beyond the $sd$ (or 8-20) shell this interaction causes the highest-$j$ state to be pushed to the shell below setting up a different structure than the harmonic oscillator. This level is called the intruder (or deserter) since it possesses a different parity ($(-1)^N$) than its partner states in such shell. This phenomenon can be visualized in figure \ref{shellmixing} and in table \ref{table1} are shown the intruder states up to $\eta$ = 6.

In terms of algebra and symmetries this causes the breaking of Wigner's $SU(4)$ supermultiplet and the harmonic oscillator $SU(3)$ in medium-mass and heavy nuclei. All the theory developed by Elliott is then valid as long as protons and neutrons occupy essentially the same states and where spin-orbit interaction is weak enough so that the nuclear energy levels resemble those of the harmonic oscillator, condition only fulfilled by light nuclei. Such a rich and physically insightful theory is worth adapting to heavier mass regions such that, at least approximately, the $SU(3)$ symmetry can be restored.

To achieve this, a few corrections have been proposed taking advantage of partial recovery of degeneration due to deformation and spatial overlap of certain nuclear states \cite{kota20203}. These are called the pseudo-$SU(3)$, quasi-$SU(3)$ and proxy-$SU(3)$ models. The later was proposed in 2017 by D. Bonatsos $et$ $al.$ \cite{PhysRevC.95.064325} as a novel approach to restoring $SU(3)$ symmetry in heavy nuclei. This particular scheme will be treated in this chapter covering its physical and mathematical foundations along with its main results so far.

\begin{table}[h]
\begin{center}
\begin{tabular}{ |c | c| c |c| } 
  \hline
  Principal quantum number $\eta$ & Shell & Intruder state & Parity \\ 
  \hline
   3 &28-50 & $1g^{9/2}$ & $+$ \\ 
  \hline
  4 & 50-82& $1h^{11/2}$ & $-$ \\ 
  \hline
  5 & 82-126& $1i^{13/2}$ & $+$ \\ 
  \hline
  6 &126-184 & $1j^{15/2}$ & $-$ \\ 
  \hline
\end{tabular}
  \caption{Intruder states in the shell model with spin-orbit interaction.}
    \label{table1}
\end{center}
\end{table}

\section{The 0[110] Pairs}

\begin{figure}[h!]
    \centering 
\includegraphics[width=1\textwidth]{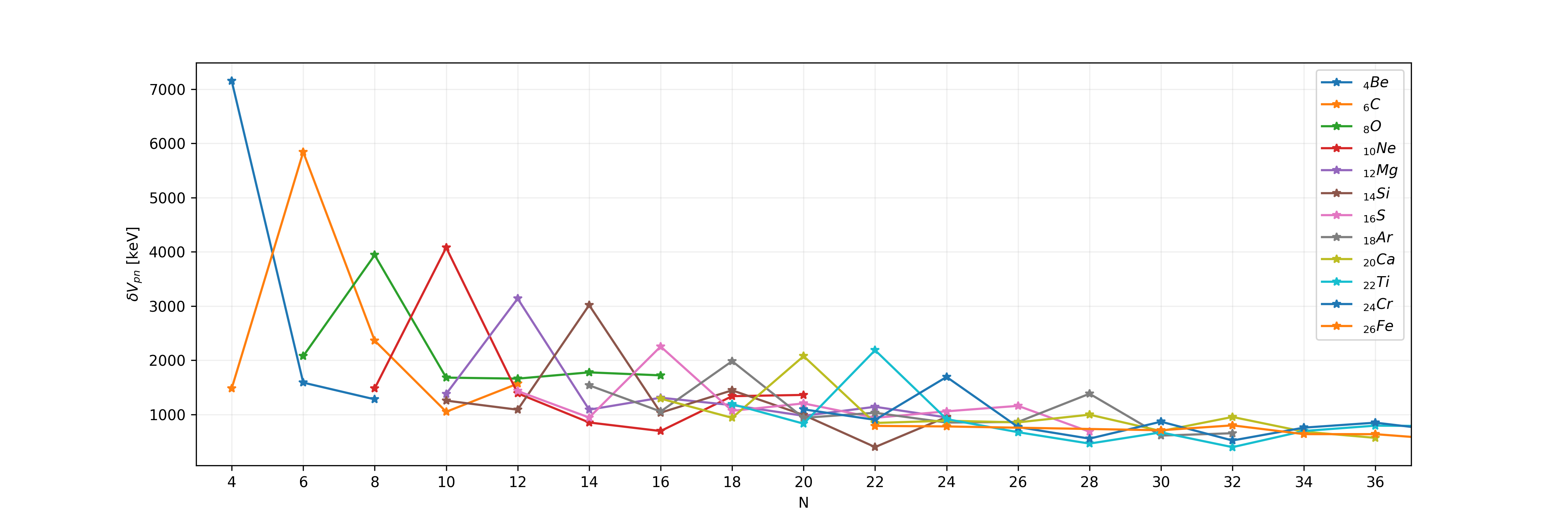}
    \caption{$\delta V_{pn}$ for even-even light elements where the maximization is observed at $N = Z$. Data retrieved from \cite{NNDC} and figure generated using Matplotlib.}
    \label{VpnLight}
\end{figure}

\begin{figure}[h!]
    \centering 
\includegraphics[width=1\textwidth]{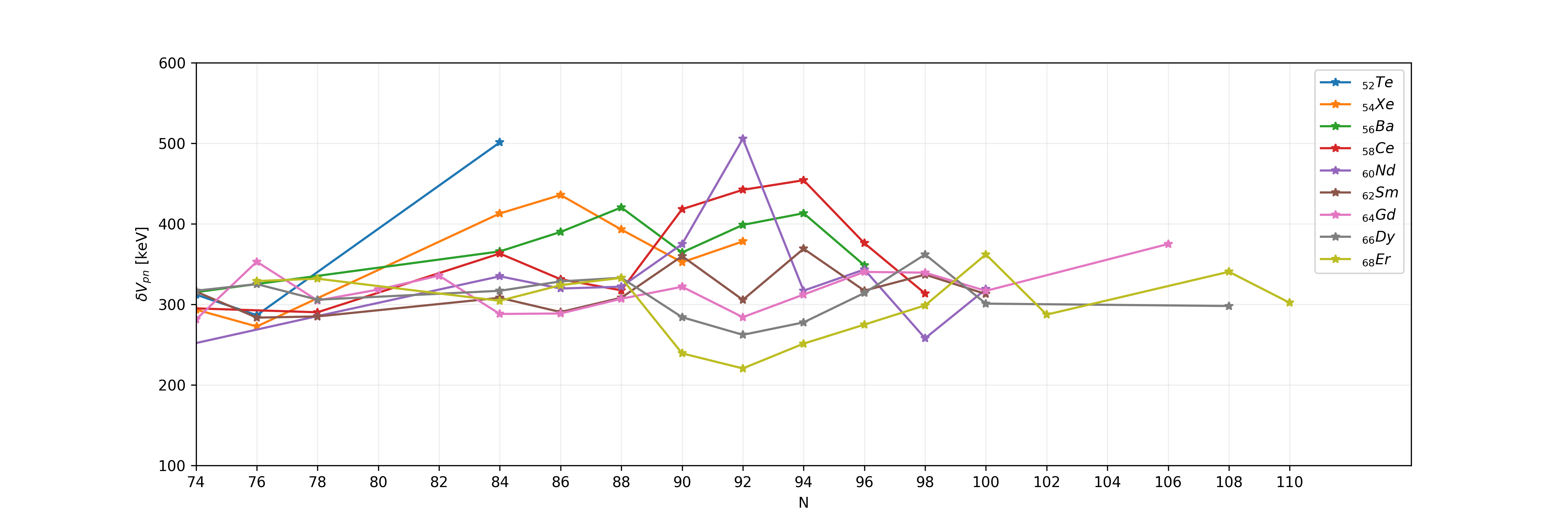}
    \caption{$\delta V_{pn}$ for some even-even elements in proton shell 50-82 where the maximization is observed around $N_{valence} \approx Z_{valence}$. Data retrieved from \cite{NNDC} and figure generated using Matplotlib.}
    \label{VpnMedium}
\end{figure}

\begin{figure}[h!]
    \centering 
\includegraphics[width=1\textwidth]{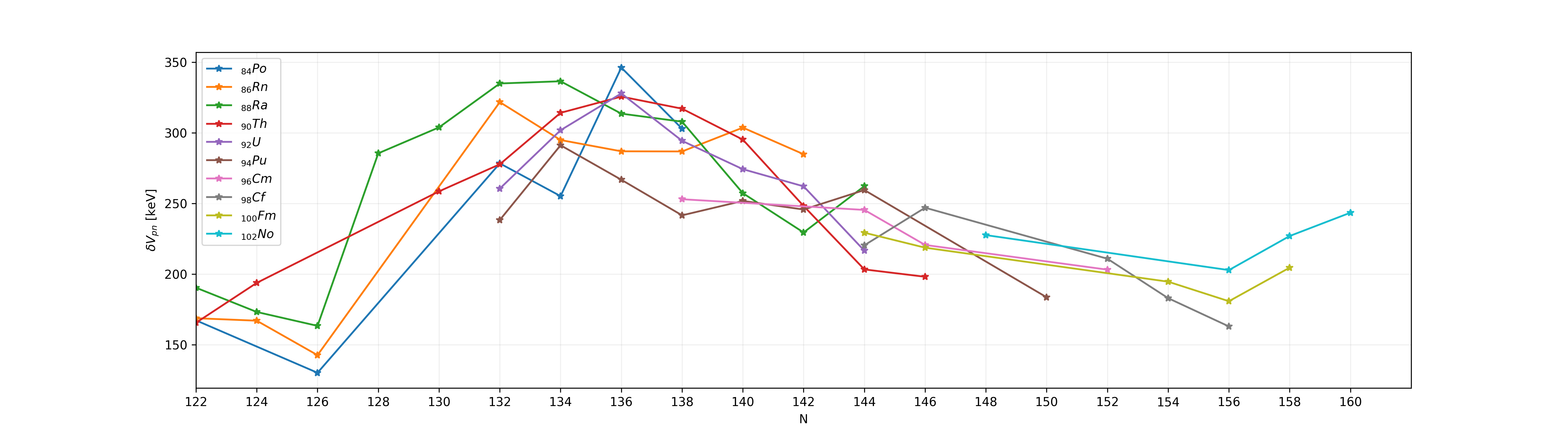}
    \caption{$\delta V_{pn}$ for some even-even  elements in proton shell 82-126 where the maximization is also observed around $N_{valence} \approx Z_{valence}$. Data retrieved from \cite{NNDC} and figure generated using Matplotlib.}
    \label{VpnHeavy}
\end{figure}

In many-body quantum systems, correlations between the constituent particles frequently emerge from the microscopic interactions which directly influence the structure and characteristics of the system. It is also the case for atomic nuclei where the different orbitals and states might interfere to escalate certain nuclear properties. Consider the double binding energy differences defined for even-even nuclei as \cite{TestWigner}
\begin{equation} 
\begin{aligned}
\delta V_{pn} = \frac{1}{4}\left(B(Z,N)+B(Z-2,N-2)-B(Z-2,N)-B(Z,N-2)\right),
\label{dVpn}
\end{aligned}
\end{equation}

it expresses the strength of the interaction between the least bound two protons and two neutrons. It has been shown \cite{Wignerlikeenergy} that this nuclear property maximizes at $Z = N$ for light nuclei because of the large overlap between the proton and neutron wavefunctions since in this region the spin-isospin $SU(4)$ symmetry is approximately maintained. This is shown for a few elements in figure \ref{VpnLight}. However, this condition is not fulfilled for medium and heavy nuclei since the $Z=N$ isotopes lie far beyond the  proton drip line. For these nuclei it has been discovered that the maximization of $\delta V_{pn}$ is found around $N_{valence}\approx Z_{valence}$, at least locally, given the complexity of the nuclear interaction of these isotopes. A relevant property is that maximization of $\delta V_{pn}$ is highly correlated to the onset of collectivity \cite{PhysRevC.88.054309}. The plots of $\delta V_{pn}$ for shells 50-82 and 82-126 can be found  in figures \ref{VpnMedium} and \ref{VpnHeavy} respectively.

\begin{figure}[h!]
    \centering 
\includegraphics[width=0.4\textwidth]{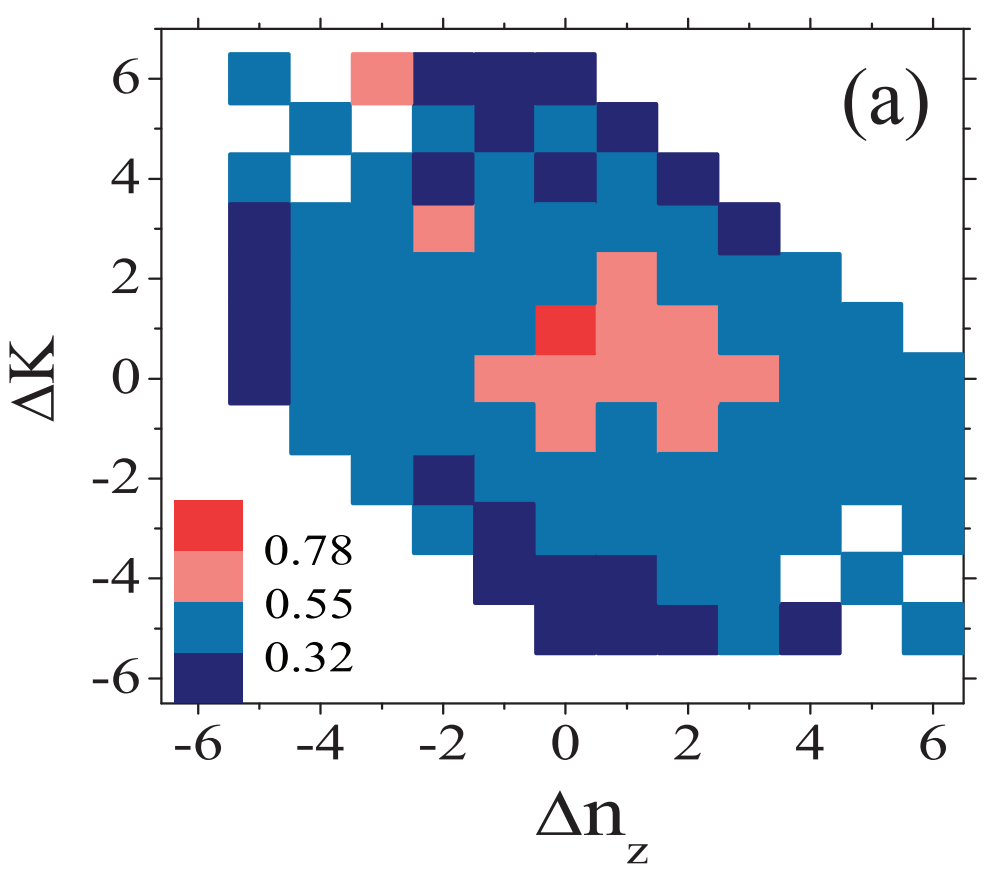}
    \caption{Average spatial overlap for proton and neutron states in $Z = $50-82 and $N =$82-126 shells. Figure taken from \cite{PhysRevC.88.054309}.}
    \label{Overlapin5082}
\end{figure}

Investigations trying to explain this phenomena \cite{PhysRevC.88.054309} using Nilsson wavefunctions as
\begin{equation} 
\begin{aligned}
\chi_{\eta\Omega} = \sum_{l,\Lambda}a^{\Omega}_{l\Lambda}\left|\eta l\Lambda\Sigma\right>,
\label{NilssonCoeffs}
\end{aligned}
\end{equation}
with coefficients $a^{\Omega}_{l\Lambda}$ obtained by solving the Nilsson Hamiltonian for medium deformations and wavefunctions overlaps defined as
\begin{equation} 
\begin{aligned}
\mathcal{O} = \int (\chi^*_{\eta_1\Omega_1}\chi_{\eta_1\Omega_1})(\chi^*_{\eta_2\Omega_2}\chi_{\eta_2\Omega_2})dV,
\label{Overlap}
\end{aligned}
\end{equation}
found that proton-neutron orbitals related by $\Delta K[\Delta \eta \Delta n_z \Delta\Lambda] = 0[110]$ have large $\mathcal{O}$ values and thus isotopes with simultaneous occupations of states following that relations show a $\delta V_{pn}$ increase (recall that for axially symmetric nuclei $K = \Omega$ and the notation $n_z = n_{\parallel}$ is adopted). The average spatial overlap for isotopes in the proton 50-82 and neutron 82-126 shell are shown in figure \ref{Overlapin5082}. This effect of high spatial overlaps of these states has been corroborated by other methodologies such as density functional theory \cite{PhysRevLett.98.132502} and Monte-Carlo shell model calculations \cite{MCSM}. These $0[110]$ states are also called deShalit-Goldhaber pairs and were initially discovered studying $\beta$ transition probabilities showing that the presence of these configurations provided a stabilizing effect \cite{PhysRev.92.1211}.

These pairs posses very similar spatial shapes, identical angular momentum and spin properties. Such similarity holds as well for proton-proton and neutron-neutron $0[110]$ pairs so, in principle, they could replace each other inflicting minimal changes on the system. This is the foundation of the proxy-$SU(3)$ model which will be explained in the following section along with its application to the 82-126 proton and 126-184 neutron shells.

\section{Explanation of the Scheme}

 This scheme describes deformed nuclei where the spherical degeneracy has been lifted and the nuclear single-particle dynamics is ruled by the Nilsson Hamiltonian of equation \ref{shellmodelhnilssonlarge}. Focusing on neutron-neutron and proton-proton $0[110]$ pairs, the model will be explained taking as example the shells 82-126 and 126-184 respectively since these include the isotopes studied in this thesis. These shells comprehend the spherical states $3p^{1/2}$, $3p^{3/2}$, $2f^{5/2}$, $2f^{7/2}$, $1h^{9/2}$, $1i^{13/2}$ for 82-126 and $4s^{1/2}$, $3d^{3/2}$, $3d^{5/2}$, $2g^{7/2}$, $2g^{9/2}$, $1i^{11/2}$, $1j^{15/2}$ for 126-184 where the intruders are $1i^{13/2}$ and $1j^{15/2}$ respectively. Notice that the Nilsson states of $1h^{11/2}$ of shell 50-82 form $0[110]$ pairs with those of $1i^{13/2}$ where only the $ 13/2[606]$ remains unpaired. The same happens between $1j^{15/2}$ and $1i^{13/2}$ with the $15/2[707]$ remaining unpaired. In general, the states $K_{max}[\eta+1\hspace{1mm} 0\hspace{1mm}\eta+1]$ will not have a $0[110]$ pair because of the additional angular momentum unit not present in the lower shell. This is summarized in table \ref{table2} for shell 82-126 only. 

After this identification, the idea is to proceed with the replacement of the intruder Nilsson levels by their corresponding $0[110]$ pairs excluding the $K_{max}[\eta+1\hspace{1mm} 0\hspace{1mm}\eta+1]$ states where the new shell orbitals will be called proxy orbitals. This causes a change in one unit in both $\eta$, $n_z$ and parity inversion of sign which will affect matrix elements of $\boldsymbol{L}\cdot\boldsymbol{S}$ and $\boldsymbol{L}^2$ as shown in  \cite{PhysRevC.95.064325}. This way the original 82-126 and 126-184 shells with orbitals  $pfh$ and $sdgi$ respectively of the harmonic oscillator are recovered along with their corresponding spatial symmetries $U(21)$ and $U(28)$ each one having $SU(3)$ subalgebras. 

Since this method relies on the Nilsson asymptotic states, the proxy orbitals of the lower shell require a correction term of $1-2\epsilon/3$ in the energy eigenvalue \ref{energylarge} to be pushed up to the energy values of the replaced orbitals. The non-diagonal elements contributed by the terms $\boldsymbol{L}\cdot\boldsymbol{S}$ and $\boldsymbol{L}^2$ must be dealt with numerical diagonalization. Concerning the excluded orbitals, it can be argued that for medium deformations these states will be among the highest in energy of the shell, as can be see in Nilsson diagrams like the one in figure \ref{nilssonmedium}, so they will come into play only near the shell closure where deformation will be very small and microscopic calculations are more appropriate. Thus, the proxy-$SU(3)$ is more suitable for nuclei away from magic numbers possessing medium and large deformation.

\begin{table}[h]
\begin{center}
\begin{tabular}{ |c|c|c|c| } 
\hline
\centering
Shell & Original spherical orbitals & Original Nilsson states & Proxy states \\
\hline
\centering
\multirow{21}{4em}{82-126} & $2f^{7/2}$ & 1/2[541] & 1/2[541] \\
& & 3/2[532] & 3/2[532]  \\ 
& & 5/2[523] & 5/2[523] \\ 
& & 7/2[514] & 7/2[514] \\
& $1h^{9/2}$ & 1/2[530] & 1/2[530] \\ 
& & 3/2[521] & 3/2[521] \\ 
& & 5/2[512] & 5/2[512] \\ 
& & 7/2[503] & 7/2[503] \\ 
& & 9/2[505] & 9/2[505] \\ 
& $\boldsymbol{1i^{13/2}}$ & $\boldsymbol{1/2[660] }$& $\boldsymbol{1/2[550] }$\\ 
& & $\boldsymbol{3/2[651] }$& $\boldsymbol{3/2[541] }$\\ 
& & $\boldsymbol{5/2[642] }$& $\boldsymbol{5/2[532] }$\\ 
& & $\boldsymbol{7/2[633] }$& $\boldsymbol{7/2[523] }$\\ 
& & $\boldsymbol{9/2[624] }$& $\boldsymbol{9/2[514] }$\\ 
& & $\boldsymbol{11/2[615]}$ & $\boldsymbol{11/2[505]}$ \\ 
& & $\boldsymbol{13/2[606]}$ & $\boldsymbol{-}$ \\ 
& $3p^{3/2}$ & 1/2[521] & 1/2[521] \\ 
& & 3/2[512] & 3/2[512] \\ 
& $2f^{5/2}$ & 1/2[510] & 1/2[510] \\ 
& & 3/2[501] & 3/2[501] \\ 
& & 5/2[503] & 5/2[503] \\ 
& $3p^{1/2}$ & 1/2[501] & 1/2[501] \\
\hline
\end{tabular}
  \caption{Original orbitals and corresponding subspace of proxy orbitals for shell 82-126. The replaced ones are shown bold.}
    \label{table2}
\end{center}
\end{table}

Since the main purpose of this model is the restoration of $SU(3)$ algebra, the greatest advantage should be obtained in the framework of symmetries. The algebra $U((\eta+1)(\eta+2)/2)$ for spatial wavefunctions in each shell is recovered with the corresponding ground state representations of the form $[2^n]$ for even and $[2^n 1]$ for odd number of nuclides where $n$ represents the number of pairs. These will allow the calculation of corresponding $(\lambda,\mu)$ irreps of $SU(3)$ by means of the branching problem where an important question emerges concerning which of these irreps  is associated to the nuclear ground state. As mentioned in the last chapter, the general procedure is finding the irrep with the largest order two Casimir value, also called leader irrep, because it will minimize the energy of the Hamiltonian in equation \ref{Helliott}. However, as has been studied extensively \cite{Martinou2021, MARTINOUARXIV, MARTINOUARXIV2}, the nuclear interaction favours the most symmetric spatial wavefunctions which can be identified by the largest value of 
\begin{equation} 
\begin{aligned}
r = \frac{\lambda+\mu}{\lambda+2\mu}.
\label{h.w.}
\end{aligned}
\end{equation}

These are known as highest weight irreps and may not always coincide with the largest Casimir value. Physically it represents a particular filling of the orbitals favouring those that contribute to a prolate cylindrical deformation. For proxy-$SU(3)$ applications, the highest weight representation will be associated to the ground state. This physically supported convention is appropriate to explain the phenomenology of nuclear deformation as will be shown in the next section.

\section{Corroboration Through Nilsson Model Calculations, Prolate Dominance and Prolate-Oblate Shape Transition}

\begin{figure}[h!]
    \centering 
\includegraphics[width=0.8\textwidth]{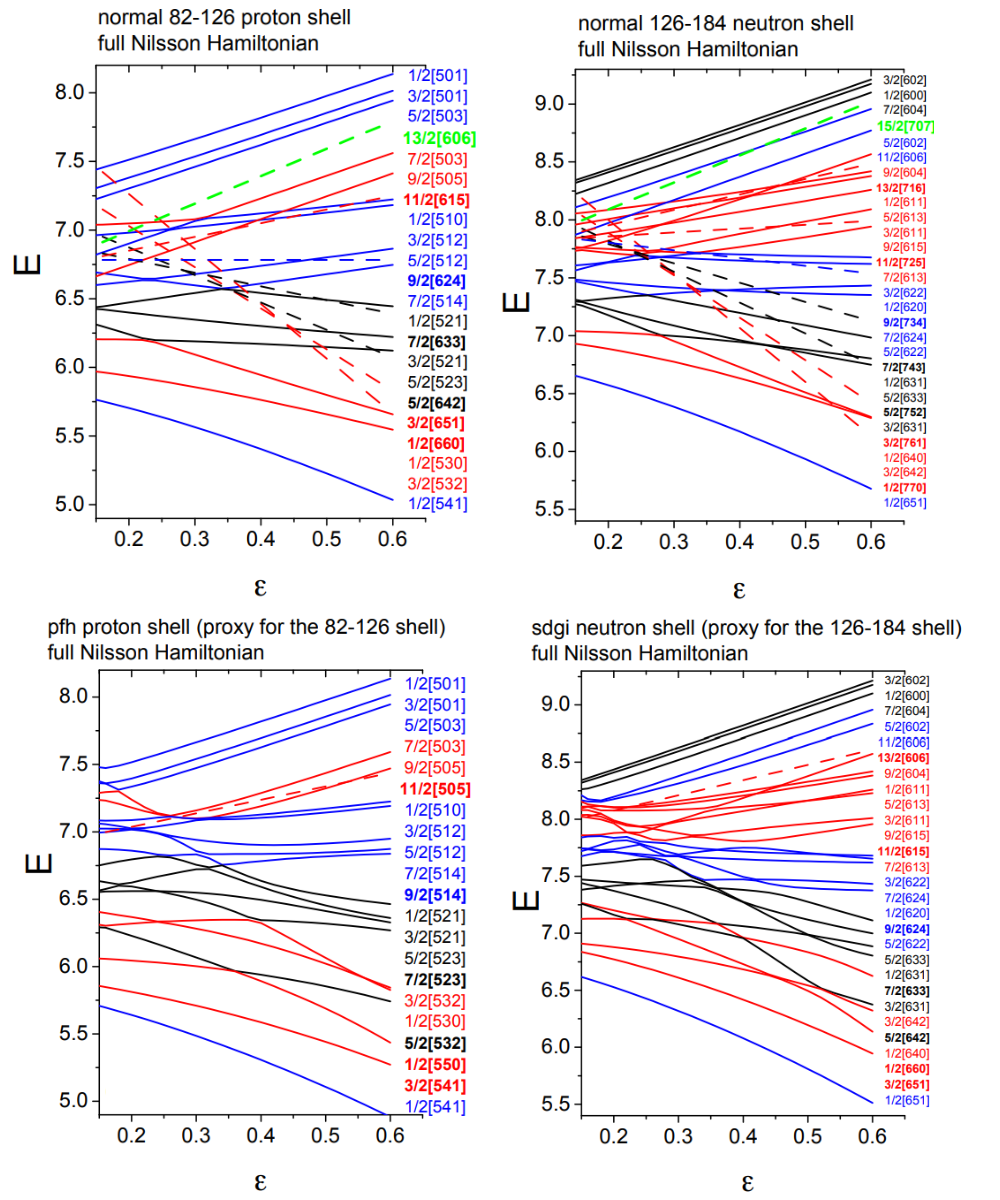}
    \caption{Nilsson diagrams of proton 82-126 and neutron 126-184 shells. Excluded states are shown in green. The proxy replaced and replacing states are denoted bold in the state labels. Up: Original orbitals. Down: Proxy orbitals. Figure taken from \cite{PhysRevC.95.064325}.}
    \label{DiagNilssonPN}
\end{figure}

In reference \cite{PhysRevC.95.064325} and its supplemental material is shown that the replacement of the intruder states causes the appearance of spurious matrix elements not present before. These entries represent the "damage" made by the approximation of the proxy-$SU(3)$ but, in general, they are at least one order of magnitude smaller than the diagonal elements so are considered as small corrections. The full diagonalization of the Nilsson Hamiltonian for large deformation of equation \ref{shellmodelhnilssonlarge} will result in the energy eigenvalues of the system with dependence on the deformation parameter $\epsilon.$ The results for proton 82-126 and neutron 126-184 shells  are shown as Nilsson diagrams and compared with the original orbitals in figure \ref{DiagNilssonPN}. Notice that since medium and large deformations are considered, then the degeneracy is not recovered for small values of $\epsilon$ and thus only $\epsilon \geq 1.5$ is shown.

Another important result concerns the prediction of shape observables $\beta$, $\gamma$ and the prolate over oblate shape dominance observed along the whole nuclear chart as seen in figure \ref{Quadr}. By means of the highest weight irrep and equations \ref{betagammalambdamu}, it is possible to obtain the $\beta$ and $\gamma$ values for deformed nuclei in a parameter independent way. In references \cite{PhysRevC.95.064326, MARTINOUARXIV2} are shown tables of the highest weight irreps for the ground state of several nuclei having $Z=50-126$ where it can be seen that most of the representations correspond to prolate shapes in agreement with the experimental facts and their deformation parameters agree reasonably well with the measured values.

The choosing of highest weight irrep as the ground state has the consequence of breaking the particle-hole symmetry and displacing the prolate to oblate shape transition boundary. This particular symmetry states that if a nucleus has $n_1$ valence protons with corresponding irrep $(\lambda_{\pi},\mu_{\pi})$, $n_2$ valence neutrons with corresponding irrep $(\lambda_{\nu},\mu_{\nu})$ and corresponding ground state $(\lambda_{\pi}+\lambda_{\nu},\mu_{\pi}+\mu_{\nu})$, then its particle-hole conjugate should posses a $(\mu_{\pi}+\mu_{\nu}, \lambda_{\pi}+\lambda_{\nu})$ irrep. As an example, the isotope $^{224}$Th has eight valence protons and neutrons and its corresponding  particle-hole conjugate $^{294}$Og has eight proton- and neutron-holes. The later has an irrep (12,42) and the former a (60,8) which shows clearly the particle-hole symmetry breaking. This phenomenon causes the appearance of the prolate-oblate shape transition on the nuclear chart in agreement with the observations. Without this breaking, the boundary would be at the middle of the shell which is in disagreement with experimental evidence. It is worth mentioning that the symmetry is restored for few isotopes near the closure of the shell.

In the next chapter the proxy-$SU(3)$ model will be applied to the shell-like quarteting in nuclei for proton shell 82-126. This model will be used to study shape observables, excitation energy spectra and reduced electromagnetic transition probabilities. Since this region is poorly recorded in nuclear databases, results will be compared with the few measured values to test the goodness of the model, and if not possible, they will be left as predictions to be tested when new data is available.

\chapter{Quartet Nuclei Spectra and Electromagnetic Transition Predictions}
\InitialCharacter{T}he atomic nucleus is a difficult system in physics because of its many body character and having not few enough particles to be treated from \textit{ab initio} in a computationally feasible way (at least not yet) nor many enough for statistics to describe it. All this added to the fact that there is not a precise knowledge of the residual nucleon-nucleon interaction, the combination of collective, single particle dynamics and radioactive decay instability makes a phenomenological modelling the best option  for its understanding. As long as the model predictions fit to the experimental data in an acceptable way, the ideas involved on it will not be far from the actual complex processes inside the nucleus.

This chapter will integrate the theoretical framework of the previous three in a phenomenological approach applied to even-even heavy nuclei in the proxy-$SU(3)$ scheme. It consists in coupling the spatial wave functions of protons and neutrons such that can form a specific representation composed of what is known as quartets. These particular representations are supported by experimental evidence on the binding energies of single nucleons compared to sets of two protons and two neutrons in the even-even nuclei. The nuclear region studied is currently very unexplored and its data is very scarce, so in preparation of future experimental efforts, the theoretical background and a few predictions are prepared for some nuclear properties concerning deformation, energy spectra and electromagnetic transitions. 

The methodology exposed in this section is based mainly on \cite{CsehTh} where the author J. Cseh, states that the performance of his model for the low-lying spectra is expected to behave similar in many other cases than the isotope $^{224}$Th. Thus, an extension of its reach to other quartet nuclei and predictions involving different physical pictures is proposed in this chapter. The methodology and procedure are explained in the text body along with the content in the appendices.

\section{Quarteting in the Atomic Nucleus}

\begin{figure}[h!]
    \centering 
\includegraphics[width=\textwidth]{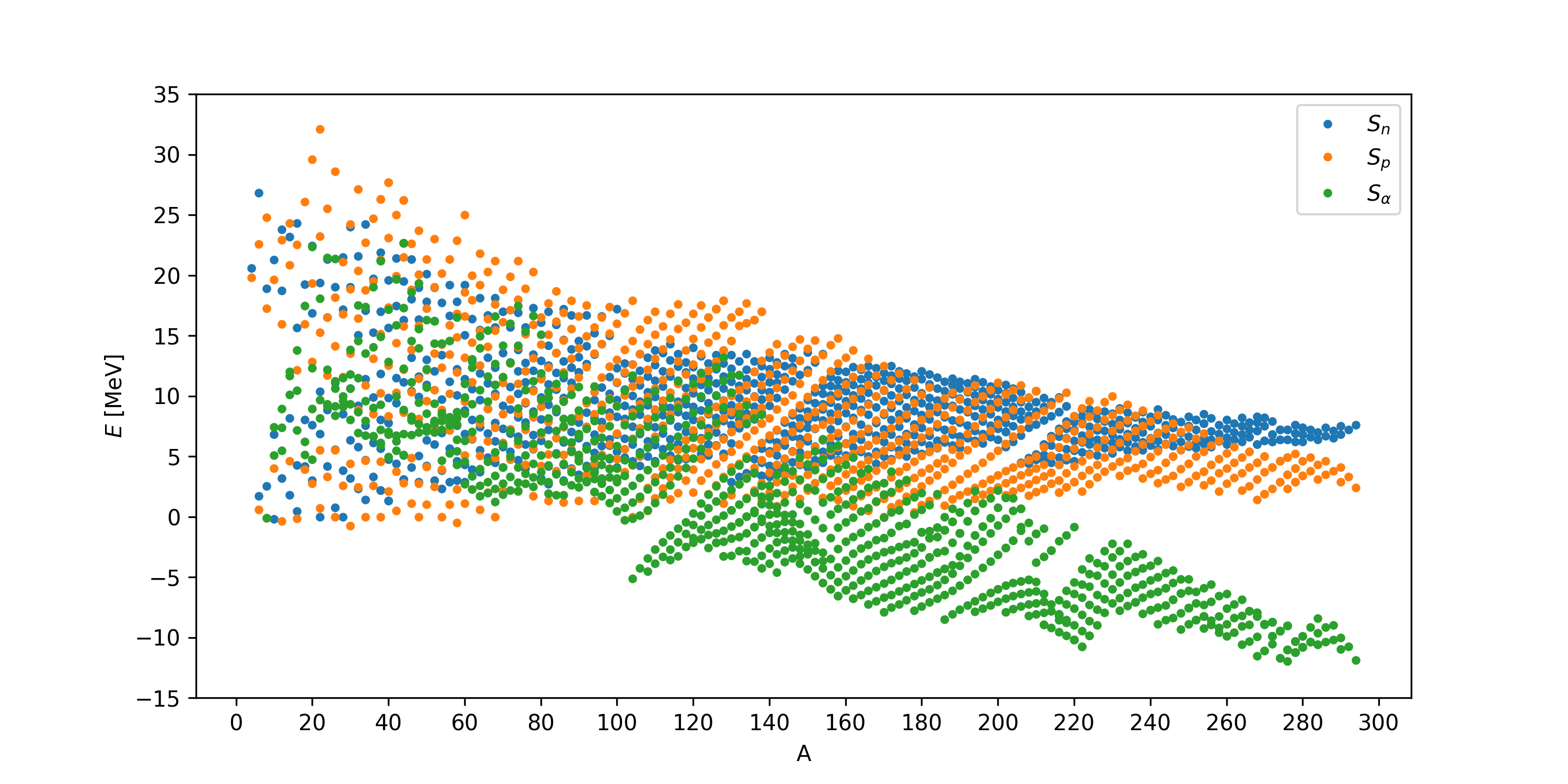}
    \caption{Binding energies of neutrons $S_n$, protons $S_p$ and alpha particles $S_{\alpha}$ in even-even nuclei. Data retrieved from \cite{NNDC} and figure generated using Matplotlib.}
    \label{BEalpha}
\end{figure}

The pairing phenomena in nuclei along with the observations of double binding energy differences and alpha particle separation energies indicate that subsets of the nucleus consisting of four nucleons, two protons and two neutrons, happen to be interacting strongly and influence several nuclear properties. For heavy isotopes, these subsets of alpha particles are more bound between them than to the whole nucleus as shown in figure \ref{BEalpha}. This property has been worked theoretically by Wigner supermultiplet theory \cite{Wignerspinisospin} and alpha-clustering models \cite{alphaclustering,Ren2018} .

This led to the definition of a quartet as two protons and two neutrons in a single orbital without specification of any coupling scheme \cite{PhysRevLett.25.1043}. The quartet symmetry was proposed in \cite{HARVEY1973191} as the two protons and neutrons having a permutational symmetry representation $[4]$ and spin-isospin $[1^4]$. This can be extended to define the heavy quartet nuclei as composed of $Z = 2I$ protons and $N = 2I$ neutrons in subsequent valence shells ($I$ being an integer) with associated ground state permutational representations $[2^I]_\pi$ and $[2^I]_\nu$ respectively that hold Pauli exclusion principle. Since spin-isospin is no longer appropriate, the proton-neutron scheme will be considered where the permutational symmetries for each nucleon kind in the ground or excited states must combine to $[4^I]$.

These impositions in the structure of the atomic nucleus require to be coordinated with the shell model in a 
shell-model-like quarteting which was worked in \cite{CSEH2015213, CSEH2016312} resulting in the phenomenological algebraic quartet model (PAQM) and semimicroscopic algebraic quartet model (SAQM). The former considers the quartets as structureless objects, while the latter deals with its constituent nucleons explicitly and will be applied in this thesis. In order to construct the model space of a quartet nuclei in the SAQM for states corresponding to fundamental and excitation of one and two particles, the next steps are followed for which an example can be found in appendix F:
\begin{enumerate}
  \item Selection of core to be worked on. In the case for heavy nuclei it will be $^{208}_{82}$Pb$_{126}$.
  \item The shell states will be labelled by the proxy-$SU(3)$ scheme. This condition truncates all the possible states into a subset where the $SU(3)$ symmetry is restored.
  \item The fundamental state permutational symmetry will be given by $[2^I]_\nu\otimes[2^I]_\pi$. This representation will be the same for algebra $U(\mathcal{N}_\nu)\otimes U(\mathcal{N}_\pi)$ where $\mathcal{N} = (\eta+1)(\eta+2)/2$ for each corresponding nucleon shell. In the case of this thesis it is $U(21)\otimes U(28)$. Excitation of one neutron or proton will result in the $1\hbar\omega$ representations $[2^{I-1}1]_\nu\otimes[1]_\nu\otimes[2^I]_\pi$ in $U(28)\otimes U(36)\otimes U(21)$  and  $[2^{I}]_\nu\otimes[2^{I-1}1]_\pi\otimes[1]_\pi$ in $U(28)\otimes U(21)\otimes U(28)$ respectively. For the $2\hbar\omega$ representations one can excite two neutrons, two protons or one neutron and one proton resulting in $[2^{I-1}]_\nu\otimes[2]_\nu\otimes[2^I]_\pi$, $[2^{I}]_\nu\otimes[2^{I-1}]_\pi\otimes[2]_\pi$ and $[2^{I}1]_\nu\otimes[1]_\nu\otimes[2^{I}1]_\pi\otimes[1]_\pi$ respectively, each in their corresponding algebras. Notice that each outer product can be coupled to $[4^I]$ whether it is ground or excited state.
  \item These representations must each be decomposed independently to obtain the allowed quartet $SU_\nu(3)\otimes SU_\pi(3)$ irreps for each excitation. These was performed by the branching  $U(\mathcal{N})\supset SU(3)$ by means of specialized software as shown in appendix F.
  \item Couple the leading irreps of $SU_\nu(3)$ and $SU_\pi(3)$ into the final product space $SU_{\nu+\pi}(3)$, i.e. $(\lambda,\mu)_\pi\otimes(\lambda,\mu)_\nu = \sum (\lambda,\mu)_{\pi+\nu}$. This task was also performed by specialized software.
  \item Discard of spurious states. These appear in the model space as irreps corresponding to excitations of the center of mass in the formulation of the model. As was demonstrated in \cite{HECHT197134}, the center of mass operator is an $SU(3)$ tensor of rank (1,0). This implies that $1\hbar\omega$ spurious states can be identified by coupling $0\hbar\omega$ irreps as  $(1,0)\otimes(\lambda,\mu)_{\pi+\nu}^{0\hbar\omega}$. Similarly, for the $2\hbar\omega$ one computes products $(1,0)\otimes(\lambda,\mu)_{\pi+\nu}^{1\hbar\omega}$ and $(2,0)\otimes(\lambda,\mu)_{\pi+\nu}^{0\hbar\omega}$. Deletion of this states will result in the intrinsic excitation model space.
\end{enumerate}

\section{Nuclear Region and Model Space}

\begin{figure}[h!]
    \centering 
\includegraphics[width=0.8\textwidth]{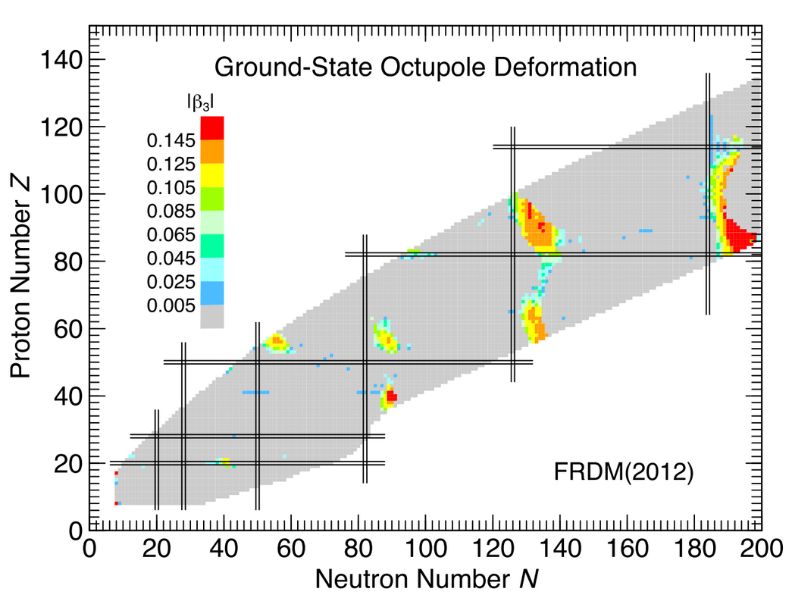}
    \caption{Octupole deformation parameter $\beta_{30}$ predictions. Figure taken from \cite{FRDM}.}
    \label{FRDMOctupole}
\end{figure}

The region studied in this thesis was the quartet nuclei in shell 82-126 up to $A=248$, explicitly these are the isotopes $^{212}$Po, $^{216}$Rn, $^{220}$Ra, $^{224}$Th, $^{228}$U, $^{232}$Pu, $^{236}$Cm, $^{240}$Cf, $^{244}$Fm and $^{248}$No. The $A = 248$ limit was chosen since it is the currently achieved experimentally proton drip line for heavy quartets. This region has gained interest because some of them might posses an octupole axial deformation $\alpha_{30}$ (according to equation \ref{Rexpansion}) which is of relevance to investigations about nuclear shapes and structure \cite{pear,Butler}. The finite range droplet model \cite{FRDM} predicts a non-zero octupole deformation $\beta_{30}$ for $^{220}$Ra and $^{224}$Th as can be seen in figure \ref{FRDMOctupole} and supported experimentally by their excitation spectra as explained in section 2.2.2. 

Focusing on symmetry considerations, since spin-isospin symmetry is badly broken in this shell, the algebra $U(4\mathcal{N})$ with $\mathcal{N} = (\eta+1)(\eta+2)/2$ is no longer useful. However, given that quartet nuclei have total spin $S=0$, one can focus on spatial algebras separating them from spin symmetries while dealing with protons and neutrons degrees of freedom independently. The algebra chain to be considered will be given by  
\begin{equation}
\resizebox{0.9\hsize}{!}{$
\begin{aligned}
&\Bigg[ U\left(2\mathcal{N}_{\pi}\right)\supset\bigg[U\left(\mathcal{N}_{\pi}\right)\supset SU_{\pi}(3)\bigg]\otimes SU_{\pi}(2)\Bigg]\otimes \Bigg[ U\left(2\mathcal{N}_{\nu}\right) \supset\bigg[U\left(\mathcal{N}_{\nu}\right)\supset SU_{\nu}(3)\bigg]\otimes SU_{\nu}(2)\Bigg]\\
&\supset \Bigg[ \bigg[SU_{\pi}(3)\otimes SU_{\nu}(3)\bigg]\supset SU_{\pi+\nu}(3) \supset SO_{\pi+\nu}(3)\Bigg]\otimes \Bigg[ \bigg[ SU_{\pi}(2)\otimes SU_{\nu}(2)\bigg]\supset SU_{\pi+\nu}(2)\Bigg],
\end{aligned} 
\label{algebraModelSpace}
$}
\end{equation}
and the corresponding state
\begin{equation}
\begin{aligned}
&\Bigg|[f]_{\pi}(\lambda,\mu)_{\pi}S_\pi,\hspace{0.7mm}[f]_{\nu}(\lambda,\mu)_{\nu}S_\nu;\hspace{0.7mm} (\lambda,\mu)_{\pi+\nu}\hspace{0.4mm} K_{\pi+\nu}\hspace{0.4mm} L_{\pi+\nu} \hspace{0.4mm}S_{\pi+\nu}\hspace{0.4mm} J_{\pi+\nu}\Bigg>.
\end{aligned} 
\label{algebraicstate}
\end{equation}

Where the emerging $SU(3)$ representations will be obtained considering the branching problem $U(\mathcal{N})\supset SU(3)$ in the proxy-$SU(3)$ scheme. For these particular isotopes, the leader and highest weight irreps coincide as shown in \cite{MARTINOUARXIV2} so this issue will not be of concern.

In order to construct the model space, i.e., the irreps of the nuclear system the steps explained in the last section must be followed. The excitation states are computed for $n\hbar\omega$ where $n=$0,1,2 for each of the isotopes mentioned. A few results are shown in table \ref{tableofirreps} organized by decreasing value of $\mathcal{C}_2[SU(3)]$ along with deformation parameters computed from leading irreps using equation \ref{betagammalambdamu}.

\renewcommand{\arraystretch}{1.3}
\begin{table}[p]
\begin{center}
\begin{tabular}{c@{\hskip 3mm}c@{\hskip 3mm}c@{\hskip 1mm}c@{\hskip 3mm}c} \toprule\toprule
     {Isotope} & {$n$} & {$(\lambda,\mu)$} & {$\beta$} & {$\gamma$ [°]} \\ \midrule
    $^{212}$Po & 0: & $( 22 , 0 ),( 20 , 1 ),( 18 , 2 ),( 16 , 3 ),( 14 , 4 ),( 12 , 5 ),( 10 , 6 ),...$ & 0.0613 &  0.0000  \\
           & 1:  & $( 23 , 0 )$ & E: 0.0303(19) & \\
           & 2:  & $( 24 , 0 )$ & &  \\ \midrule
    $^{216}$Rn  & 0:   & $( 36 , 4 ),( 37 , 2 ),( 38 , 0 ),( 34 , 5 ),( 35 , 3 )^2,( 36 , 1 )^2,(32,6),...$& 0.1037 & 5.20872 \\
           & 1:  & $( 39 , 3 )^2,( 40 , 1 )^2,( 37 , 4 ),( 38 , 2 )^3,( 39 , 0 ),( 1 , 22 ),( 2 , 20 ),...$&  &  \\
           & 2:  & $( 42 , 2 )^3,( 43 , 0 ),( 40 , 3 ),( 41 , 1 )^2$& &  \\ \midrule
    $^{220}$Ra & 0:   & $( 54 , 0 ),( 52 , 1 ),( 50 , 2 ),( 48 , 3 ),( 46 , 4 ),( 44 , 5 ),( 42 , 6 ),...$& 0.1432 & 0.0000 \\
           & 1:  & $( 56 , 1 )^2,( 54 , 2 )^2,( 55 , 0 ),( 52 , 3 )^2,( 50 , 4 )^2,( 48 , 5 )^2,( 46 , 6 )^2,...$&  &  \\
           & 2:  & $( 58 , 2 )^3,( 59 , 0 ),( 56 , 3 )^3,( 57 , 1 )^2,( 54 , 4 )^3,( 52 , 5 )^3,( 50 , 6 )^3,...$& &  \\ \midrule
    $^{224}$Th  & 0:   & $( 60 , 8 ),( 61 , 6 ),( 62 , 4 ),( 63 , 2 ),( 64 , 0 ),( 58 , 9 ),( 59 , 7 )^2,...$& 0.1667 & 6.1784 \\
           & 1:  & $( 65 , 6 )^2,( 66 , 4 )^2,( 67 , 2 )^2,( 63 , 7 )^2,( 64 , 5 )^4,( 65 , 3 )^4,( 66 , 1 )^2,...$& E: 0.174(6) & \\
           & 2:  &$( 70 , 4 )^3,( 71 , 2 ),( 72 , 0 ),( 68 , 5 )^3,( 69 , 3 )^4,( 70 , 1 )^2,( 66 , 6 ),...$ & &  \\ \midrule
    $^{228}$U & 0:   &$( 70 , 8 ),( 71 , 6 ),( 72 , 4 ),( 73 , 2 ),( 74 , 0 ),( 68 , 9 ),( 69 , 7 )^2,...$ & 0.1880 & 5.3487 \\
           & 1:  & $( 74 , 8 )^2,( 75 , 6 )^2,( 76 , 4 )^2,( 77 , 2 )^2,( 78 , 0 )^2,( 72 , 9 )^2,( 73 , 7 )^4,...$&  &  \\
           & 2:  & $( 78 , 8 )^3,( 79 , 6 )^3,( 80 , 4 )^3,( 81 , 2 )^3,( 82 , 0 )^3,( 76 , 9 )^3,( 77 , 7 )^6,...$& &  \\ \midrule
    $^{232}$Pu & 0:   & $( 84 , 0 ),( 82 , 1 ),( 80 , 2 ),( 78 , 3 ),( 76 , 4 ),( 74 , 5 ),( 72 , 6 ),...$& 0.2076 & 0.0000  \\
           & 1:  & $( 87 , 2 )^2,( 85 , 3 )^2,( 86 , 1 )^2,( 83 , 4 )^2,( 84 , 2 )^2,( 85 , 0 ),( 81 , 5 )^2,,...$&  &  \\
           & 2:  & $( 90 , 4 )^3,( 91 , 2 ),( 92 , 0 ),( 88 , 5 )^3,( 89 , 3 )^4,( 90 , 1 )^2,( 86 , 6 )^3,...$& &  \\ \midrule
    $^{236}$Cm & 0:   & $( 82 , 12 ),( 83 , 10 ),( 84 , 8 ),( 85 , 6 ),( 86 , 4 ),( 87 , 2 ),( 88 , 0 ),...$& 0.2140 & 6.7351  \\
           & 1:  & $( 89 , 9 )^2,( 90 , 7 )^2,( 91 , 5 )^2,( 92 , 3 )^2,( 87 , 10 )^2,( 88 , 8 )^4,( 89 , 6 )^4,...$&  &  \\
           & 2:  & $( 96 , 6 )^3,( 97 , 4 ),( 98 , 2 ),( 99 , 0 ),( 94 , 7 )^3,( 95 , 5 )^4,( 96 , 3 )^2,...$& &  \\ \midrule
    $^{240}$Cf & 0:   & $( 84 , 16 ),( 85 , 14 ),( 86 , 12 ),( 87 , 10 ),( 88 , 8 ),( 89 , 6 ),( 90 , 4 ),...$& 0.2198 & 8.5651 \\
           & 1:  & $( 90 , 15 )^2,( 91 , 13 )^2,( 92 , 11 )^2,( 93 , 9 )^2,( 94 , 7 )^2,( 95 , 5 )^2,...$&  &  \\
           & 2:  & $( 96 , 14 )^3,( 97 , 12 )^3,( 98 , 10 )^3,( 99 , 8 )^3,( 100 , 6 )^3,( 101 , 4 )^3,...$& &  \\ \midrule
    $^{244}$Fm & 0:   & $( 90 , 12 ),( 91 , 10 ),( 92 , 8 ),( 93 , 6 ),( 94 , 4 ),( 95 , 2 ),( 96 , 0 ),...$& 0.2231 & 6.1784 \\
           & 1:  & $( 95 , 13 )^2,( 96 , 11 )^2,( 97 , 9 )^2,( 98 , 7 )^2,( 99 , 5 )^2,( 100 , 3 )^2,...$&  &  \\
           & 2:  & $( 100 , 14 )^3,( 101 , 12 )^3,( 102 , 10 )^3,( 103 , 8 )^3,( 104 , 6 )^3,...$& &  \\ \midrule
    $^{248}$No & 0:   & $( 100 , 0 ),( 98 , 1 ),( 96 , 2 ),( 94 , 3 ),( 92 , 4 ),( 90 , 5 ),( 88 , 6 ),...$& 0.2261 & 0.0000  \\
           & 1:  & $( 104 , 3 )^2,( 102 , 4 )^2,( 103 , 2 )^2,( 100 , 5 )^2,( 101 , 3 )^2,( 102 , 1 )^2,...$&  &  \\
           & 2:  & $( 108 , 6 )^3,( 109 , 4 ),( 110 , 2 ),( 111 , 0 ),( 106 , 7 )^3,( 107 , 5 )^4,...$& &  \\ \bottomrule \bottomrule
\end{tabular}
\caption{Irreducible representations ($\lambda,\mu$) for ground and $n = $0,1,2 excited states for each isotope. The super-index denotes multiplicity. Deformation parameters are also shown where the E represents experimental value retrieved from \cite{NNDC}.}
\label{tableofirreps}
\end{center}
\end{table}

\section{Hamiltonian and Nuclear Spectra}

The phenomenological Hamiltonian considered was worked initially in reference \cite{CsehTh}. It is based on Hamiltonians \ref{hrotrew}, \ref{Helliott} and reads as
\begin{equation}
\begin{aligned}
H = \hbar\omega n + a\mathcal{C}_2[SU(3)] + b\mathcal{C}_3[SU(3)] + c\mathbf{L}^2 - H_{\text{g.s.}},  
\end{aligned} 
\label{HCseh}
\end{equation}

where $\hbar\omega$ is the harmonic oscillator strength as calculated by \cite{BLOMQVIST1968545}, the operator $n$ is the $U(3)$ linear Casimir, $\mathcal{C}_2[SU(3)]$ and $\mathcal{C}_3[SU(3)]$ represent the Casimir operators of $SU(3)$ arising from quadrupole and collective interaction respectively, $\textbf{L}$ is the angular momentum operator and $H_{\text{g.s.}}$ represents the energy eigenvalue of the ground state having the purpose of normalizing it to zero. The parameters $a$, $b$ and $c$ are related to the moments of inertia of the nucleus and must be fitted from experimental data. The $a$, $b$ constrain the energy distance between band heads while $c$ describes the intraband energy levels spacing. An aspect worth mentioning is that representations corresponding to oblate and prolate deformations will be distinguished by $\mathcal{C}_3[SU(3)]$ but not by $\mathcal{C}_2[SU(3)]$.

Since the experimental data is very poor in this region, most of the $a$ parameters could not be estimated due to the fact that no excited rotational bands have been measured yet, so only intraband separation energy parameter $c$ was obtained for all isotopes by using $E_{2_1^+}$ experimental value (when existing) or its estimation by equation \ref{BE2ANDE2}. For the isotopes  $^{220}$Ra and $^{224}$Th the value of parameter $b$ was fixed to -0.004$a$ because of their dependence between each other in the fit process according to \cite{CsehTh}.

\begin{figure}[h!]
\centering 
\includegraphics[width=0.9\textwidth, trim={1cm 0 0 0}, clip]{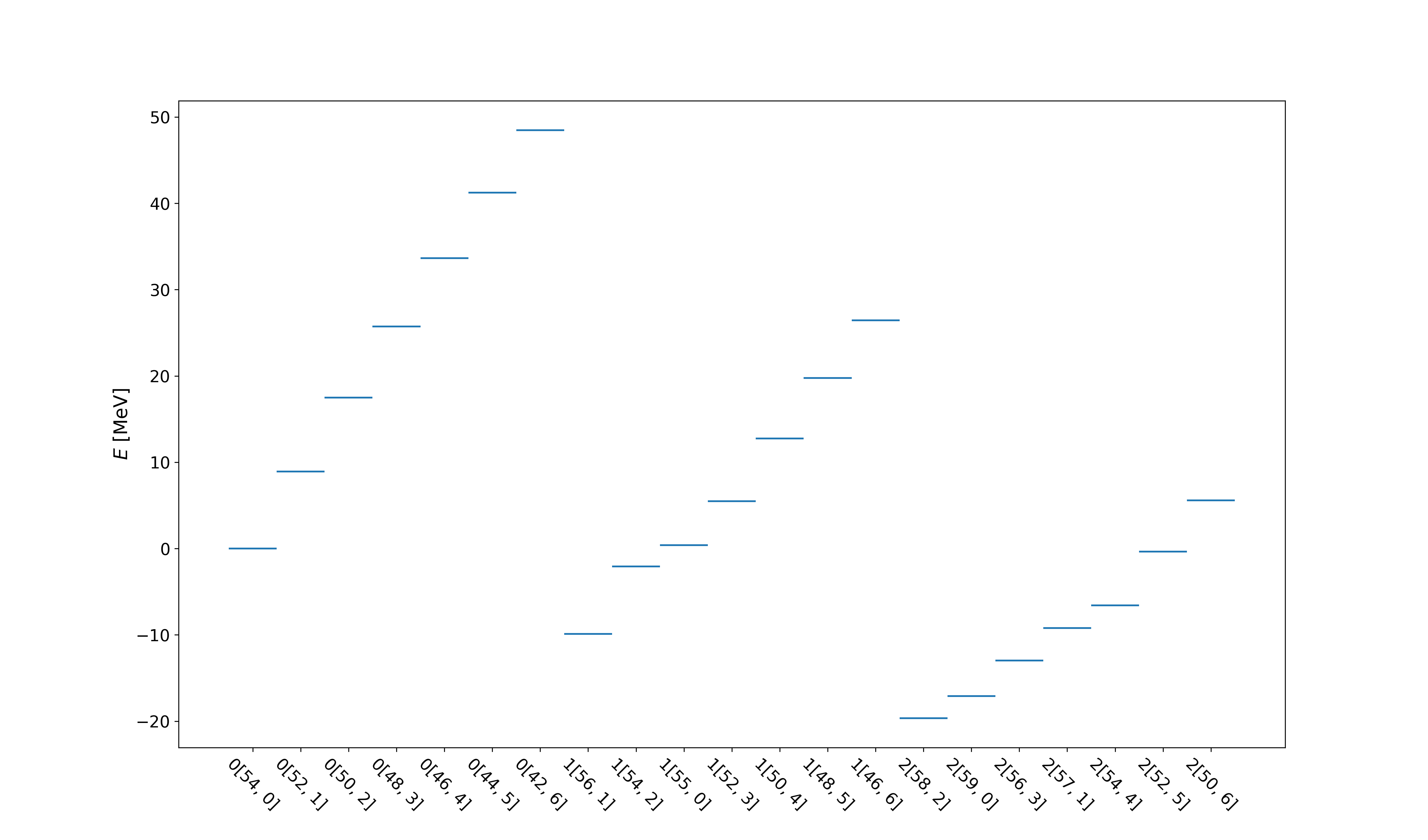}
\includegraphics[width=0.9\textwidth, trim={1cm 0 0 0}, clip]{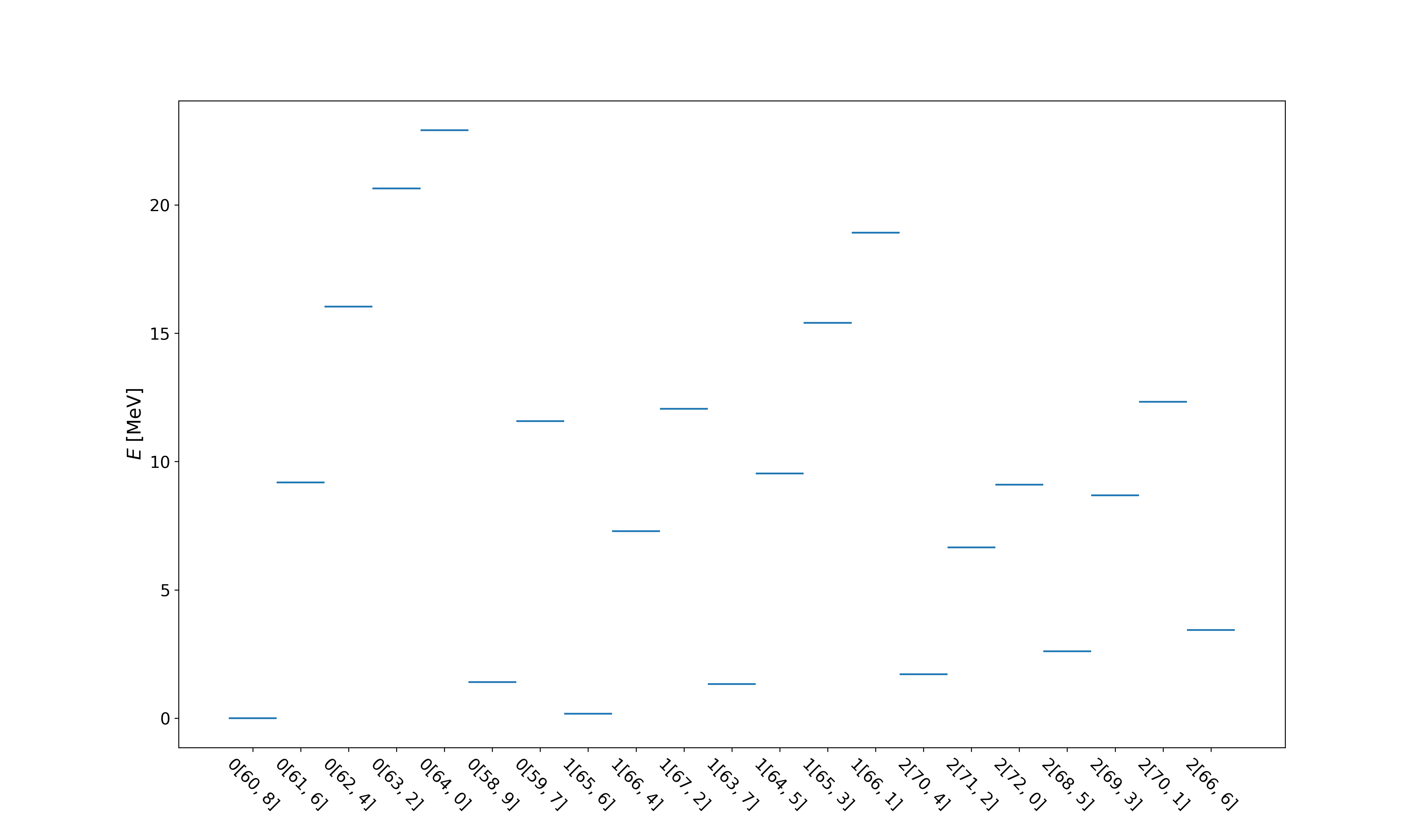}
\caption{Above: Band heads of $^{220}$Ra. Above: Band heads of $^{224}$Th. These were calculated with the parameters of fit shown in table \ref{tableofHparameters}. The labels are in the notation $n[\lambda,\mu]$. Figures generated with Matplotlib.}
\label{BandHeads}
\end{figure}

The nuclear spectra obtained after the fitting process can be seen in figure \ref{Spectra} for each isotope. The nuclei $^{220}$Ra and $^{224}$Th which posses the most complete experimental data, fit to root mean squares (RMS) of 0.3937 and 0.1180 respectively.  For $^{216}$Rn, two bands starting at $L = $12$^{+}$ and 13$^{-}$ have been detected but cannot be associated to any representation generated by the $SU(3)$ model. Additionally, the band heads for the first few irreps of isotopes $^{220}$Ra and $^{224}$Th are shown in figure \ref{BandHeads}. The $^{220}$Ra band heads display some of them below $E=$ 0 MeV, thus, more experimental data must be gathered for its  proper physical explanation.

\begin{table}[h!]
\begin{center}
\begin{tabular}{cccc} \toprule\toprule
    {Isotope} & {$\hbar\omega$ [MeV]}  & {$a$ [MeV]} & {$c$ [MeV/$\hbar^2$]} \\ \midrule
    $^{216}$Rn  & 6.8056 & - & 0.02566\\ \midrule
    $^{220}$Ra  & 6.7683 & -0.1739 & 0.00674 \\ \midrule
    $^{224}$Th  & 6.7318 & -0.1486 & 0.00872\\ \midrule
    $^{228}$U  & 6.6962 & - & 0.00983\\ \midrule
    $^{232}$Pu  & 6.6613 & - & 0.01071 \\ \midrule
    $^{236}$Cm  & 6.6272 & - & 0.00750\\  \midrule
    $^{240}$Cf  & 6.5938 & - & 0.00876 \\ \midrule
    $^{244}$Fm  & 6.5611 & - & 0.00815 \\ \midrule
    $^{248}$No  & 6.5291 & - & 0.00763 \\ \bottomrule \bottomrule
\end{tabular}
\caption{Parameters of Hamiltonian \ref{HCseh} for each isotope. Since excited rotational bands have not been measured in many of the isotopes, parameters $a$ could not be estimated and are left to when these data are obtained.}
\label{tableofHparameters}
\end{center}
\end{table}

\section{$E2$ Electromagnetic Intraband Transitions}
Since rotational bands follow the angular momentum patterns $L = 0^+$, $2^+$, $4^+$, $6^+$,... and $L = 1^-$, $3^-$, $5^-$, $7^-$,..., according to selection rules and knowledge of electromagnetic transitions, the dominant mechanism will be the $E2$ type for these intraband $\gamma$-decays. The theoretical treatment of reduced transition probabilities $B(E2)$ relevant for the following calculations can be found in appendix G and references therein.

The processes $L \xrightarrow{} L-2$ which will be denoted as $\downarrow$ and $L-2\xrightarrow{} L$ denoted as $\uparrow$ can both be observed experimentally and despite their obvious difference, its reduced transition probabilities are related as  
\begin{equation}
\begin{aligned}
B(E2; 	\downarrow) = 
 \frac{2L-3}{2L+1}B(E2; 	\uparrow),
\end{aligned} 
\label{Redtransitionprobabilityrelation}
\end{equation}
by making use of equation \ref{Redtransitionprobability} (under the convention of including the factors $1/\sqrt{2L_f-1}$ in the application of the Wigner-Eckart theorem). The $B(E2)$ are usually reported in units of $e^2\text{b}^2$, where $e$ is the elementary charge and $\text{b} = 10^{-28}\text{m}^2$  is an area unit called barn. There is a unit system where the numerical values of $B(E2; 	\downarrow)$ are usually reported. This was developed first in \cite{PhysRev.83.1073} where the Weisskopf unit (W. u.) was defined. This particular unit comes from the evaluation of a $\sigma\lambda$ transition for an extreme application of the independent-particle model (actually a single proton). A value in W. u. is interpreted as the number of nucleons involved in the transition. The mathematical expression of one Weisskopf unit for electric and magnetic transitions are given by
\begin{equation}
\begin{aligned}
&B_{\text{W. u.}}(E\lambda) = \frac{1.2^{2\lambda}}{4\pi}\left(\frac{3}{\lambda+3}\right)^2A^{2\lambda/3}10^{-2\lambda}\hspace{1mm}e^2\text{b}^{\lambda},\\
&B_{\text{W. u.}}(M\lambda) = \frac{1.2^{2\lambda-2}}{\pi}\left(\frac{3}{\lambda+3}\right)A^{(2\lambda-2)/3}10^{2\lambda-1}\hspace{1mm}(\mu_N/c)^2\text{b}^{\lambda-1}.
\end{aligned} 
\label{w.u.expr}
\end{equation}

 This also serves as a criteria for classifying nuclear collectivity since for transitions between $2_1^+$ and $0_1^+$ states a value of $B(E2) = 30-40$ W.u. indicates approximately vibrational collectivity and $B(E2) = 100-200$ W.u. indicates approximately rotational excitations \cite{Heyde_2016}.

As explained in appendix G, the theoretical expressions to be used are 
\begin{equation}
\begin{aligned}
B(E2; L_i\hspace{1.5mm} \text{g.s.}\xrightarrow{} L_f\hspace{1.5mm}\text{g.s.})& = 
\left( \frac{3ZR_0^2}{4\pi} \right)^2\left( 
 \beta+\frac{2}{7}\sqrt{\frac{5}{\pi}}\beta^2\right)^2\langle L_i020|L_f0\rangle^2 \hspace{0.8mm} 10^{-4} e^2\text{b}^2,
 \end{aligned} 
\label{BE2deformation}
\end{equation}
\begin{equation}
\begin{aligned}
 B\big(E2; (\lambda,\mu)KL_i\xrightarrow\ (\lambda,\mu)KL_f\big) &= 
 \hspace{0.8mm}\frac{2L_f+1}{2L_i+1}\hspace{0.8mm}\mathcal{C}_2[SU(3)]\hspace{0.8mm}\alpha^2|\langle(\lambda,\mu)KL_i;(11)2||(\lambda,\mu)KL_f\rangle_{\rho = 1}|^2,
\end{aligned} 
\label{BE2SU3}
\end{equation}
where the first equation will be used to predict the $B(E2; 0^+_1\hspace{1.5mm} \text{g.s.}\xrightarrow{} 2^+_1\hspace{1.5mm}\text{g.s.})$ with the deformation parameter $\beta$ calculated from proxy-$SU(3)$ ground state representations in case no experimental data is available as is for all isotopes except $^{224}$Th. Then, by means of this $B(E2)$ value in W.u. the value of effective transition parameter $\alpha$ is fitted and the other intraband transition rates can be predicted. This methodology is a combination of liquid drop and $SU(3)$ models whose results should be consistent and help to understand the nuclear dynamics of these isotopes.

\begin{figure}[h!]
\centering 
\includegraphics[width=0.44\textwidth, trim={1cm 1cm 0 0.2cm}, clip]{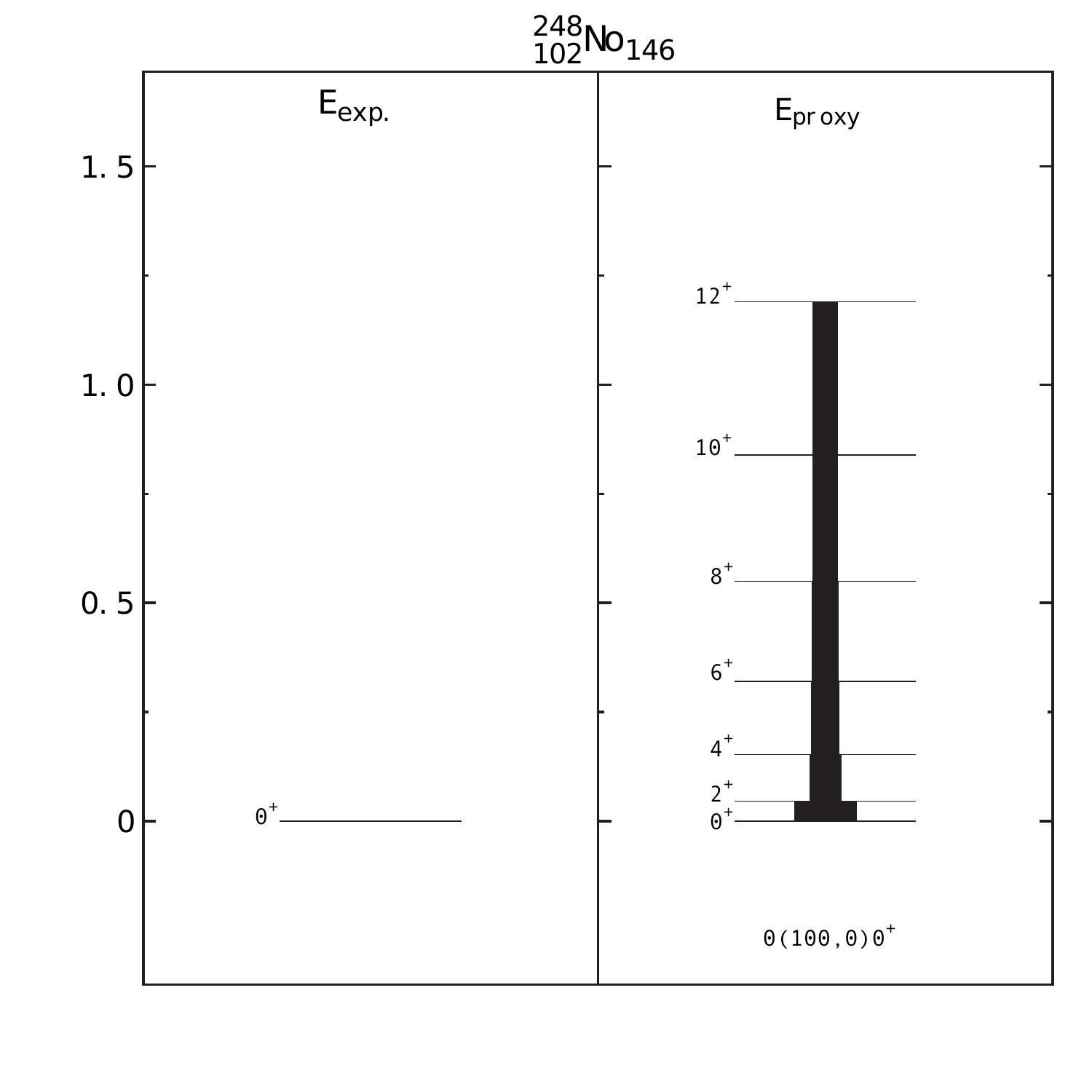}
\caption{Experimental and theoretical spectra of heavy quartet nuclei and reduced transition probabilities $B(E2)$ represented by wide lines between states. Each band is labelled as $n(\lambda,\mu)K^{\pi}$. Figures generated with DataGraph \cite{DataGraph}.}
\label{Spectra}
\end{figure}

It has been known since a while that $B(E2)$ and $E_{2^+_1}$ exhibit a uniform behaviour for a wide span of nuclei \cite{GRODZINS196288}. This regularity follows the empirical rule expressed by the equation \cite{takigawa}
\begin{equation}
\begin{aligned}
E_{2^+_1}\hspace{1mm}B(E2;  2^+_1\hspace{1.5mm} \text{g.s.}\xrightarrow{} 0^+_1\hspace{1.5mm}\text{g.s.}) \approx (25\pm 8)\frac{Z^2}{A} \text{MeV}e^2\text{fm}^4,
\end{aligned} 
\label{BE2ANDE2}
\end{equation}
which can be used to estimate $E_{2^+_1}$ and thus to obtain the parameter $c$ of \ref{HCseh} to predict the ground state band energy levels. The results of this methodology are in table \ref{Tableofelectromagnetictransition} and figures \ref{Spectra}.

\begin{table}[h!]
\begin{center}
\begin{tabular}{ccccc} \toprule\toprule
    {Isotope} & {$\alpha^2$ [W.u.]} & {$B(E2; 2^+_1 \xrightarrow{} 0^+_1)$ [W.u.]} & {$E_{2^+_1}$ [MeV]} \\ \midrule
    $^{212}$Po  & 0.004673 & 11.0264  &   1.00499 $\pm$ 0.32159\\
           & & E: 2.57$\left(  \substack{+38\\-29}  \right)$ & E: 0.727330(9)\\ \midrule
    $^{216}$Rn  & 0.021865 &  34.0750 &  0.32632$ \pm$ 0.10443 \\
           & & & E: 0.4614(2) \\ \midrule
    $^{220}$Ra  & 0.022717 &  69.9239 &  0.15953 $\pm$ 0.05105 \\
           & & & E: 0.17847(12)\\ \midrule
    $^{224}$Th  & 0.022383 &  100.6625  & 0.11114 $\pm$ 0.03556\\
           & & E: 96(7)  & E: 0.0981(3) \\ \midrule
    $^{228}$U  & 0.023811 &  135.6923 & 0.08267 $\pm$ 0.02645 \\
           & & & E: 0.059(14) \\ \midrule
    $^{232}$Pu  & 0.023951 &  175.0361 & 0.06424 $\pm$ 0.02056 \\
           & & & \\ \midrule
    $^{236}$Cm  & 0.025559 &  204.7149 &  0.05505 $\pm$ 0.01762 \\
           & & & E: 0.0474(4) \\ \midrule
    $^{240}$Cf  & 0.024589 &  214.9608 &  0.05253 $\pm$ 0.01681 \\
           & & & \\ \midrule
    $^{244}$Fm  & 0.024325 &  231.2175 &  0.04893 $\pm$ 0.01566 \\
           & & & \\ \midrule
    $^{248}$No  & 0.024034 &  247.5446 & 0.04578 $\pm$ 0.01465 \\
           & & & \\ \bottomrule \bottomrule
\end{tabular}
\caption{Electromagnetic transition parameters $\alpha^2$ of equation \ref{BE2SU3} for the transition between $0^+_1$ and $2^+_1$ used to predict the other interband transitions. The $B(E2)$ values shown are computed from equation \ref{BE2deformation} and the experimental values denoted by E were retrieved from \cite{NNDC}. Additionally the predictions of first $E_{2^+_1}$ from \ref{BE2ANDE2} are shown where the uncertainties are not to be taken as experimental errors so several more significant figures are displayed.}
\label{Tableofelectromagnetictransition}
\end{center}
\end{table}

\section{Perspectives}

Some interesting topics arise as a continuation of this thesis on which progress is being made by the author and his advisor. These are investigations about the contribution of the octupole deformation to the nuclear region studied from a group theoretical perspective as introduced in \cite{Isacker_2016}, with the required adaptations to proxy-$SU(3)$. The calculation of a nuclear mass equation is also a topic to work on because of its relevance to nuclear structure and limits of existence.

Concerning the nuclear stability, there is another electromagnetic transition observed in these isotopes: the interband $E1$. This one requires the $E1$ operator to be expressed as an $SU(3)$ tensor and then proceed with the calculation of matrix elements based on the reference \cite{J2007ReducedME}. In a similar way, the alpha decay can be studied and happens to be interesting due to the fact that isotopes from $^{244}$Fm down to $^{212}$Po are connected in the sense that they can decay by this mechanism going through the quartet nuclei between them.

These open questions conclude the thesis and will be worked in the near future by the author. The methodology of the model and procedure of the calculations were presented up to a certain degree of detail for its uses in similar projects and extensions to the current models. The main difficulty for a phenomenological model like this is its dependence on parameters obtained from experimental data, which are very deficient in the isotopes studied. Given the relevance and open questions in this nuclear region, there is a need for experimental efforts in the synthetization and measurement of radioactive decay observables which are expected to confirm the goodness of the model in explaining their nuclear structure.

\begin{tabular}{cc}
\includegraphics[width=0.45\textwidth, trim={1.5cm 0 0 0}, clip]{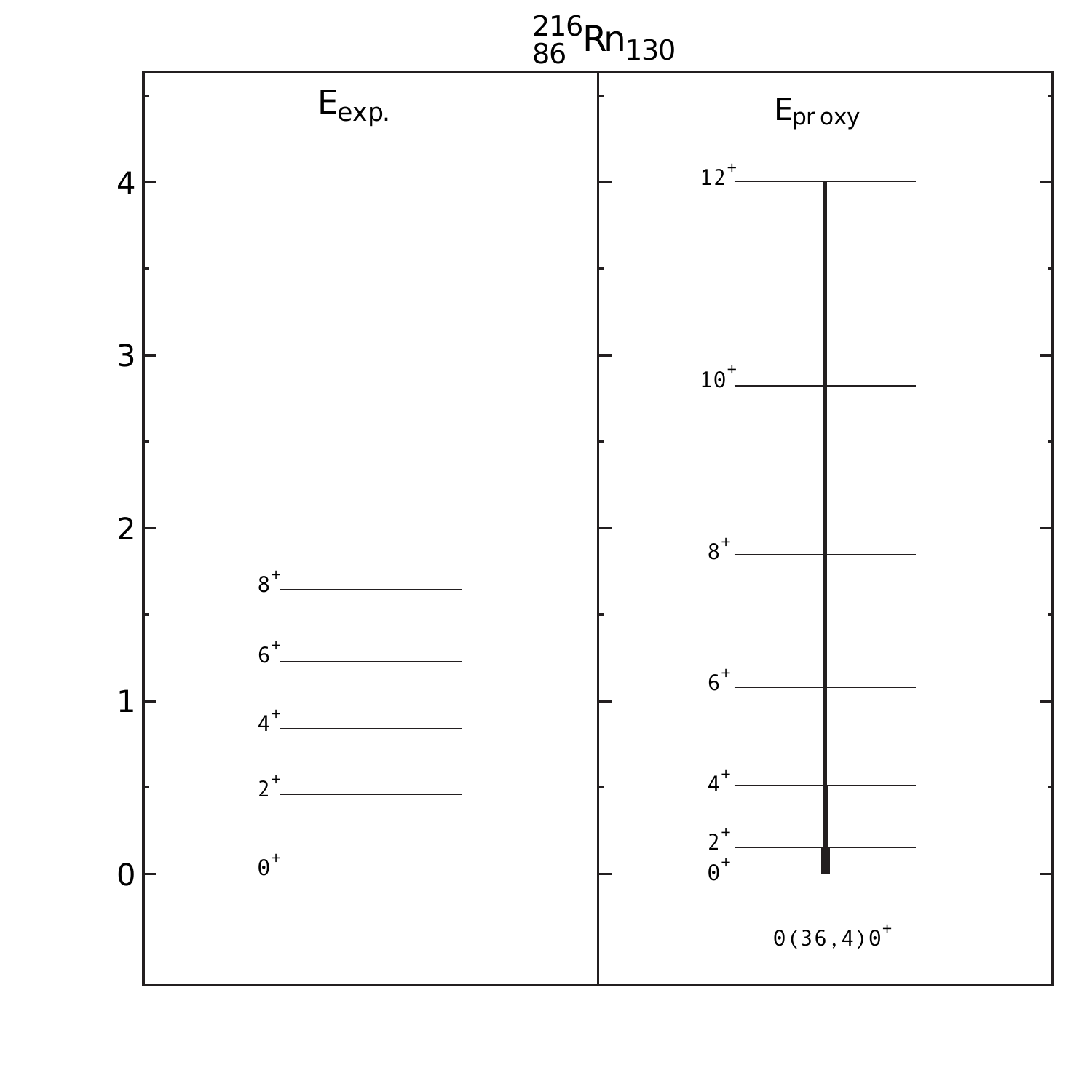}
\includegraphics[width=0.5\textwidth]{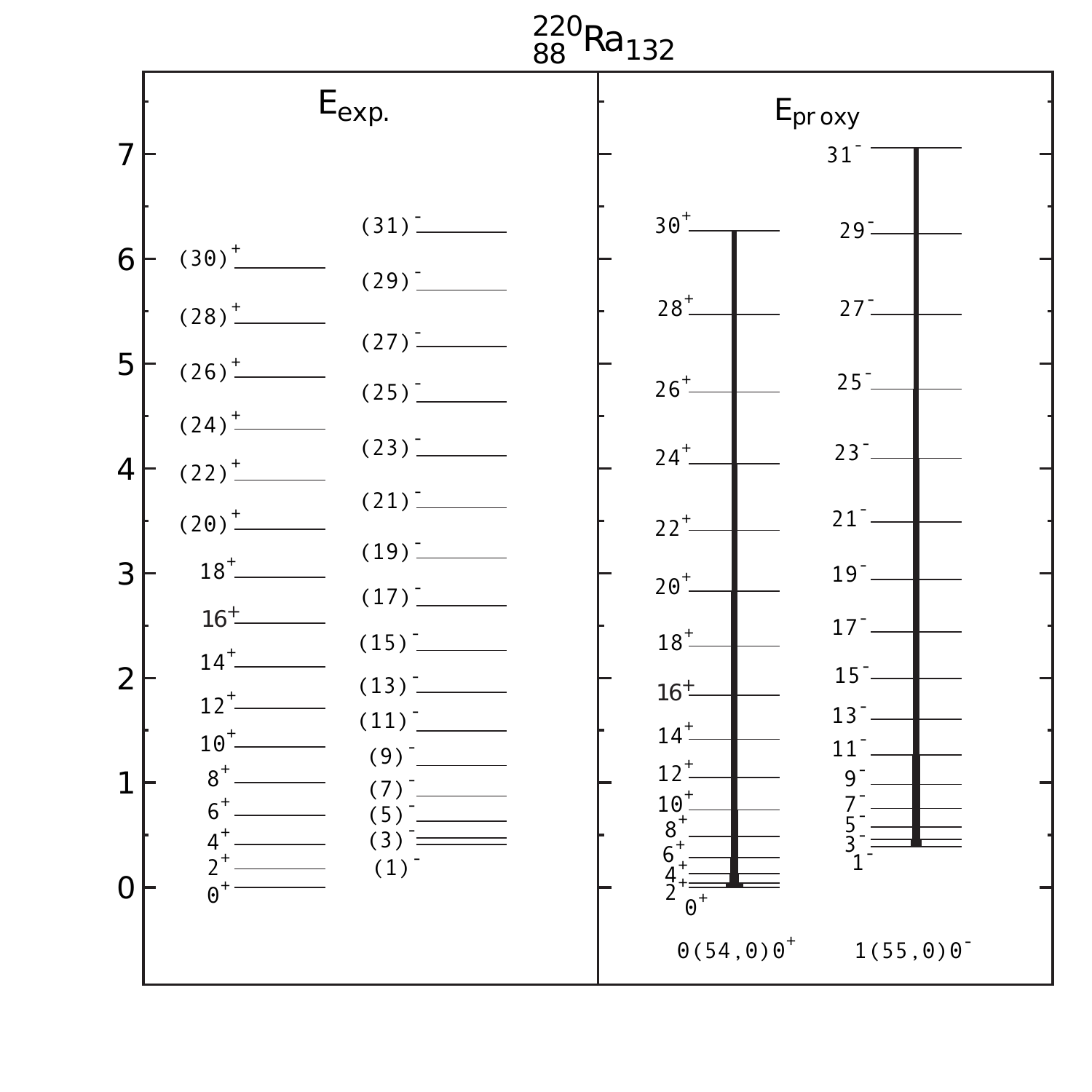}
\end{tabular}

\begin{tabular}{cc}
\includegraphics[width=0.45\textwidth, trim={1.5cm 0 0 0}, clip]{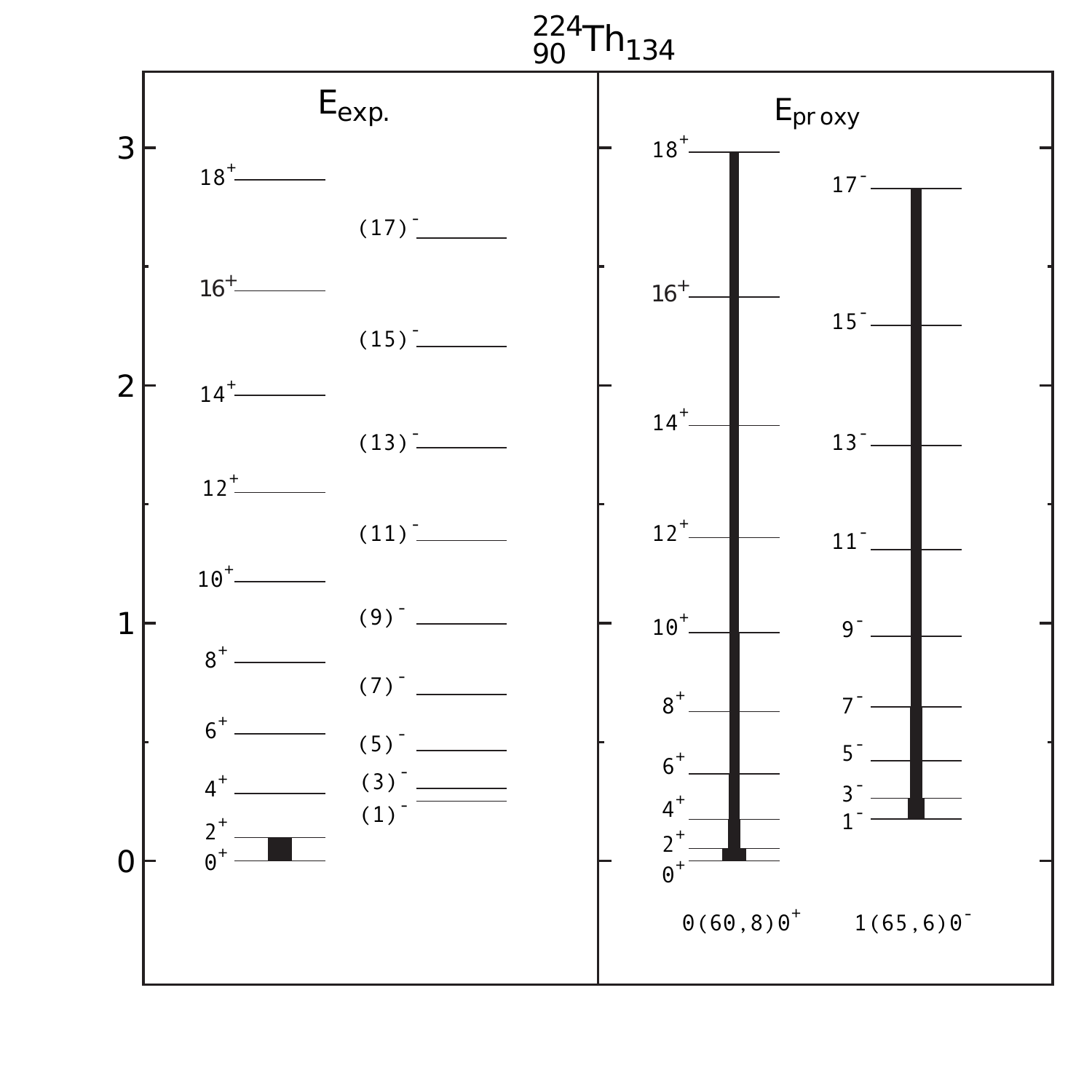}
\includegraphics[width=0.5\textwidth]{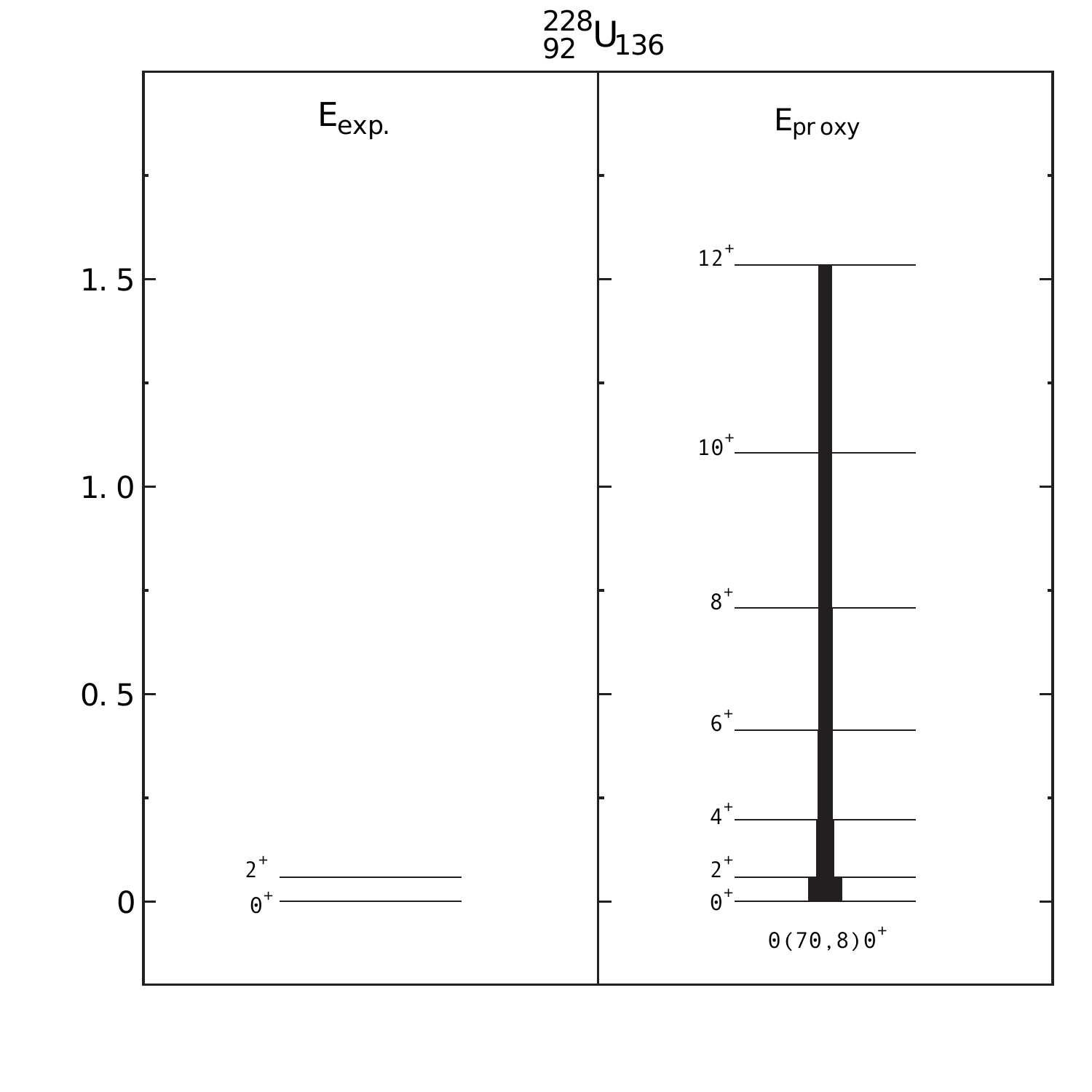}
\end{tabular}

\begin{tabular}{cc}
\includegraphics[width=0.45\textwidth, trim={1cm 1cm 0 0.2cm}, clip]{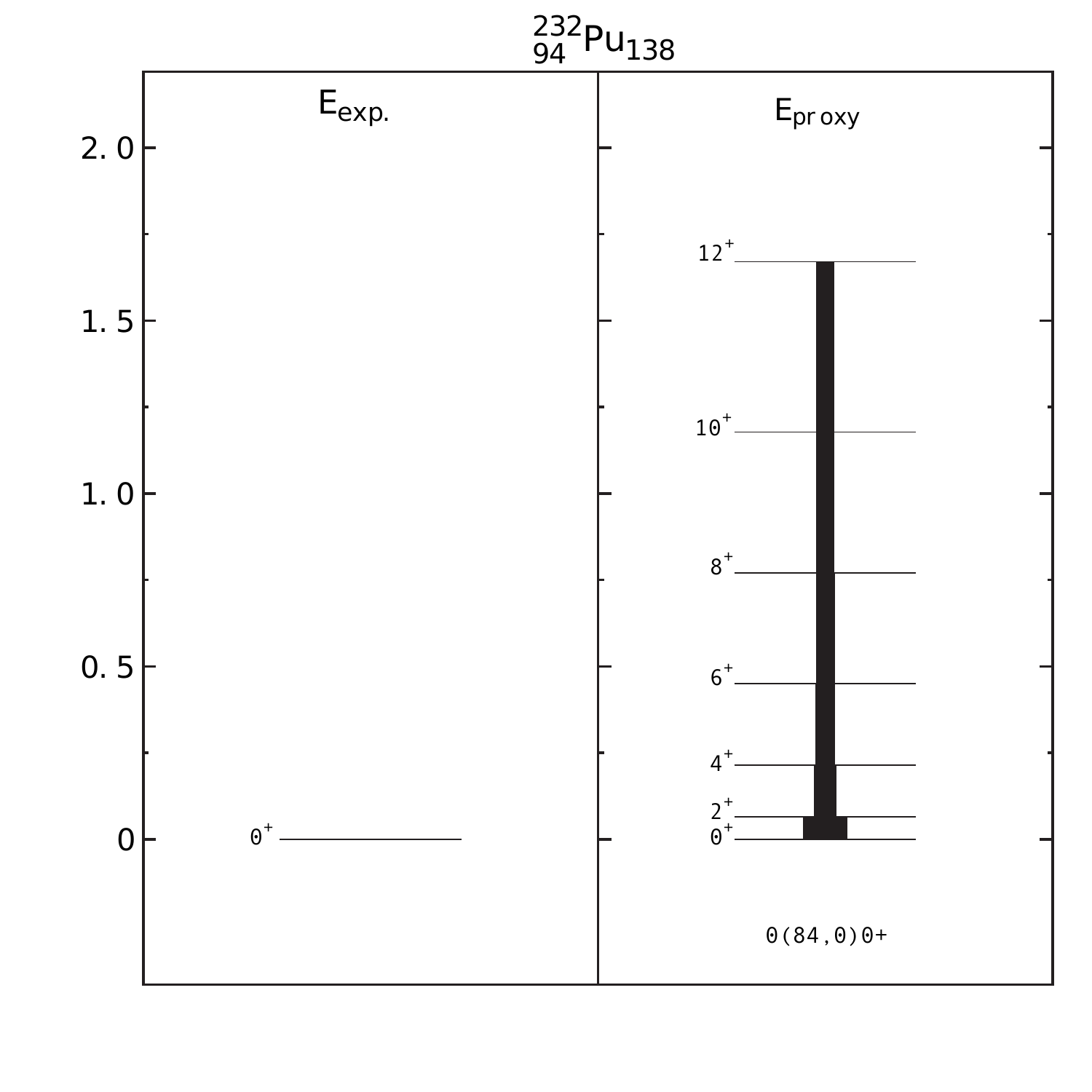}
\includegraphics[width=0.48\textwidth, trim={0 1cm 0 0.2cm}, clip]{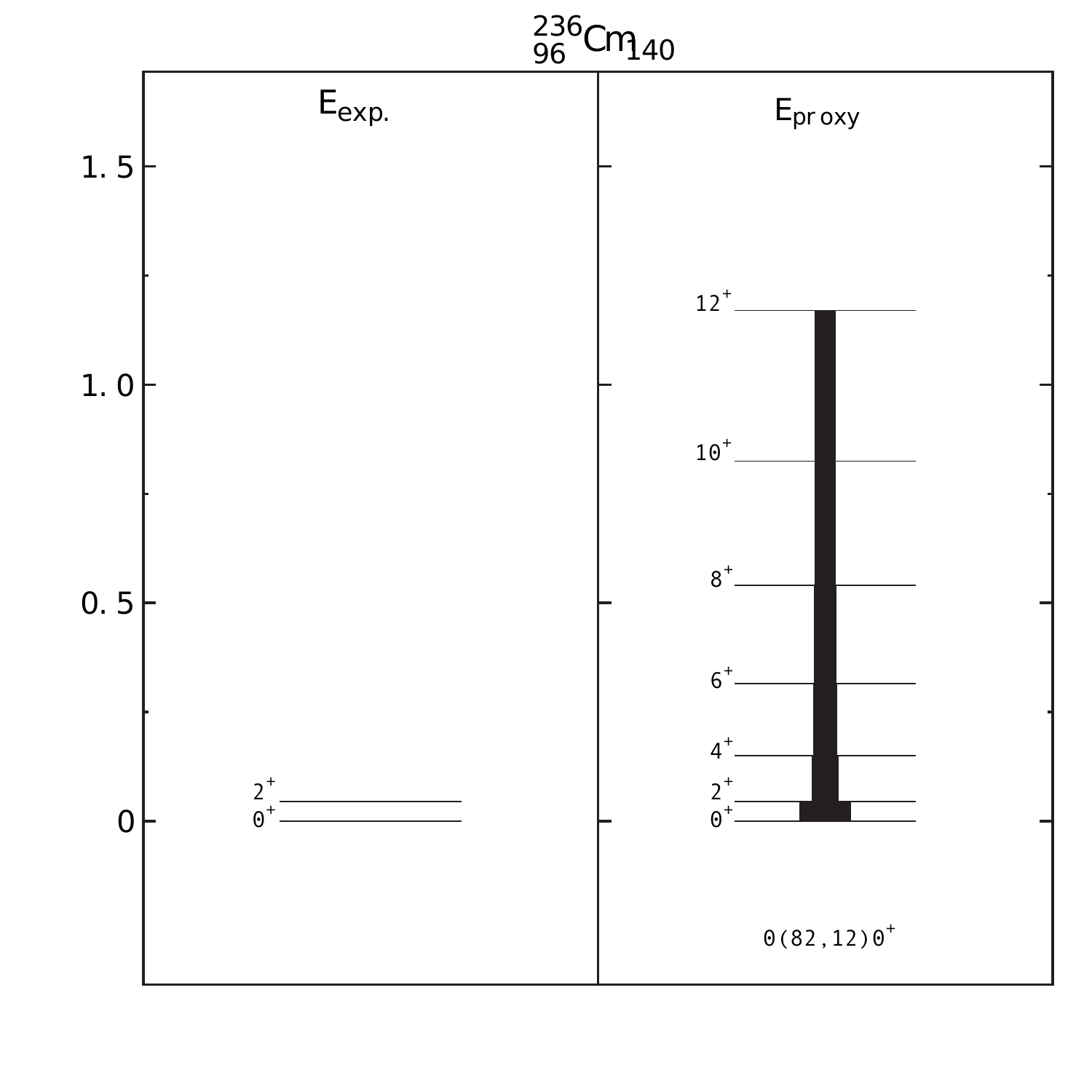}
\end{tabular}

\begin{tabular}{cc}
\includegraphics[width=0.45\textwidth, trim={1cm 1cm 0.5cm 0.2cm}, clip]{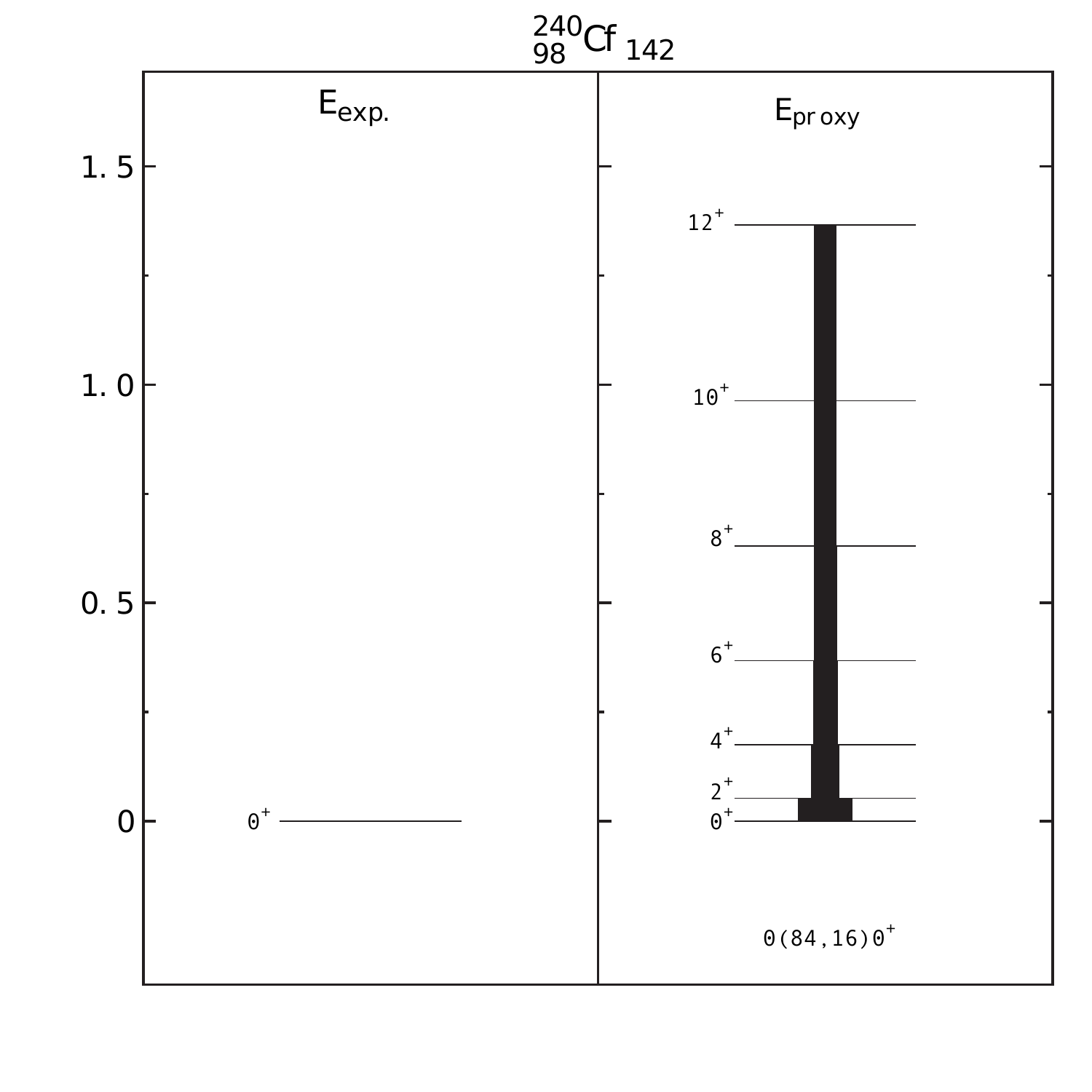}
\includegraphics[width=0.48\textwidth, trim={0 1cm 0 0.2cm}, clip]{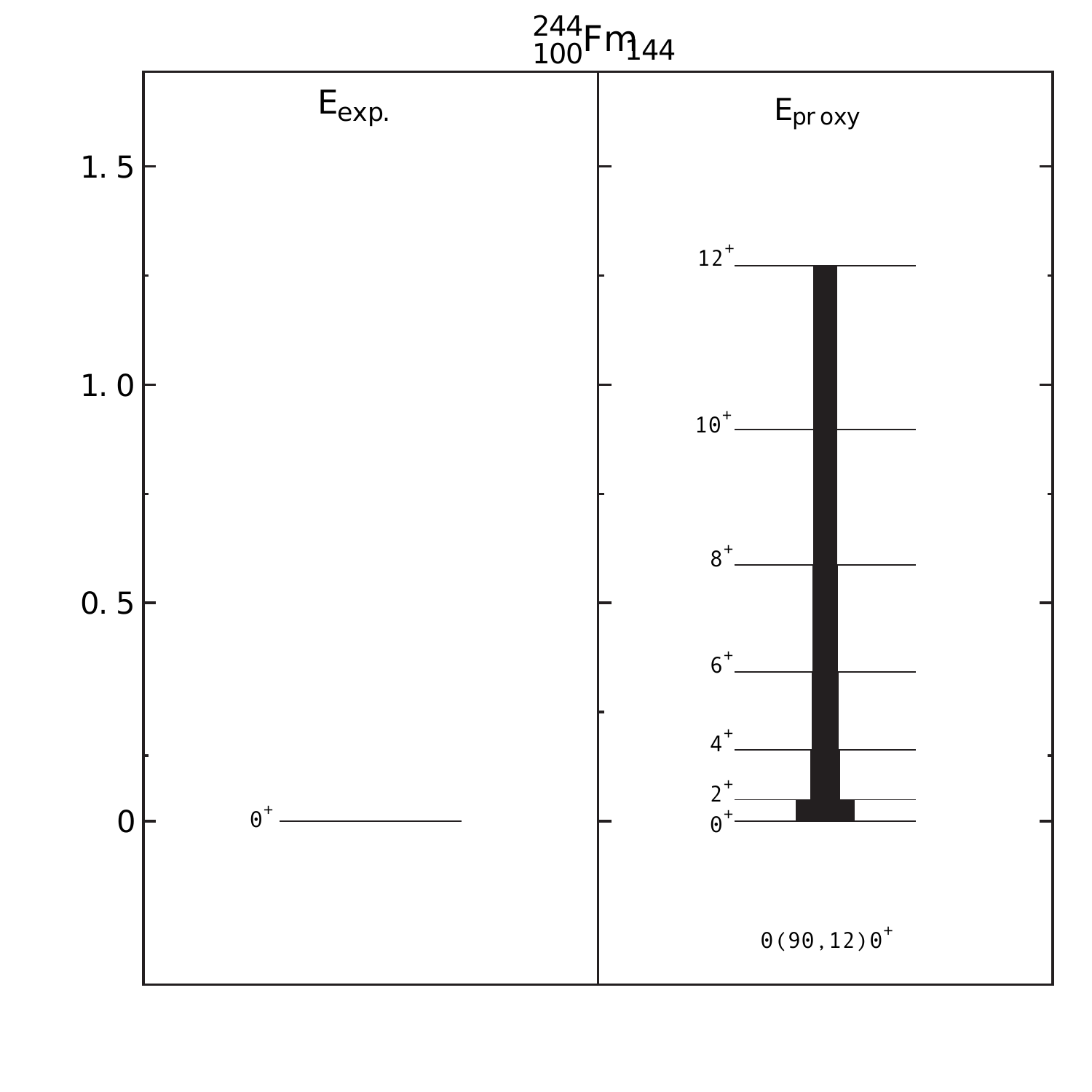}
\end{tabular}

\chapter*{Conclusions\markboth{Conclusions}{Conclusions}}
\label{sec:Conclusions}
\addcontentsline{toc}{chapter}{Conclusions}
%taking from \section*{adapted from the the SDFDM model with scalars conclusions}

A recompilation of some of the most relevant topics for heavy nuclei models is condensed into a more self-complete text from the knowledge scattered in several books and research papers. These topics included single-particle models, collective motion of the nucleus, algebraic models, symmetry breaking and symmetry restoring in atomic nuclei. The relations between them were stated and shown how can be complemented to a deeper understanding of the atomic nucleus. In the appendices, some explicit calculations were done  which are generally skipped in the literature and left for the reader to verify. 

The semimicroscopic algebraic quartet model in the proxy-$SU(3)$ scheme devised originally by J. Cseh in \cite{CsehTh} was extended to other heavy nuclei obtaining  irreducible representations, predictions of deformation parameters, excitation spectra and $E2$ electromagnetic transition probabilities. Despite of the symmetry impositions and truncation of the total states, the model works good with the few experimental data available. In hopes of future experimental efforts, the theoretical background was prepared where future measurements will judge the goodness of the model and possibly suggest improvements and extensions. 

The combination of several models from different theoretical frameworks and aid of specialized software was explained with certain degree of detail so that can be reproduced and tested.  This methodology happens to be an alternative to the yet computationally unfeasible \textit{ab initio} models and their results reflect that despite of the simplifications, it approaches considerably well to the actual dynamics between the nucleons. A more complete experimental characterization of this nuclear region will allow for a better comprehension of the most relevant dynamics and to improve the current models.

During the development of the thesis a few ideas arose which were mentioned as perspectives in the last chapter. These are related to alpha decay between the quartet nuclei for which some experimental data is available, $E1$ transition probabilities and octupole contributions to the nuclear spectra. These topics will be worked by the author in the near future and it is expected to provide further corroboration of the suitability of the model to describe  these isotopes.
%
% Appendices
	\appendices
\chapter{Shell Model Quantum Numbers and Energy Levels}

%%%%%%%%%%%%%%%%%%%%%%%%%%%%%%%%%%%%%%%%%%%%%%%%%%%%%%%%%%%%%%%%%%%%%%%%%%%%%%%%%%%%%%%%%%%%%%%%%%%%%%%%%%%%%
The single particle shell model Hamiltonian is
\begin{equation}
    H = -\frac{\hbar^2}{2M}\nabla^2 + V(r) -\mathcal{C}(r)\boldsymbol{L}\cdot \boldsymbol{S} - \mathcal{D}\left(\boldsymbol{L}^2 -\left<\boldsymbol{L}^2\right>_N\right).
    \label{smhamiltonina}
\end{equation}
For the treatment presented in this section, $V(r)$ represents only nuclear interaction, i.e., only the neutron energy levels will be shown. First consider $V(r)$ as the Woods-Saxon (W-S) potential
\begin{equation}
    V_{WS}(r) = -\frac{V_0}{1+e^{(r-R)/a}},
    \label{wsp}
\end{equation} 
which is based on the observed nuclear mass density by scattering experiments \cite{prussin}. The parameters are isotope dependent with $V_0$ the well depth in the range 40-50 MeV, $a\approx $ 0.5 fm is the nuclear surface thickness, $R = r_0 A^{1/3}$ is the nuclear radius approximation with $r_0 \approx $ 1.25 fm and for this potential $\mathcal{D} = $ 0. Another commonly used potential is the isotropic harmonic oscillator (IHO)
\begin{equation}
    V_{IHO}(r) = -V_0 + \frac{1}{2}M\omega^2 r^2,
    \label{wsp}
\end{equation} 
where $\hbar\omega = \text{41}A^{-1/3}$MeV \cite{takigawa} or a more refined $\hbar\omega = \text{45}A^{-1/3} -\text{25}A^{-2/3}$MeV \cite{BLOMQVIST1968545}. The term $\mathcal{D}\left(\boldsymbol{L}^2 -\left<\boldsymbol{L}^2\right>_N\right)$ with $\mathcal{D} \neq 0$ has the purpose of reproducing the W-S radial shape. This potential is called modified oscillator (MO) \cite{nilsson1995shapes}.

 The state labels will be given by the complete set of commuting observables (for both W-S and IHO potentials) which are $\left[H, \boldsymbol{L}^2, \boldsymbol{S}^2, \boldsymbol{J}^2, J_z, \boldsymbol{\pi} \right]$ where $\boldsymbol{\pi}$ is the parity operator. The states will then be in the angular momentum spherical coupled basis $\left|n l \frac{1}{2} j m_j \pi\right>$ with $\pi = (-1)^l$ and $n$ the radial quantum number. By means of the relation $N = 2n +l$ with $N$ the principal quantum number of the IHO  they can be labeled $\left|N l \frac{1}{2} j m_j \pi\right>$. It is worth mentioning that nuclear structure databases usually label the states only by their energy, total angular momentum and parity in the notation $j^\pi$.

\begin{figure}[h]
    \centering 
\includegraphics[width=0.9\textwidth]{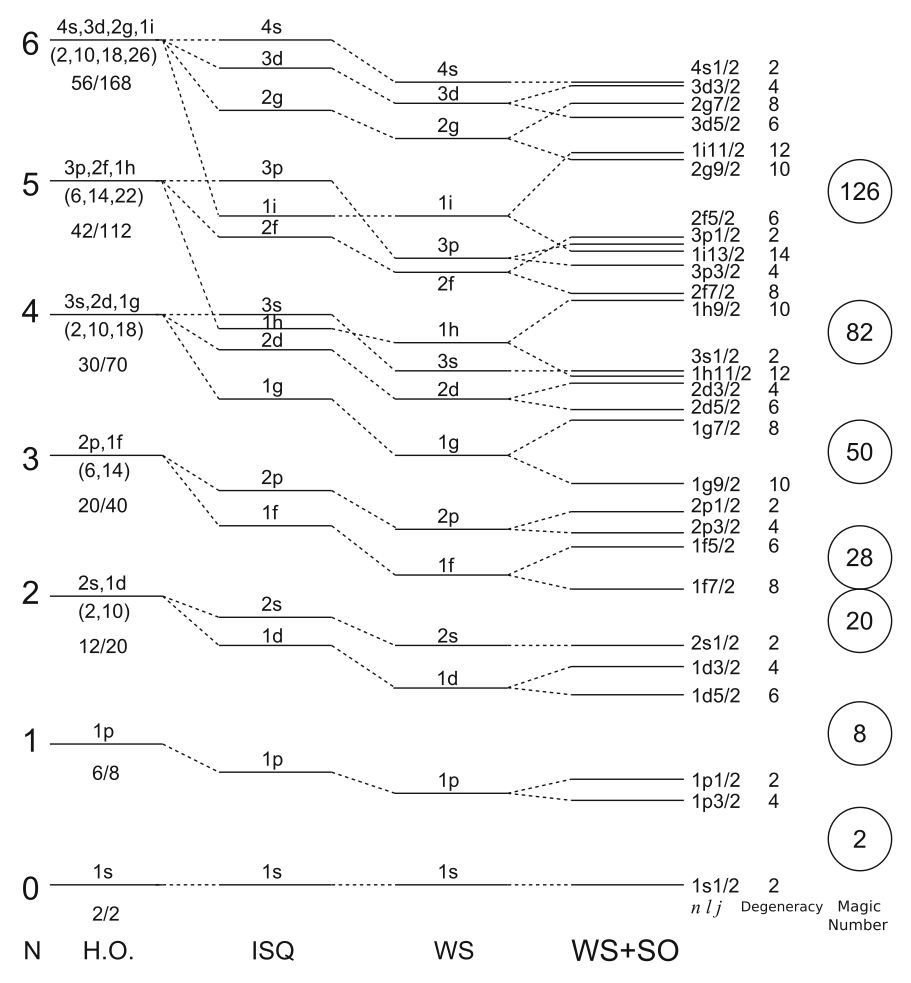}
    \caption{Single particle nucleon energy levels mean fields for harmonic oscillator (H.O.), infinite square-well (ISQ), Woods-Saxon potential (WS) and Woods-Saxon potential plus spin-orbit interaction (WS+SO). Notice the state labels, the degeneracy and appearance of shell structure along with magic numbers. Notice here $N$ starts at 0, in some references $N$ start at 1. Figure taken from \cite{takigawa}}
    \label{levels}
\end{figure}

The energy levels for the IHO can be computed exactly but those of W-S potential require numerical methods. These can be seen in figure \ref{levels} where the appearance of shell structure, splitting of levels, degeneracy and magic numbers are indicated. It is not the purpose of this section to go further than the already stated, so the reader is encouraged to review \cite{suhonen, moshinsky1996harmonic, takigawa} for details about eigenfunctions, diagonalization of the Hamiltonian and the extension to the proton states which is not very different.

\chapter{ Tensor Operators}

%%%%%%%%%%%%%%%%%%%%%%%%%%%%%%%%%%%%%%%%%%%%%%%%%%%%%%%%%%%%%%%%%%%%%%%%%%%%%%%%%%%%%%%%%%%%%%%%%%%%%%%%%%%%%

The reader may be familiar with the so-called spherical tensors $T^k_q$ which are the special case of tensor operators for the $SO(3)$ algebra. They hold the commutation relations
\begin{equation}
\begin{aligned}
&[J_z, T^k_q] = \hbar q  T^k_q,\\
&[J_{\pm}, T^k_q] = \hbar\sqrt{(k\mp q)(k\pm q + 1)} T^k_{q\pm1}.
\label{so3commutation}
\end{aligned}
\end{equation}

A generalization is now presented considering an algebra chain and corresponding irrep labels basis as 
\begin{equation}
\begin{aligned}
&\mathcal{G}\supset \mathcal{G'},\\
&\left|\Lambda\hspace{1mm}\Lambda'\right>,
\label{algebrachainappendix}
\end{aligned}
\end{equation}
then the tensor operators $T^{\Lambda}_{\Lambda'}$ with respect to $\mathcal{G}$ are defined by the relation
\begin{equation}
\begin{aligned}
&[g_k, T^{\Lambda}_{\Lambda'}] = \sum_{\Lambda''}\left< \Lambda\Lambda''\right| g_k\left| \Lambda \Lambda' \right> T^{\Lambda}_{\Lambda''}.
\label{algebrachainappendix}
\end{aligned}
\end{equation}

In a similar way $SO(3)$ and $SU(2)$ basis vectors $\left|JM\right>$ can be coupled by means of tensor product to a sum of irreps, so can the vectors $\left|\Lambda\hspace{1mm}\Lambda'\right>$ as 
\begin{equation}
\begin{aligned}
&\left|\Lambda_1 \Lambda_2; \rho \Lambda_{12}\Lambda'_{12}\right> = \sum_{\Lambda'_1,\Lambda'_2} \left< \Lambda_1\Lambda'_1 \Lambda_2\Lambda'_2|\rho\Lambda_{12}\Lambda'_{12} \right> \left|\Lambda_1\Lambda'_1\right>\otimes \left|\Lambda_2\Lambda'_2\right>,
\label{coupling}
\end{aligned}
\end{equation}
and the inverse relation.
\begin{equation}
\begin{aligned}
&\left|\Lambda_1\Lambda'_1\right>\otimes \left|\Lambda_2\Lambda'_2\right> = \sum_{\rho,\Lambda_{12},\Lambda'_{12}} \left<\rho \Lambda_{12}\Lambda'_{12} |\Lambda_{1}\Lambda'_{1} \Lambda_{2}\Lambda'_{2} \right>^{*}\left|\Lambda_1 \Lambda_2; \rho \Lambda_{12}\Lambda'_{12}\right>.
\label{uncoupling}
\end{aligned}
\end{equation}

The label $\rho$ denotes possible multiplicity since a tensor product may couple to a certain representation more than once and $\left< \Lambda_1\Lambda'_1 \Lambda_2\Lambda'_2|\rho\Lambda_{12}\Lambda'_{12} \right>$ represents the Clebsch-Gordan coefficients for $\mathcal{G}$. Equally important is the generalized Wigner-Eckart theorem
\begin{equation}
\begin{aligned}
& \left<\Lambda_1\Lambda'_1\right|T^{\Lambda}_{\Lambda'}\left|\Lambda_2\Lambda'_2\right> = \sum_{\rho}\left<\rho\Lambda_1\Lambda'_1|\Lambda\Lambda'\Lambda_2\Lambda'_2\right>^{*}\left<\rho\Lambda_1||T^{\Lambda}|| \Lambda_2 \right>,
\label{WEtheorem}
\end{aligned}
\end{equation}
where $\left<\rho\Lambda_1||T^{\Lambda}|| \Lambda_2 \right>$ are the reduced matrix elements. Similar generalizations of recoupling coefficients 6$-j$ and 9$-j$ can be found in \cite{iachello2014lie}. Tensor coupling is also possible as 
\begin{equation}
\begin{aligned}
& [T^{M}\otimes U^{N} ]^{\rho\Lambda}_{\Lambda'} = \sum_{\Lambda'',\Lambda'''} \left<M\Lambda''N\Lambda'''|\rho\Lambda\Lambda'\right>T^{M}_{\Lambda''}U^{N}_{\Lambda'''},
\label{Tensorcoupling}
\end{aligned}
\end{equation}
useful theorems and reduction formulas can be found in \cite{iachello2014lie} and \cite{suhonen}. 

Before finishing this appendix, it is worth mentioning the Racah factorization lemma which allows the reduction of coupling coefficients into simpler factors. Consider an algebra chain and states of the type 
\begin{equation}
\begin{aligned}
& \mathcal{G}_1\oplus \mathcal{G}_2\supset\mathcal{G}_{12}\supset\mathcal{G}_{12}'\supset\mathcal{G}_{12}''\\
&\left|\Lambda_1\hspace{0.5mm}\Lambda_2\hspace{0.5mm}\rho\hspace{0.5mm}\Lambda_{12}\hspace{0.5mm}\lambda_{12}\hspace{0.5mm}\mu_{12}\hspace{0.5mm}\right>,
\label{racahalgebrachain}
\end{aligned}
\end{equation}
and an uncoupling of states 
\begin{equation}
\begin{aligned}
&\left|\Lambda_1\hspace{0.5mm}\Lambda_2\hspace{0.5mm}\rho\hspace{0.5mm}\Lambda_{12}\hspace{0.5mm}\lambda_{12}\hspace{0.5mm}\mu_{12}\hspace{0.5mm}\right> = \sum_{\lambda_1,\lambda_2,\mu_1,\mu_2}\left<\Lambda_1\hspace{0.5mm}\lambda_1\hspace{0.5mm}\mu_1\hspace{0.5mm};\hspace{0.5mm}\Lambda_2\hspace{0.5mm}\lambda_2\hspace{0.5mm}\mu_2\hspace{0.5mm}|\hspace{0.5mm}\rho\hspace{0.5mm}\Lambda_{12}\hspace{0.5mm}\lambda_{12}\hspace{0.5mm}\mu_{12}\right>\left|\Lambda_1\lambda_1\mu_1\right>\otimes\left|\Lambda_2\lambda_2\mu_2\right>,
\label{racahalgebrachain}
\end{aligned}
\end{equation}
then the coupling coefficient can be factored in two parts corresponding to coefficients for reductions $\mathcal{G}\supset\mathcal{G'}$ and $\mathcal{G'}\supset\mathcal{G''}$ respectively as 
\begin{equation}
\begin{aligned}
\left<\Lambda_1\hspace{0.5mm}\lambda_1\hspace{0.5mm}\mu_1\hspace{0.5mm};\hspace{0.5mm}\Lambda_2\hspace{0.5mm}\lambda_2\hspace{0.5mm}\mu_2\hspace{0.5mm}|\hspace{0.5mm}\rho\hspace{0.5mm}\Lambda_{12}\hspace{0.5mm}\lambda_{12}\hspace{0.5mm}\mu_{12}\hspace{0.5mm}\right> = \left< \Lambda_1\hspace{0.5mm}\lambda_1\hspace{0.5mm}\Lambda_2\hspace{0.5mm}\lambda_2\hspace{0.5mm}|\hspace{0.5mm}\rho\hspace{0.5mm}\Lambda_{12}\hspace{0.5mm}\lambda_{12} \right> \left< \lambda_1\hspace{0.5mm}\mu_1\hspace{0.5mm}\lambda_2\hspace{0.5mm}\mu_2 \hspace{0.5mm}|\hspace{0.5mm}\lambda_{12}\hspace{0.5mm}\mu_{12}\right>.
\label{racahalgebrafact}
\end{aligned}
\end{equation}

\chapter{Racah Form of U(N) Algebra}

%%%%%%%%%%%%%%%%%%%%%%%%%%%%%%%%%%%%%%%%%%%%%%%%%%%%%%%%%%%%%%%%%%%%%%%%%%%%%%%%%%%%%%%%%%%%%%%%%%%%%%%%%%%%%

The boson operators $b^{\dagger}_{lm}$ and $b_{lm}$ satisfy the relations

\begin{equation}
\begin{aligned}
&[b_{lm},b^{\dagger}_{l'm'}]=\delta_{ll'}\delta_{mm'},\\
&[b_{lm},b_{l'm'}]=[b^{\dagger}_{lm},b^{\dagger}_{l'm'}]=0,
\label{bosonsbbdagger}
\end{aligned}
\end{equation}

and the Racah form of them $(b^{\dagger}_{l}\widetilde{b}_{l'})^k_q$ satisfies the algebra $U(N)$ with a different form as presented in \ref{uijcommutation} which is shown now.

\begin{equation}
\resizebox{0.9\hsize}{!}{$
\begin{aligned}
\left[(b^{\dagger}_{l_0}\widetilde{b}_{l_1})^k_q,(b^{\dagger}_{l_2}\widetilde{b}_{l_3})^{k'}_{q'}\right] & = \sum_{m_0,m_1,m_2,m_3}\left<l_0m_0l_1-m_1|kq\right>\left<l_2m_2l_3-m_3|k'q'\right>\\
&\times\left((-1)^{l_1+m_1}b^{\dagger}_{l_0m_0}\widetilde{b}_{l_3-m3}\delta_{l_1l_2}\delta_{m_1m_2}-(-1)^{l_3+m_3}b^{\dagger}_{l_2m_2}\widetilde{b}_{l_1-m1}\delta_{l_0l_3}\delta_{m_0m_3}\right),
\end{aligned}
$}
\label{BIGCOMMUTATOR1}
\end{equation}

where \ref{bosonsbbdagger} was used,

\begin{equation}
\resizebox{0.9\hsize}{!}{$
\begin{aligned}
=\sum_{m_0,m_1,m_2,m_3}(-1)^{l_1+m_1+l_0-m_0}\sqrt{\frac{2k+1}{2l_1+1}}\left<kql_0-m_0|l_1-m_1\right>\left<l_2m_2l_3-m_3|k'q'\right>b^{\dagger}_{l_0m_0}\widetilde{b}_{l_3-m3}\delta_{l_1l_2}\delta_{m_1m_2}\\
-\sum_{m_0,m_1,m_2,m_3}(-1)^{l_3+m_3+l_2-m_2}\sqrt{\frac{2k'+1}{2l_3+1}}\left<l_0m_0l_1-m_1|kq\right>\left<k'q'l_2-m_2|l_3-m_3\right>b^{\dagger}_{l_2m_2}\widetilde{b}_{l_1-m1}\delta_{l_0l_3}\delta_{m_0m_3},
\label{BIGCOMMUTATOR2}
\end{aligned}
$}
\end{equation}

where the identity $\left<j_1m_1j_2m_2|j_3m_3\right> = (-1)^{j_1-m_1}\sqrt{\frac{2j_3+1}{2j_2+1}}\left<j_3m_3j_1-m_1|j_2m_2\right>$ from equation 3.5.16 of \cite{edmonds1996angular} was used,

$$\begin{aligned}
=\sum_{m_0,m_2,m_3}(-1)^{l_0+k'+l_3-q}&\sqrt{\frac{2k+1}{2l_1+1}}
\sum_{k''}\sqrt{(2l_1+1)(2k''+1)}\begin{Bmatrix}
  k & l_0 & l_1 \\
  l_3 & k' & k'' 
 \end{Bmatrix}\\ &\times\left<k-qk''q'+q|k'q'\right>\left<l_0m_0l_3-m_3|k'' q'+q\right>b^{\dagger}_{l_0m_0}\widetilde{b}_{l_3-m3}\delta_{l_1l_2}
 \end{aligned}$$

\begin{equation}
\begin{aligned}
 -\sum_{m_1,m_2,m_3}(-1)^{l_1+k+l_2-q'}&\sqrt{\frac{2k'+1}{2l_3+1}}\sum_{k'''}\sqrt{(2l_3+1)(2k'''+1)}\begin{Bmatrix}
  k' & l_2 & l_0 \\
  l_1 & k & k''' \end{Bmatrix}\\ &\times\left<k'-q'k'''q+q'|kq\right>\left<l_2m_2l_1-m_1|k'''q+q'\right> b^{\dagger}_{l_2m_2}\widetilde{b}_{l_1-m1}\delta_{l_0l_3},
\label{BIGCOMMUTATOR3}
\end{aligned}
\end{equation}
where the following identity was used
$$\begin{aligned}&\left< j_1m_1j_2m_2 |j_{12}m_1+m_2 \right>\left< j_{12}m_1+m_2j_3m-m_1-m_2 |jm \right>\\ &\hspace{20mm}=\sum_{j_{23}}(-1)^{j_1+j_2+j_3+j}\sqrt{(2j_{12}+1)(2j_{23}+1)}\begin{Bmatrix}
  j_1 & j_2 & j_{12} \\
  j_3 & j & j_{23} \end{Bmatrix}\\
  &\hspace{60mm}\times\left< j_2m_2j_3m-m_1-m_2 |j_{23}m-m_1 \right>\left< j_1m_1j_{23}m-m_1|jm \right>,\end{aligned}$$
  
from equation 6.2.6 of \cite{edmonds1996angular}, along with \resizebox{0.5\hsize}{!}{$
\left< j_1m_1j_2-m_2 |j_3-m_3\right> = (-1)^{j_1+j_2-j_3}\left< j_1-m_1j_2m_2 |j_3m_3\right>$}. Coupling the creation and destruction operators

\begin{equation}
\resizebox{0.9\hsize}{!}{$
\begin{aligned}
=&\sum_{k''}(-1)^{l_0+k'+l_3-q}\sqrt{(2k+1)(2k''+1)}\begin{Bmatrix}
  k & l_0 & l_1 \\
  l_3 & k' & k'' 
 \end{Bmatrix}\left<k-qk''q'+q|k'q'\right>\left(b^{\dagger}_{l_0}\widetilde{b}_{l_3}\right)^{k''}_{q+q'}\delta_{l_1l_2}\\&-\sum_{k'''}(-1)^{l_1+k+l_2-q'}\sqrt{(2k'+1)(2k'''+1)}\begin{Bmatrix}
  k' & l_2 & l_0 \\
  l_1 & k & k''' \end{Bmatrix}\left<k'-q'k'''q+q'|kq\right>\left(b^{\dagger}_{l_2}\widetilde{b}_{l_1}\right)^{k'''}_{q+q'}\delta_{l_0l_3}.
\label{BIGCOMMUTATOR4}
\end{aligned}
$}
\end{equation}

Since $$\left<k-qk''q'+q|k'q'\right> = (-1)^{q+k'-k''}\sqrt{\frac{2k'+1}{2k''+1}}\left<kqk'q'|k''q+q'\right>,$$  $$\left<k'-q'k'''q+q'|kq\right> = (-1)^{k'+q'}\sqrt{\frac{2k+1}{2k'''+1}}\left<kqk'q'|k'''q+q'\right>,$$ the sums over $k''$ and $k'''$ both run on the same values, i.e., $\sum_{k''}=\sum_{k'''}$ and using $6-j$ symbols symmetry relations one finally obtains

\begin{equation}
\resizebox{0.9\hsize}{!}{$
\begin{aligned}
&\left[(b^{\dagger}_{l_0}\widetilde{b}_{l_1})^k_q,(b^{\dagger}_{l_2}\widetilde{b}_{l_3})^{k'}_{q'}\right]=\sum_{k''}\sqrt{(2k+1)(2k'+1)}\left<kqk'q'|k''q+q'\right>\\&\times \left[(-1)^{l_0+l_3-k''}\begin{Bmatrix}
  k & k' & k'' \\
  l_3 & l_0 & l_1 \end{Bmatrix}\left(b^{\dagger}_{l_0}\widetilde{b}_{l_3}\right)^{k''}_{q+q'}\delta_{l_1l_2}-(-1)^{l_1+l_2+k+k'}\begin{Bmatrix}
  k & k' & k'' \\
  l_2 & l_1 & l_0 \end{Bmatrix}\left(b^{\dagger}_{l_2}\widetilde{b}_{l_1}\right)^{k''}_{q+q'}\delta_{l_0l_3}\right].
\label{BIGCOMMUTATOR5}
\end{aligned}
$}
\end{equation}

Some references include a sum over $q'' = q+q'$ which is omitted here because of properties of Clebsh-Gordan coefficients about projections of angular momentum.

\chapter{Racah Form of Quadrupole and Angular Momentum Operators}

%%%%%%%%%%%%%%%%%%%%%%%%%%%%%%%%%%%%%%%%%%%%%%%%%%%%%%%%%%%%%%%%%%%%%%%%%%%%%%%%%%%%%%%%%%%%%%%%%%%%%%%%%%%%%

The Racah form of a tensor operator $T^{k}_q$ is obtained as 

\begin{equation}
\resizebox{0.9\hsize}{!}{$
\begin{aligned}
&\hspace{20mm}T^{k}_q = \mathbb{I}T^{k}_q\mathbb{I} = \sum_{l,m,l',m'}\left|lm\right>\left< lm \right|T^{k}_q\left|l'm'\right>\left< l'm'\right| = \sum_{l,m,l',m'}\left< lm \right|T^{k}_q\left|l'm'\right>b^{\dagger}_{lm}b_{l'm'}\\
& =  \sum_{l,l'}\frac{\left< l||T^{k}||l'\right>}{\sqrt{2l+1}}\sum_{m,m'}\left<l'm'kq|lm\right>b^{\dagger}_{lm}b_{l'm'} =  \sum_{l,l'}\frac{\left< l||T^{k}||l'\right>}{\sqrt{2l+1}}\sum_{m,m'}(-1)^{l'-m'}\left<lml'-m'|kq\right>\sqrt{\frac{2l+1}{2k+1}}b^{\dagger}_{lm}b_{l'm'}\\& = \frac{1}{\sqrt{2k+1}}\sum_{l,l'}\langle l||T^{k}||l'\rangle\sum_{m,m'}(-1)^{l'+m'}\left<lml'm'|kq\right>b^{\dagger}_{lm}b_{l'-m'} = \frac{1}{\sqrt{2k+1}}\sum_{l,l'}\langle l||T^{k}||l'\rangle\left(b^{\dagger}_{l}\widetilde{b}_{l'}\right)^k_q
\label{TRacah}
\end{aligned}
$}
\end{equation}

Thus for $L^1_q$ 
\begin{equation}
\begin{aligned}
&L^{1}_q = \frac{1}{\sqrt{3}}\sum_{l,l'}\langle l||L^{1}||l'\rangle\left(b^{\dagger}_{l}\widetilde{b}_{l'}\right)^1_q,
\label{LRacah}
\end{aligned}
\end{equation}

where the reduced matrix element is given by
\begin{equation}
\resizebox{0.9\hsize}{!}{$
\begin{aligned}
&\hspace{15mm}\langle l||L^{1}||l'\rangle = \frac{\sqrt{2l+1}}{\left<l'm'10|lm\right>}\langle lm|J^1_0 | l'm' \rangle = \frac{\sqrt{2l+1}m}{m\sqrt{2l+1}\sqrt{\frac{1}{l(l+1)(2l+1)}}} = \sqrt{l(l+1)(2l+1)},
\label{LRacah2q}
\end{aligned}
$}
\end{equation}
which implies that
\begin{equation}
\begin{aligned}\hspace{5mm}L^{1}_q = \sum_{l}\sqrt{\frac{l(l+1)(2l+1)}{3}}\left(b^{\dagger}_{l}\widetilde{b}_{l}\right)^1_q.
\label{LRacah2}
\end{aligned}
\end{equation}

For $Q^2_q$ as defined in \ref{qmodified}
\begin{equation}
\resizebox{0.9\hsize}{!}{$
\begin{aligned}
&Q^{2}_q = \frac{1}{\sqrt{5}}\sum_{l,l'}\langle\eta l||Q^{2}||\eta l'\rangle\left(b^{\dagger}_{l}\widetilde{b}_{l'}\right)^2_q =  \frac{\sqrt{4\pi}}{5}\sum_{l,l'}\left(\langle\eta l||r^2Y^2(\Omega)||\eta l'\rangle+ \langle\eta l||p^2Y^2(\Omega_p)||\eta l'\rangle\right)\left(b^{\dagger}_{l}\widetilde{b}_{l'}\right)^2_q,
\label{QRacah}
\end{aligned}
$}
\end{equation}
where the states require the label of the major shell $\eta$ involved because of $r$ and $p$ coordinates appearance. Both reduced matrix elements have the same value  
\begin{equation}
\begin{aligned}
&\langle\eta l||r^2Y^2(\Omega)||\eta l'\rangle = \frac{\sqrt{2l+1}}{\left< l'm'20 | lm \right>}\langle \eta l |r^2  | \eta l' \rangle \langle \eta l m |Y^2_0(\Omega) | \eta l' m' \rangle\\
& = \sqrt{\frac{5}{4\pi}}\frac{2l+1}{\sqrt{2l'+1}}\langle \eta l |r^2  | \eta l' \rangle \frac{\left< l-m20 | l'-m \right>\left< l020 | l'0 \right>}{\left< l'm20 | lm \right>},
\label{QRaca2}
\end{aligned}
\end{equation}

where the spherical harmonic addition theorem \cite{rose2013elementary} was used. The quotient of Clebsh-Gordan coefficients has the allowed values

\begin{equation}
\begin{aligned}
& \frac{\left< l-m20 | l'-m \right>\left< l020 | l'0 \right>}{\left< l'm20 | lm \right>} = \begin{cases}
        &-\sqrt{\frac{l(l+1)}{(2l-1)(2l+3)}}\hspace{3mm}\text{if}\hspace{3mm} l'=l,\\
        &\sqrt{\frac{3}{2}}\frac{1}{2l+1}\sqrt{\frac{(l+1)(l+2)(2l+5)}{2l+3}}\hspace{3mm}\text{if}\hspace{3mm} l'=l\pm 2.
        \end{cases}
\label{QRaca3}
\end{aligned}
\end{equation}

Replacing in \ref{QRacah} and recalling that radial functions are real implying that $\langle \eta l |r^2  | \eta l' \rangle = \langle \eta l' |r^2  | \eta l \rangle$ it is obtained that 

\begin{equation}
\begin{aligned}
 Q^{2}_q = &-\sum_{l}\langle \eta l |r^2  | \eta l \rangle\sqrt{\frac{l(l+1)(2l+1)}{5(2l-1)(2l+3)}}\left(b^{\dagger}_{l}\widetilde{b}_{l}\right)^2_q \\&+ \sum_{l}\langle \eta l |r^2  | \eta l +2 \rangle\sqrt{\frac{6(l+1)(l+2)}{5(2l+3)}}\left( \left(b^{\dagger}_{l}\widetilde{b}_{l+2}\right)^2_q  + \left(b^{\dagger}_{l+2}\widetilde{b}_{l}\right)^2_q  \right),
\label{QRaca4}
\end{aligned}
\end{equation}

the expected values of $r^2$ are given by \cite{LAWSON} 

\begin{equation}
\begin{aligned}
 &\langle \eta l |r^2  | \eta l \rangle = \eta+\frac{3}{2},\\
  &\langle \eta l |r^2  | \eta l+2 \rangle = \sqrt{(\eta-l)(\eta+l+3)},
\label{lawsonr2}
\end{aligned}
\end{equation}
thus it is obtained 

\begin{equation}
\begin{aligned}
 Q^{2}_q = &-\left(\eta+\frac{3}{2}\right)\sum_{l}\sqrt{\frac{l(l+1)(2l+1)}{5(2l-1)(2l+3)}}\left(b^{\dagger}_{l}\widetilde{b}_{l}\right)^2_q \\&+ \sum_{l}\sqrt{\frac{6(l+1)(l+2)(\eta-l)(\eta+l+3)}{5(2l+3)}}\left(\left(b^{\dagger}_{l}\widetilde{b}_{l+2}\right)^2_q  + \left(b^{\dagger}_{l+2}\widetilde{b}_{l}\right)^2_q  \right),
\label{QRaca4}
\end{aligned}
\end{equation}

With these definitions and the result of appendix C, the commutators of the $SU(3)$ algebra can be computed. Two of them are shown now and the third one is left as an exercise for the reader
\begin{equation}
\resizebox{0.9\hsize}{!}{$
\begin{aligned}
&\hspace{20mm}\left[L^1_q, L^1_{q'}\right] = \sum_{l,l'}\sqrt{\frac{l(l+1)(2l+1)}{3}}\sqrt{\frac{l'(l'+1)(2l'+1)}{3}}\left[\left(b^{\dagger}_{l}\widetilde{b}_{l}\right)^1_q, \left(b^{\dagger}_{l'}\widetilde{b}_{l'}\right)^1_{q'}\right]\\
&=\sum_{l,l'}\sqrt{\frac{l(l+1)(2l+1)}{3}}\sqrt{\frac{l'(l'+1)(2l'+1)}{3}}\sum_{k''}3\left< 
1q1q'|k''q+q'\right>\\&\hspace{20mm}\times\left[(-1)^{l+l'-k''}\begin{Bmatrix}
  1 & 1 & k'' \\
  l' & l & l \end{Bmatrix}\left(b^{\dagger}_{l}\widetilde{b}_{l'}\right)^{k''}_{q+q'}\delta_{ll'}-(-1)^{l+l'}\begin{Bmatrix}
  1 & 1 & k'' \\
  l' & l & l \end{Bmatrix}\left(b^{\dagger}_{l'}\widetilde{b}_{l}\right)^{k''}_{q+q'}\delta_{ll'}\right]\\&=\sum_{l,k''}l(l+1)(2l+1)\left< 
1q1q'|k''q+q'\right>\begin{Bmatrix}
  1 & 1 & k'' \\
  l & l & l \end{Bmatrix}\left[(-1)^{-k''}\left(b^{\dagger}_{l}\widetilde{b}_{l}\right)^{k''}_{q+q'}- \left(b^{\dagger}_{l}\widetilde{b}_{l}\right)^{k''}_{q+q'}\right]\\&=\sum_{l}l(l+1)(2l+1)\left< 
1q1q'|1q+q'\right>\begin{Bmatrix}
  1 & 1 & 1 \\
  l & l & l \end{Bmatrix}(-2)\left(b^{\dagger}_{l}\widetilde{b}_{l}\right)^{1}_{q+q'}\\&=\sum_{l}l(l+1)(2l+1)\left< 
1q1q'|1q+q'\right>\sqrt{\frac{2}{3}}\sqrt{\frac{(2l-1)!}{(2l+2)!}}(-2)\left(b^{\dagger}_{l}\widetilde{b}_{l}\right)^{1}_{q+q'}\\& = \sum_{l}\sqrt{\frac{l(l+1)(2l+1)}{3}}\left< 
1q1q'|1q+q'\right>(-\sqrt{2})\left(b^{\dagger}_{l}\widetilde{b}_{l}\right)^{1}_{q+q'}\\& = -\sqrt{2}\left< 
1q1q'|1q+q'\right>L^1_{q+q'}.
\label{COMMSU3}
\end{aligned}
$}
\end{equation}

The 6-j symbols were computed using Mathematica.

\begin{equation}
\begin{aligned}
&\left[L^1_q, Q^2_{q'}\right] =  -\left(\eta+\frac{3}{2}\right)\sum_{l,l'}\sqrt{\frac{l(l+1)(2l+1)}{3}}\sqrt{\frac{l'(l'+1)(2l'+1)}{5(2l'-1)(2l'+3)}}\left[\left(b^{\dagger}_{l}\widetilde{b}_{l}\right)^1_q, \left(b^{\dagger}_{l'}\widetilde{b}_{l'}\right)^2_{q'}\right]\\&+ \sum_{l,l'}\sqrt{\frac{l(l+1)(2l+1)}{3}}\sqrt{\frac{6(l+1)(l+2)(\eta-l)(\eta+l+3)}{5(2l+3)}}\\&\hspace{50mm} \times\left(\left[\left(b^{\dagger}_{l}\widetilde{b}_{l}\right)^1_q, \left(b^{\dagger}_{l'}\widetilde{b}_{l'+2}\right)^2_{q'}\right]  + \left[\left(b^{\dagger}_{l}\widetilde{b}_{l}\right)^1_q, \left(b^{\dagger}_{l'+2}\widetilde{b}_{l'}\right)^2_{q'}\right] \right),
\label{COMMSU31}
\end{aligned}
\end{equation}

both sums will be computed separately. The first is analog to the commutator for angular momentum components and has the form

\begin{equation}
\resizebox{0.9\hsize}{!}{$
\begin{aligned}
& -\left(\eta+\frac{3}{2}\right)\sum_{l}\sqrt{\frac{l(l+1)(2l+1)}{3}}\sqrt{\frac{l(l+1)(2l+1)}{5(2l-1)(2l+3)}}\sum_{k''}\sqrt{15}\left< 
 1q2q' |k''q+q' \right>\begin{Bmatrix}
  1 & 2 & k'' \\
  l & l & l \end{Bmatrix}\\&\hspace{80mm}\times\left[ (-1)^{-k''}\left(b^{\dagger}_{l}\widetilde{b}_{l}\right)^{k''}_{q+q'}+\left(b^{\dagger}_{l}\widetilde{b}_{l}\right)^{k''}_{q+q'} \right],
\label{COMMSU32}
\end{aligned}
$}
\end{equation}
where $k''= 1,2,3$ of which only the value of 2 contributes
\begin{equation}
\resizebox{0.9\hsize}{!}{$
\begin{aligned}
& -\left(\eta+\frac{3}{2}\right)\sum_{l}\frac{l(l+1)(2l+1)}{\sqrt{(2l-1)(2l+3)}}\left< 
 1q2q' |2q+q' \right>\begin{Bmatrix}
  1 & 2 & 2 \\
  l & l & l \end{Bmatrix}\times2\left(b^{\dagger}_{l}\widetilde{b}_{l}\right)^{2}_{q+q'}\\& =  -\left(\eta+\frac{3}{2}\right)\sum_{l}\frac{l(l+1)(2l+1)}{\sqrt{(2l-1)(2l+3)}}\left< 
 1q2q' |2q+q' \right>\left(-\sqrt{\frac{6}{5}}\frac{1}{2\sqrt{l(l+1)(2l+1)}}\right)\times2\left(b^{\dagger}_{l}\widetilde{b}_{l}\right)^{2}_{q+q'}\\& =  \sqrt{6}\left< 
 1q2q' |2q+q' \right>\left(\eta+\frac{3}{2}\right)\sum_{l}\sqrt{\frac{l(l+1)(2l+1)}{5(2l-1)(2l+3)}}\left(b^{\dagger}_{l}\widetilde{b}_{l}\right)^{2}_{q+q'}, 
\end{aligned}
$}
\label{COMMSU32}
\end{equation}
for the second term a few steps will be skipped but can be easily followed. It will be equal to
\begin{equation}
\resizebox{0.9\hsize}{!}{$
\begin{aligned}
\\& \sum_{l}\sqrt{\frac{l(l+1)(2l+1)}{3}}\sqrt{\frac{6(l+1)(l+2)(\eta-l)(\eta+l+3)}{5(2l+3)}}\sum_{k''}\sqrt{15}\left< 1q2q'| k'' q+q'\right>\\&\hspace{20mm}\times \begin{Bmatrix}
  1 & 2 & k'' \\
  l+2 & l & l \end{Bmatrix}\left( 
(-1)^{k''}\left(b^{\dagger}_{l}\widetilde{b}_{l+2}\right)^{k''}_{q+q'} + \left(b^{\dagger}_{l+2}\widetilde{b}_{l}\right)^{k''}_{q+q'}\right)  
\\& +\sum_{l'}\sqrt{\frac{(l'+2)(l'+3)(2l'+5)}{3}}\sqrt{\frac{6(l'+1)(l'+2)(\eta-l')(\eta+l'+3)}{5(2l'+3)}}\sum_{k''}\sqrt{15}\left< 1q2q'| k'' q+q'\right>\\&\hspace{20mm}\times 
\begin{Bmatrix}
  1 & 2 & k'' \\
  l' & l'+2 & l'+2 \end{Bmatrix}\left( 
(-1)^{k''}\left(b^{\dagger}_{l'+2}\widetilde{b}_{l'}\right)^{k''}_{q+q'} + \left(b^{\dagger}_{l'}\widetilde{b}_{l'+2}\right)^{k''}_{q+q'}\right)\\&
= \sum_{l}\sqrt{l(l+1)(2l+1)}\sqrt{\frac{6(l+1)(l+2)(\eta-l)(\eta+l+3)}{(2l+3)}}\left< 1q2q'| 2 q+q'\right>\\&\hspace{20mm}\times \sqrt{\frac{2}{15}}\sqrt{\frac{l}{(l+1)(2l+1)}}\left( 
\left(b^{\dagger}_{l}\widetilde{b}_{l+2}\right)^{2}_{q+q'} + \left(b^{\dagger}_{l+2}\widetilde{b}_{l}\right)^{2}_{q+q'}\right)  
\\& +\sum_{l'}\sqrt{(l'+2)(l'+3)(2l'+5)}\sqrt{\frac{6(l'+1)(l'+2)(\eta-l')(\eta+l'+3)}{(2l'+3)}}\left< 1q2q'| 2 q+q'\right>\\&\hspace{20mm}\times 
\left( -\sqrt{\frac{2}{15}}\sqrt{\frac{l'+3}{(2l'+5)(l'+2)}} \right)\left( \left(b^{\dagger}_{l'+2}\widetilde{b}_{l'}\right)^{2}_{q+q'} + \left(b^{\dagger}_{l'}\widetilde{b}_{l'+2}\right)^{2}_{q+q'}\right)\\&
= \sum_{l}\frac{\sqrt{6}}{3}l\sqrt{\frac{6(l+1)(l+2)(\eta-l)(\eta+l+3)}{5(2l+3)}}\left< 1q2q'| 2 q+q'\right>\left( 
\left(b^{\dagger}_{l}\widetilde{b}_{l+2}\right)^{2}_{q+q'} + \left(b^{\dagger}_{l+2}\widetilde{b}_{l}\right)^{2}_{q+q'}\right)  
\\& -\sum_{l'}\frac{\sqrt{6}}{3}(l'+3)\sqrt{\frac{6(l'+1)(l'+2)(\eta-l')(\eta+l'+3)}{5(2l'+3)}}\left< 1q2q'| 2 q+q'\right>\left( \left(b^{\dagger}_{l'+2}\widetilde{b}_{l'}\right)^{2}_{q+q'} + \left(b^{\dagger}_{l'}\widetilde{b}_{l'+2}\right)^{2}_{q+q'}\right)\\&=-\sqrt{6}\left< 1q2q'| 2 q+q'\right>\sum_{l}\sqrt{\frac{6(l+1)(l+2)(\eta-l)(\eta+l+3)}{5(2l+3)}}\left( 
\left(b^{\dagger}_{l}\widetilde{b}_{l+2}\right)^{2}_{q+q'} + \left(b^{\dagger}_{l+2}\widetilde{b}_{l}\right)^{2}_{q+q'}\right) ,
\label{COMMSU33}
\end{aligned}
$}
\end{equation}
where in the last equality it was considered the fact that $\sum_l=\sum_{l'}$. Adding \ref{COMMSU32} and \ref{COMMSU33} it is seen that 

\begin{equation}
\begin{aligned}
&\left[L^1_q, Q^2_{q'}\right] = -\sqrt{6}\left< 1q2q'| 2 q+q'\right>Q^2_{q+q'}.
\label{COMMSU35}
\end{aligned}
\end{equation}

\chapter{Quadrupole and Angular Momentum Operators as $SU(3)$ Tensors}

%%%%%%%%%%%%%%%%%%%%%%%%%%%%%%%%%%%%%%%%%%%%%%%%%%%%%%%%%%%%%%%%%%%%%%%%%%%%%%%%%%%%%%%%%%%%%%%%%%%%%%%%%%%%%

In this appendix the expansion of operators $L^1_q$ and $Q^2_q$ as $SU(3)$ tensors is done. Consider the creation operator in the spherical basis $b^\dagger_{lm}$ which creates a state 
\begin{equation}
\begin{aligned}
&b^\dagger_{lm}|0\rangle = |l,m\rangle,
\label{bdaggercreation}
\end{aligned}
\end{equation}
corresponding to an irrep $[1]$ of $U((\eta+1)(\eta+2)/2)$ whose branching to $SU(3)\supset SO(3)\supset SO(2)$ corresponds to $(\eta,0)$, with $K = 0$, $l = \eta, \eta-2,\eta-4,...0$ or 1 and $m = -l,...,l$. Thus $b^\dagger_{lm}$ is a $(\eta,0)$ tensor in $SU(3)$ and
\begin{equation}
\begin{aligned}
b^{\dagger}_{(\eta,0),K=0,l,m}|0\rangle = |[1],(\eta,0),K=0,l,m\rangle.
\label{bdecompinsu3}
\end{aligned}
\end{equation}

The $SU(3)$ labels are required for the adjoint destruction operator $\widetilde{b}_{lm} = (-1)^{l+m}b_{l-m}$ as well. To obtain them consider the number operator which must be a tensor of kind
\begin{equation}
\begin{aligned}
n_{(0,0),K=0,l=0,m=0} = \sum_{l,m}b^{\dagger}_{(\eta,0),K=0,l,m}b_{(\lambda,\mu),K,l,m},
\label{numberoperator}
\end{aligned}
\end{equation}
which imposes $b_{(\eta,0),K=0,l,m}$ and for the adjoint $\widetilde{b}_{(0,\eta),K=0,l,m} = (-1)^{\phi+l+m}b_{(\eta,0),K=0,l,-m}$ where $\phi$ is a phase to be determined by coupling the operators as follows
\begin{equation}
\begin{aligned}
n_{(0,0),K=0,l=0,m=0} &= \sum_{l,m}b^{\dagger}_{(\eta,0),K=0,l,m}b_{(0,\eta),K =0,l,m} \\
&= \sum_{l,m}(-1)^{\phi+l+m}b^{\dagger}_{(\eta,0),K=0,l,m}\widetilde{b}_{(0,\eta),K=0,l,-m}\\
&= \sum_{l,m}(-1)^{\phi+l+m} \langle(\eta,0),K=0,l;(0,\eta),K=0,l||(0,0),K=0,0\rangle \\ &\hspace{30mm}\times
\langle l m l -m|00\rangle
\left[b^{\dagger}_{(\eta,0),l}\widetilde{b}_{(0,\eta),l}\right]_{(0,0),K=0,l=0,m=0},
\label{phasecalculation}
\end{aligned}
\end{equation}
where no additional summations where added because of the coupling to $(0,0)K=0,l=0, m=0$. The values of the Clebsch-Gordan coefficients can be determined to be \cite{kota20203}
\begin{equation}
\begin{aligned}
\langle(\eta,0),K=0,l;(0,\eta),&K=0,l||(0,0),K=0,0\rangle = (-1)^\eta\sqrt{\frac{2l+1}{\text{dim}(\eta,0)}},\\
&\langle l m l -m|00\rangle = (-1)^{l-m}\sqrt{\frac{1}{2l+1}},
\label{Clebschbothalgebras}
\end{aligned}
\end{equation}
thus
\begin{equation}
\begin{aligned}
n_{(0,0),K=0,l=0,m=0}  = \sum_{l,m} (-1)^{\phi+\eta}\sqrt{\frac{1}{\text{dim}(\eta,0)}}
\left[b^{\dagger}_{(\eta,0),l}\widetilde{b}_{(0,\eta),l}\right]_{(0,0),K=0,l=0,m=0},
\label{numbercontinuation}
\end{aligned}
\end{equation}
where $\phi=\eta$ and $\sqrt{1/\text{dim}(\eta,0)}$ is a normalization factor.

In order to obtain the Racah form of the coupling between creation and destruction operators consider
\begin{equation}
\resizebox{0.9\hsize}{!}{$
\begin{aligned}
&\left(b^{\dagger}_{l}\widetilde{b}_{l'}\right)_{(\lambda,\lambda),K,L,q}= \sum_{\lambda,K}(-1)^{\eta}\langle(\eta,0),K=0,l;(0,\eta),K=0,l'||(\lambda,\lambda),K,L\rangle \left[b^{\dagger}_{(\eta,0),l}\widetilde{b}_{(0,\eta),l'}\right]_{(\lambda,\lambda),K,L,q},
\label{Racahexpansiontermsll}
\end{aligned}
$}
\end{equation}
where the coupling must be of the form $(\lambda,\lambda)$ because of the product
\begin{equation}
\begin{aligned}
&(\eta,0)\otimes(0,\eta) = (0,0)+(1,1)+(2,2)+...+(\eta-1,\eta-1)+(\eta,\eta).
\label{prdetazero}
\end{aligned}
\end{equation}

Notice that the $(1,1),K=1$ representation generates $l=1,2$ corresponding to the labels of the operators 
\begin{equation}
\begin{aligned}\hspace{5mm}L^{1}_q = \sum_{l}\sqrt{\frac{l(l+1)(2l+1)}{3}}\left(b^{\dagger}_{l}\widetilde{b}_{l}\right)^1_q,
\label{LRacah2appendix5}
\end{aligned}
\end{equation}

\begin{equation}
\begin{aligned}
 Q^{2}_q = &-\left(\eta+\frac{3}{2}\right)\sum_{l}\sqrt{\frac{l(l+1)(2l+1)}{5(2l-1)(2l+3)}}\left(b^{\dagger}_{l}\widetilde{b}_{l}\right)^2_q \\&+ \sum_{l}\sqrt{\frac{6(l+1)(l+2)(\eta-l)(\eta+l+3)}{5(2l+3)}}\left(\left(b^{\dagger}_{l}\widetilde{b}_{l+2}\right)^2_q  + \left(b^{\dagger}_{l+2}\widetilde{b}_{l}\right)^2_q  \right),
\label{QRaca4appendix5}
\end{aligned}
\end{equation}
where the following couplings must be computed
\begin{equation}
\resizebox{0.9\hsize}{!}{$
\begin{aligned}
&\left(b^{\dagger}_{l}\widetilde{b}_{l}\right)_{(1,1),K=1,l=1,q}= (-1)^{\eta}\langle(\eta,0),K=0,l;(0,\eta),K=0,l||(1,1),K=1,1\rangle \left[b^{\dagger}_{(\eta,0),l}\widetilde{b}_{(0,\eta),l}\right]_{(1,1),K=1,l=1,q},\\
&\left(b^{\dagger}_{l}\widetilde{b}_{l}\right)_{(1,1),K=1,l=2,q}= (-1)^{\eta}\langle(\eta,0),K=0,l;(0,\eta),K=0,l||(1,1),K=1,2\rangle \left[b^{\dagger}_{(\eta,0),l}\widetilde{b}_{(0,\eta),l}\right]_{(1,1),K=1,l=2,q},\\
\label{racahcoupledterms11}
\end{aligned}$}
\end{equation}
using analytic expressions found in \cite{10.1063/1.1666267} and \cite{kota20203}. This shows that angular momentum and quadrupole operators are $SU(3)$ tensors of kind $L_{(1,1),K=1,l=1,q}$ and $Q_{(1,1),K=1,l=2,q}$.

\chapter{Lie Algebras Software Implementation}

%%%%%%%%%%%%%%%%%%%%%%%%%%%%%%%%%%%%%%%%%%%%%%%%%%%%%%%%%%%%%%%%%%%%%%%%%%%%%%%%%%%%%%%%%%%%%%%%%%%%%%%%%%%%%

The software used in this thesis were Schur: An interactive programme for calculating properties of Lie groups and symmetric functions \cite{schurprogram}, UNTOU3 \cite{DRAAYER1989279}, UNtoU3 \cite{LANGR2019442} and SU3lib \cite{DYTRYCH2021108137}, all of them of public access. The UNTOU3 and UNtoU3 compute the irreps of $U(3)$ contained in an irrep of $U(\mathcal{N})$ exclusively for $\mathcal{N} = (\eta+1)(\eta+2)/2$ where $\eta = $1,2,3,4,5,6,7, that is, the branching problem $U((\eta+1)(\eta+2)/2)\supset SU(3)$. UNTOU3 code written in FORTRAN is more general than UNtoU3 since it allows irreps $[f_1,f_2,f_3,...,f_{\mathcal{N}}]$ for any value of $f_i$ and additionally prints the Gel’fand patterns \cite{iachello2014lie}, Casimir values and dimensions as well, while the UNtoU3 written in C++ only allows irreps of the form $[2^a,1^b,0^c]$ but has the advantage of a more robust architecture with the possibility of parallel computations which makes it a lot faster. A new software that combines the advantages of these two might be a necessity for nuclear physics. During the writing of this thesis an extension of UNtoU3 was developed by P. Fasano which admits irreps $[4^a, 3^b, 2^c,1^d,0^e]$ and can be found in \cite{UNtoU3improved}.
\begin{figure}[h!]
    \centering 
\includegraphics[width=0.7\textwidth]{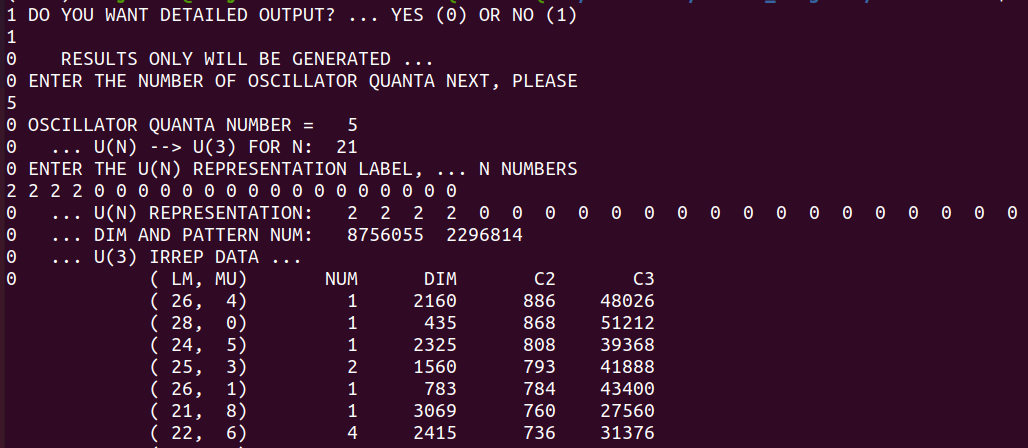}
\includegraphics[width=0.7\textwidth]{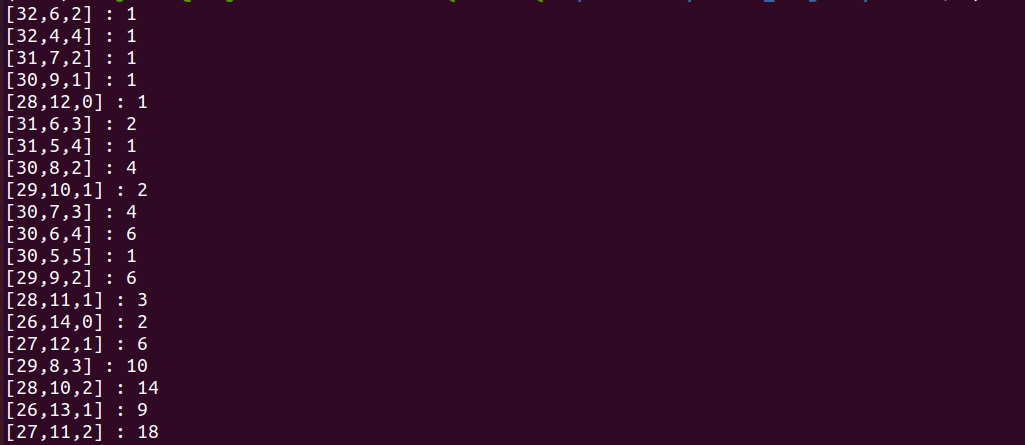}
    \caption{Above: Partial output of UNTOU3 software performing the irreps of $U(3)$ included in $[2^4]$ of $U(21)$. Below: Partial output of the same calculation of UNtoU3.}
    \label{UNTOU3}
\end{figure}

An example of their outputs is seen partially in figure \ref{UNTOU3}. Notice that UNTOU3 prints the $SU(3)$ irreps $(\lambda,\mu)$ obtained form $[f_1,f_2,f_3]$ of $U(3)$ as $(f_1-f_2,f_2-f_3)$ and requires to enter the $\mathcal{N}$ numbers separated by spaces while for UNtoU3 it must be specified in a .cpp file of which some examples are provided in the source of the code along with details on the parallelization of the computations. In order to organize the irreps obtained from UNtoU3 by decreasing value of Casimir operator a proper code must be written by the user.

\begin{figure}[h!]
    \centering 
\includegraphics[width=0.7\textwidth]{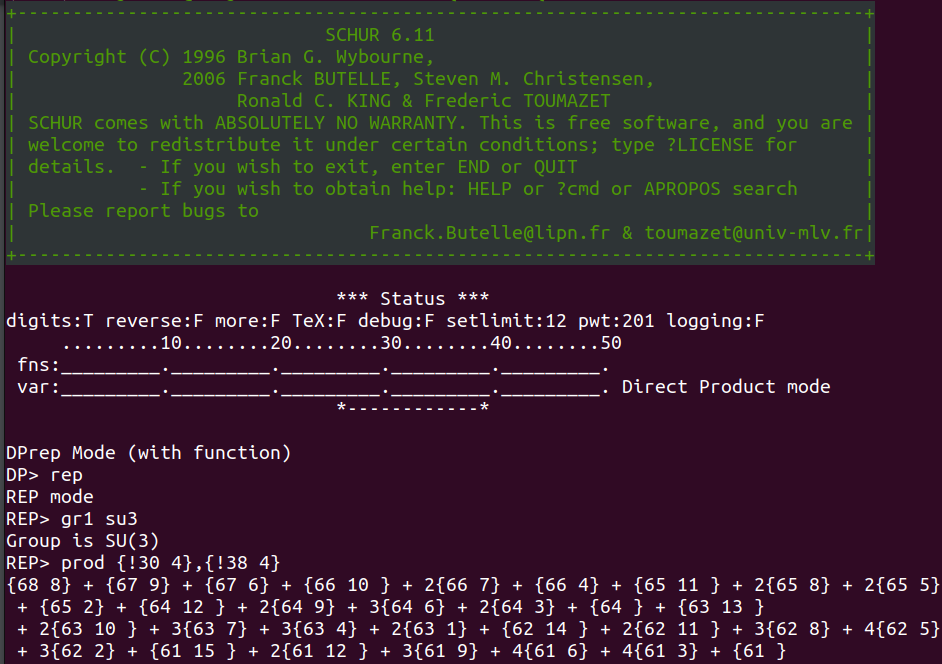}
    \caption{Partial output of calculation $(26,4)\otimes(34,4)$ in software Schur.}
    \label{Schur}
\end{figure}

Since only the leading irreps are considered for each proton $SU_\nu(3)$ and neutron $SU_\pi(3)$ spaces, to obtain the coupled irreps of $SU_{\pi+\nu}(3)$ it was used the Schur software in the current version 6.11. Among several many functionalities it provides, one of them is the possibility to calculate tensor (Kronecker) products between irreps of a same algebra. Schur has four modes of operation, by default it starts in direct product mode (DP) but the one of interest is the representation mode (REP) where operations on a single Lie algebra can be performed. Schur is case insensitive so in order to change to REP mode one can type the command "REP" or "rep" once the program starts. Then it requires the group to be specified which for $SU(3)$ can be done by the command "gr1 su3". The command to calculate tensor products is "prod". For the group $SU(3)$, the irreps must be entered between curly brackets and in $U(3)$ notation. Integers of the irreps must be separated by spaces and if it has two digits or more one must type it with a previous "!" character. As an example, for $(26,4)\otimes(34,4)$ it must be entered "prod \{!30 4\},\{!38 4\}" as shown in figure \ref{Schur}. The resulting irreps of $SU(3)$ are obtained by subtracting the second digit from the first one. The reader is encouraged to obtain and read the Schur program manual where its many uses are explained in detail.

As an illustration of how these three software can be applied, the 0$\hbar\omega$, 1$\hbar\omega$ and 2$\hbar\omega$ model space are computed for the isotope $^{220}$Ra. The core is $^{208}$Pb and possesses three quartets. These are described by the irrep $[2^3]_\pi\otimes[2^3]_\nu$ in $U(21)\otimes U(28)$ for the 0$\hbar\omega$ ground state. Using UNTOU3 or UNtoU3 it is found that the leader irrep of $U(21)\supset SU(3)_\pi$ contained in $[2^3]_\pi$ is $(24,0)_\pi$. Similarly, for $U(28)\supset SU(3)_\nu$ the leader irrep contained in $[2^3]_\nu$ is $(30,0)_\nu$. Now, coupling $SU(3)_\pi\otimes SU(3)_\nu$ irreps to $SU(3)_{\pi+\nu}$, the total model space is obtained. This is, $(24,0)_\pi\otimes(30,0)_\nu = ( 54 , 0 ) + ( 52 , 1 )+( 50 , 2 )+( 48 , 3 )+( 46 , 4 )+( 44 , 5 )+( 42 , 6 )+...$ The last product is performed by the Schur software. For 1$\hbar\omega$, the excitation of one proton corresponds to $\{[2^2,1]\otimes[1]\}_\pi\otimes[2^3]_\nu$ in $\{U(21)\otimes U(28)\}\otimes U(28)$ and taking its leader irreps under decomposition will result in $\{(20,1)\otimes(6,0)\}_\pi\otimes(30,0)_\nu$. Coupling the proton part and considering its leader only results in $(26,1)_\pi\otimes (30,0)_\nu$. For the excitation of one neutron it corresponds to $[2^3]_\pi\otimes\{[2^2,1]\otimes[1]\}_\nu$ in $U(21)\otimes \{U(28)\otimes U(36)\}$. Similarly, it results in $(24,0)_\pi\otimes\{(25,1)\otimes(7,0)\}_\nu$ and under coupling to  $(24,0)_\pi\otimes (32,1)_\nu$. The total model space for 1$\hbar\omega$ will be $(26,1)_\pi\otimes (30,0)_\nu + (24,0)_\pi\otimes (32,1)_\nu$ which after deleting spurious states it is obtained $( 56 , 1 )^2 + ( 54 , 2 )^2 + ( 55 , 0 )+( 52 , 3 )^2+( 50 , 4 )^2+( 48 , 5 )^2+( 46 , 6 )^2+...$ Last, for 2$\hbar\omega$ the possibilities are excitation of two protons corresponding to $\{[2^2]\otimes[2]\}_\pi\otimes[2^3]_\nu$, excitation of two neutrons corresponding to  $[2^3]_\pi\otimes\{[2^2]\otimes[2]\}_\nu$ and excitation of one proton and one neutron corresponding to $\{[2^2,1]\otimes[1]\}_\pi\otimes \{[2^2,1]\otimes[1]\}_\nu$. Repeating the same procedure results in $\{(16,2)\otimes(12,2)\}_\pi\otimes(30,0)_\nu$, $(24,0)_\pi\otimes\{(20,2)\otimes(14,0)\}_\nu$ and $\{(20,1)\otimes(6,0)\}_\pi\otimes\{(25,1)\otimes(7,0)\}_\nu$ respectively, whose total model space will be  $(28,2)_\pi\otimes (30,0)_\nu + (24,0)_\pi\otimes (24,2)_\nu + (26,1)_\pi\otimes (32,1)_\nu$. After deletion of spurious states the real model space will be $( 58 , 2 )^3+( 59 , 0 )+( 56 , 3 )^3+( 57 , 1 )^2+( 54 , 4 )^3+( 52 , 5 )^3+( 50 , 6 )^3+...$ This completes the model space for the ground and first excited states of $^{220}$Ra.

\begin{figure}[h!]
    \centering 
\includegraphics[width=0.8\textwidth]{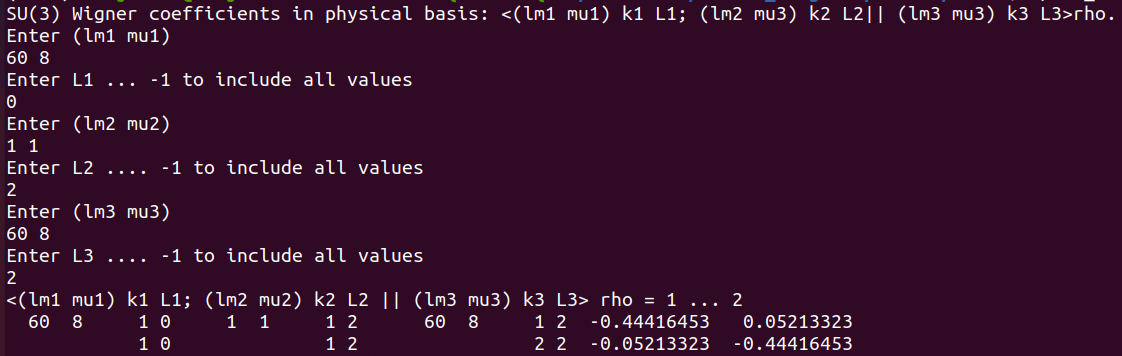}
    \caption{Calculation of coefficient $\langle(60,8),1,0; (11),1,2||(60,8),1,2\rangle_{\rho}$ in software SU3lib.}
    \label{SU3lib}
\end{figure}

The other computational necessity was the calculation of $SU(3)\supset SO(3)$ reduced Wigner coefficients required for calculating $B(E2)$ transitions in equation \ref{BE2SU3} which was solved by the $SU(3)$ library SU3lib, in particular the  module SU3\_SO3\_WignerCoeffs.cpp programmed in C++. The use of the program is very straightforward, once it starts it asks for the labels $(\lambda,\mu)$ and $L$ of the coefficient which after being entered fills the allowed labels $K$ and shows the values of the coefficients for each multiplicity coupling $\rho$. An example of the calculation of coefficient $\langle(60,8),1,0; (11),1,2||(60,8),1,2\rangle_{\rho}$ used in the electromagnetic transition of $^{224}$Th is shown in figure \ref{SU3lib}. The reader is again encouraged to obtain these free software and deepen his knowledge in their  many more uses than those exposed in this appendix.

\chapter{$B(E2)$ in Liquid Drop and $SU(3)$ Models}

%%%%%%%%%%%%%%%%%%%%%%%%%%%%%%%%%%%%%%%%%%%%%%%%%%%%%%%%%%%%%%%%%%%%%%%%%%%%%%%%%%%%%%%%%%%%%%%%%%%%%%%%%%%%%

In this section are presented the expressions for the reduced transition probability $B(E2)$ in the liquid drop and $SU(3)$ which are used in calculations on the main text. Both expressions must be coherent in their predictions and can be complemented in order to expand the understanding of electromagnetic interactions in atomic nuclei. 

The transition probability per unit time (also called transition rate) from an initial nuclear state $i$ to a final state $f$ is described by Fermi's golden rule and reads as \cite{suhonen,greiner1996nuclear}
\begin{equation}
\begin{aligned}
T_{i\xrightarrow{}f}^{\sigma\lambda\mu} = \frac{2}{\epsilon_0\hbar}\frac{\lambda+1}{\lambda((2\lambda+1)!!)^2}\left(\frac{E_{\gamma}}{\hbar c}\right)^{2\lambda+1}|\langle \Xi_f L_f M_f|\mathcal{M}_\mu^{\sigma\lambda}|\Xi_i L_i M_i\rangle|^2,
\end{aligned} 
\label{Tfermigoldenrule}
\end{equation}
where $\sigma = E,M$ to identify electric or magnetic transitions, 
 $\lambda,\mu$ represent the multipole component, $\mathcal{M}_\mu^{\sigma\lambda}$ the electromagnetic interaction operator, $E_{\gamma}$ the energy of the gamma ray emitted and the $\Xi $ represent any additional labels for the quantum state. No spin (or isospin) will be considered in this section so it will be assumed that $L = J$. By averaging over initial states, summing over final states and components $\mu$ it is obtained the expression   
\begin{equation}
\begin{aligned}
&T_{i\xrightarrow{}f}^{\sigma\lambda} = \frac{1}{2L_i+1}\sum_{M_f,\mu,M_i}T_{i\xrightarrow{}f}^{\sigma\lambda\mu} \\ &= \frac{2}{\epsilon_0\hbar}\frac{\lambda+1}{\lambda((2\lambda+1)!!)^2}\left(\frac{E_{\gamma}}{\hbar c}\right)^{2\lambda+1} \frac{|\langle \Xi_f L_f||\mathcal{M}^{\sigma\lambda}||\Xi_i L_i\rangle|^2}{2L_i+1} \sum_{M_f,\mu,M_i} |\langle L_i M_i\lambda\mu | L_f M_f
 \rangle|^2 \\ 
  & = \frac{2}{\epsilon_0\hbar}\frac{\lambda+1}{\lambda((2\lambda+1)!!)^2}\left(\frac{E_{\gamma}}{\hbar c}\right)^{2\lambda+1} B(\sigma\lambda; \Xi_i L_i\xrightarrow{}\Xi_f L_f),
\end{aligned} 
\label{Tsums}
\end{equation}
where Wigner-Eckart theorem without the factor $1/\sqrt{2L_f+1}$ convention was used and the reduced transition probability is defined as 
 \begin{equation}
\begin{aligned}
B(\sigma\lambda; \Xi_i L_i\xrightarrow{}\Xi_f L_f) = 
 \frac{2L_f+1}{2L_i+1} |\langle \Xi_f L_f||\mathcal{M}^{\sigma\lambda}||\Xi_i L_i\rangle|^2.
\end{aligned} 
\label{Redtransitionprobability}
\end{equation}

For the particular case of interest, the $B(E2)$ transition will involve the quadrupole operator $\mathcal{M}_\mu^{E2} = Q_\mu^{2}$ as
 \begin{equation}
\begin{aligned}
B(E2; \Xi_i L_i\xrightarrow{}\Xi_f L_f) = 
 \frac{2L_f+1}{2L_i+1} |\langle \Xi_f L_f||Q^{2}||\Xi_i L_i\rangle|^2,
\end{aligned} 
\label{BE2transitionprobability}
\end{equation}
where the mathematical form of $Q_\mu^2$ and its reduced matrix element will depend on the theoretical framework used. For the liquid drop model it will come from classical electrodynamics expression 
\begin{equation}
\begin{aligned}
Q_\mu^2 = \int d^3\mathbf{r}\hspace{1.5mm}r^2 Y_{2\mu}(\Omega) = \int \hspace{1.5mm}d\Omega Y_{2\mu}(\Omega)\int_0^{R(\Omega)} dr\hspace{1.5mm}r^4,
\end{aligned} 
\label{Qliquiddrop}
\end{equation}
where $R(\Omega)$ will be given by equation \ref{R2}. After performing the integral considering terms up to order two in deformation parameters $\alpha_{2\mu}$ it is obtained
\begin{equation}
\begin{aligned}
Q_\mu^2 = \frac{3ZR_0^2}{4\pi}\left(\alpha_{2\mu}-\frac{10}{\sqrt{70\pi}}[\alpha_2\times\alpha_2]_{2\mu}\right).
\end{aligned} 
\label{QliquiddropExpanded}
\end{equation}

The calculation of $\langle L_f||Q^{2}|| L_i\rangle$ in this framework \cite{greiner1996nuclear} for the ground state (g.s.) band results in the expression 
\begin{equation}
\begin{aligned}
B(E2; L_i\hspace{1.5mm} \text{g.s.}\xrightarrow{} L_f\hspace{1.5mm}\text{g.s.}) = 
\left( \frac{3ZR_0^2}{4\pi} \right)^2\left( 
 \beta+\frac{2}{7}\sqrt{\frac{5}{\pi}}\beta^2\right)^2\langle L_i020|L_f0\rangle^2,
\end{aligned} 
\label{BE2transition}
\end{equation}
where $R_0 = 1.2A^{1/3} \hspace{0.5mm}\text{fm}$, only matrix elements with respect to $SO(3)\supset SO(2)$ were considered and $\beta$ is the deformation parameter introduced in chapter 2. 

In the $SU(3)$ scheme the considered algebra chain is $SU(3)\supset SO(3)\supset SO(2)$ and its corresponding basis is $|(\lambda\mu)KLM\rangle$. The quadrupole operator is a $(11)K=1L=2$ tensor as shown in appendix E and $B(E2)$ for intraband transitions will be
 \begin{equation}
\begin{aligned}
B\big(E2; (\lambda\mu)KL_i\xrightarrow\ (\lambda\mu)KL_f\big) = 
 \frac{2L_f+1}{2L_i+1} |\langle (\lambda\mu)KL_f||Q^{(11)1,2}||(\lambda\mu)KL_i\rangle|^2,
\end{aligned} 
\label{BE2transitionprobabilitySU3}
\end{equation}
the reduced matrix element is given by \cite{casimirsSU3}
 \begin{equation}
\begin{aligned}
\langle (\lambda\mu)KL_f||Q^{(11)1,2}||(\lambda\mu)KL_i\rangle = (-1)^\phi\sqrt{4\mathcal{C}_2[SU(3)]}\langle(\lambda\mu)KL_i;(11)2||(\lambda\mu)KL_f\rangle_{\rho = 1},
\end{aligned} 
\label{reducedQSU3}
\end{equation}
where $\phi = 0$ for $\mu=0$ and $\phi = 1$ for $\mu\neq 0$, the reduced Wigner coefficients can be computed with the software \cite{su3lib}. Finally
 \begin{equation}
\begin{aligned}
B\big(E2; (\lambda\mu)KL_i\xrightarrow\ (\lambda\mu)KL_f\big) = 
 4\hspace{0.8mm}\frac{2L_f+1}{2L_i+1}\hspace{0.8mm}\mathcal{C}_2[SU(3)]\hspace{0.8mm}|\langle(\lambda\mu)KL_i;(11)2||(\lambda\mu)KL_f\rangle_{\rho = 1}|^2.
\end{aligned} 
\label{BE2SU3afterreduced}
\end{equation}

Since the chain $SU_{\pi}(3)\otimes SU_{\nu}(3)\supset SU_{\pi+\nu}(3)$ is considered for the modelling in this thesis, the quadrupole operator will be a coupling between proton and neutron quadrupole operators so a parameter $\alpha$ of the effective transition must be introduced \cite{SUN2002130} which is fitted to the the experimental value of the $2^+_1\xrightarrow{} 0^+_1$ transition \cite{CSEH2015213}. The resulting expression is
\begin{equation}
\begin{aligned}
B\big(E2; (\lambda\mu)KL_i\xrightarrow\ (\lambda\mu)KL_f\big) = 
 \hspace{0.8mm}\frac{2L_f+1}{2L_i+1}\hspace{0.8mm}\mathcal{C}_2[SU(3)]\hspace{0.8mm}\alpha^2|\langle(\lambda\mu)KL_i;(11)2||(\lambda\mu)KL_f\rangle_{\rho = 1}|^2.
\end{aligned} 
\label{YSUNBE2}
\end{equation}

%
% Acknowledgements (content in `acknowledgements.tex')
%%%%%%%%%%%%%%%%%%%%%%%%%%%%%%%%%%%%%%%%%%%%%%%%%%%%%%%%%%%%%%%%%
%%
%% use the starred version of the "acknowledgements" environment
%% to omit signatures from this section, e.g.:
%% \begin{acknowledgements*} ... \end{acknowledgements*}
%% 
%%%%%%%%%%%%%%%%%%%%%%%%%%%%%%%%%%%%%%%%%%%%%%%%%%%%%%%%%%%%%%%%%
\begin{acknowledgements}

I would like to thank all the people that in some way have helped me with the elaboration of this thesis. First to my advisor José Patricio Valencia for the many meetings along the two semesters in which this thesis was written. Thanks for your patience, time, willingness to teach me and help resolve the many questions I formulated about these interesting topics covered during this time. Thanks to my family for their support and company while I was an undergraduate. I am grateful that my parents Diego and Mónica encouraged me in attending university and helped me as much as possible so that I could pursue a career of my choosing. Special thanks to my sisters Carolina and Nataly for their support and help organizing the sections of the manuscript.  Special thanks to my younger brother Jerónimo for letting me have so many light-on night hours. Thanks to my university and its dedicated teachers for inspiring and helping me to be a better version of myself through education. Thanks to my friends and colleagues who listened to my ideas on various subjects and were eager to discuss them. Thanks to Diana for the great work in reconstructing the plots extracted from old books which improved the presentation of this thesis. Last but not least, special thanks to the authors J. Cseh, A. Martinou, D. Jenkins, P. van Isacker, J. G. Hirsch, V. K. B. Kota and F. Toumazet for their time, helpful discussions and material they shared with me via email which boosted the development of this work.

\end{acknowledgements}
	
%
%%%%% Bibliography/References (BiBTeX items in `references.bib')
%%%%	\bibliography{IEEEabrv,references}
%\bibliographystyle{jhep}
\bibliography{references}
% Abbreviations (content in `abbreviations.tex') 
% Glossary (content in `glossary.tex') 
	%\includeglossary{glossary}
%%%%%%%%%%%%%%%%%%%%%%%%%%%%%%%%%%%%%%%%%%%%%%%%%%%%
% Index
	%\printindices
%
%%%%%%%%%%%%%%%%%%
%%%%%%%%%%%%%%%%%%

%% Create CD labels:

	\definecdlabeloffsets{0}{-0.65}{0}{0.55} % upper label x offset [cm] (default=0) /  upper label y offset [cm] (default=0) /  lower label x offset [cm] (default=0) /  lower  label y offset [cm] (default=0) -- For Q-Connect KF01579 labels use the following offset values: {0}{-0.65}{0}{0.55}

	\createcdlabel{Structure of Heavy Nuclei Based on
Nucleon Quartet Condensates
 \\ Algebraic Approach in the Proxy-SU(3) Scheme}{Alejandro Restrepo Giraldo}{January}{2023}{1} % thesis title / author name / month / year / title area width in cm (recommended value: 8) 

%%
%% Create CD case covers:

	\createcdcover{Structure of Heavy Nuclei Based on
Nucleon Quartet Condensates
 \\ Algebraic Approach in the Proxy-SU(3) Scheme}{Alejandro Restrepo Giraldo}{January}{2023}{1} % thesis title / author name / month / year / title area width in cm (recommended value: 10) 

\end{document}